\documentclass[a4,10pt]{article}
\usepackage{graphicx}
\usepackage{natbib}
\usepackage{psfrag}
\usepackage{bm}
\usepackage{color}
\usepackage{amsmath}
\usepackage{amssymb}
\usepackage{amsfonts}
\usepackage{fancyhdr}
\usepackage{float}

\usepackage{wasysym}
\usepackage{multirow}
\usepackage{array}
\usepackage{mathrsfs}
\providecommand\bnabla{\boldsymbol{\nabla}}
\providecommand\bcdot{\boldsymbol{\cdot}}

\newcommand\Rey{\mbox{\textit{Re}}}  
\newcommand\She{\textit{Sh}}  

\newcommand\etal{\mbox{\textit{et al.}}}

\newcommand\eg{e.g.}
\newcommand\ie{i.e.}

\newsavebox{\astrutbox}
\sbox{\astrutbox}{\rule[-5pt]{0pt}{20pt}}

\newcommand{\twod}{two-dim\-en\-sion\-al}
\newcommand{\Twod}{Two-dim\-en\-sion\-al}
\newcommand{\threed}{three-dim\-en\-sion\-al}
\newcommand{\Threed}{Three-dim\-en\-sion\-al}
\newcommand{\Fourier}{Fou\-rier}
\renewcommand{\exp}[1]{\mathrm{e}^{#1}}

\newcommand{\vect}[1]{\bm{#1}}

\newcommand{\pde}[2]{\frac{\partial #1}{\partial #2}}

\newcommand{\BEQ}{\begin{equation}}
\newcommand{\EEQ}{\end{equation}}
\newcommand{\BER}{\begin{eqnarray}}
\newcommand{\EER}{\end{eqnarray}}
\newcommand{\BEQN}{\begin{equation*}}
\newcommand{\EEQN}{\end{equation*}}
\newcommand{\BERN}{\begin{eqnarray*}}
\newcommand{\EERN}{\end{eqnarray*}}

\newcommand{\abs}[1]{\left|#1\right|}

\makeatletter
\newcommand*{\rom}[1]{\expandafter\@slowromancap\romannumeral #1@}
\makeatother

\begin{document}

\fancyhf{}
\pagestyle{fancy}
\rhead{\textit{Stability of flow around a 180-degree bend}}
\lhead{\textit{A.\ M.\ Sapardi \etal}}

\title{Linear stability of confined flow around a 180-degree sharp bend}
%
%
%
\author{Azan M.\ Sapardi$^{1,2}$, Wisam K.\ Hussam$^{1,3}$, Alban Poth\'{e}rat$^4$ \\
and Gregory J.\ Sheard$^1$\\ 
$^1$The Sheard Lab, Department of Mechanical and Aerospace Engineering,\\
                Monash University, Victoria 3800, Australia\\
 Greg.Sheard@monash.edu\\
             $^2$Department of Mechanical Engineering, 
International Islamic University Malaysia,\\ Kuala Lumpur 53300, Malaysia\\
             $^3$School of Engineering, Australian College of Kuwait,\\ Safat 13015, Kuwait\\
             $^4$Applied Mathematics Research Centre,\\
             Coventry University, Coventry CV1 5FB, United Kingdom}

%
\maketitle
%
\begin{abstract}
This study seeks to characterise the breakdown of the steady \twod solution in the flow around a 180-degree sharp bend to infinitesimal \threed\ disturbances using a linear stability analysis. The stability analysis predicts that \threed\ transition is via a synchronous instability of the steady flows. A highly accurate global linear stability analysis of the flow was conducted with Reynolds number $\Rey<1150$ and bend opening ratio (ratio of bend width to inlet height) $0.2\leq\beta\leq5$. This range of $\Rey$ and $\beta$ captures both steady-state \twod\ flow solutions as well as the inception of unsteady \twod\ flow. For $0.2\leq\beta\leq1$, the \twod\ base flow transitions from steady to unsteady at higher Reynolds number as $\beta$ increases. The stability analysis shows that at the onset of instability, the base flow becomes three-dimensionally unstable in two different modes, namely spanwise oscillating mode for $\beta=0.2$, and spanwise synchronous mode for $\beta \geq 0.3$. The critical Reynolds number and the spanwise wavelength of perturbations increase as $\beta$ increases. For $1<\beta\leq2$ both the critical Reynolds for onset of unsteadiness and the spanwise wavelength decrease as $\beta$ increases. Finally, for $2<\beta\leq5$, the critical Reynolds number and spanwise wavelength remain almost constant. The linear stability analysis also shows that the base flow becomes unstable to different \threed\ modes depending on the opening ratio. The modes are found to be localised near the reattachment point of the first recirculation bubble.
\end{abstract}

\section{Introduction}\label{sec:introduction}
An important geometric feature of the ductwork carrying liquid-metal coolant fluid within prototype blankets in magnetic confinement fusion reactors is the presence of sharp 180-degree bends \citep{boccaccini2004materials,kirillov1995present,barleon1991mhd,barleon1996mekka,buhler2007liquid}. The transport of heat from the far side-wall of the bend and the flow around the bend are critical aspects for the efficient transport of heat from the reactor for power generation \citep{boccaccini2004materials}. Despite its relative simplicity, few studies have been conducted on the hydrodynamic flow in this geometry, and they have focused mostly on heat transfer. An understanding of this flow underpins the duct flow problem with application to fusion reactor blankets.

Fundamentally, the \twod\ flow around the sharp bend creates recirculation structures that resemble those seen in several canonical confined flow problems, including the backward-facing step flow \citep{armaly1983experimental,ghia1989analysis,barkley2002three,blackburn2008convective}.  These flows exhibit a streamline emerging from the upstream separation point that divides regions of reversed flow from the main bulk flow, and is termed the dividing streamline. When the dividing streamline reattaches to the wall downstream, a closed separation bubble is formed. Specifically, the flow first passes over a large recirculation bubble attached to the downstream side of the inner corner of the bend, and under same conditions, subsequently a second recirculation bubble develops on the opposite wall a little further downstream.  Separating and reattaching flows play an important role in numerous engineering applications, especially in heat transportation \citep{krall1966turbulent,abu2008application,larson2012heat}.

There have been a number of studies on the effect of different geometric parameters on the efficiency of heat transfer in flows around a bend. The effects of a duct with a 180-degree bend with three different turning configurations: sharp corner, rounded corner and circular turn were studied by \citet{wang1994heat}. In their study, all walls were heated and cold fluid was supplied from the inlet. Their results show that the sharp bend had the strongest turn-induced heat transfer enhancement by approximately 30\% as compared to the circular turn which has the weakest among those three configurations. Heat transfer was found to be optimum downstream of the sharp bend turn (experimental studies by \citet{metzger1986heat} and \citet{astarita2000thermofluidynamic}). In an experimental study on the effect of the size of bend openings in ducts with sharp bend by \citet{hirota1999heat}, with a channel cross-section of $50\times 25$ mm, bend openings of 70 mm and 50 mm were found to have almost the same value of Sherwood number $\She$ (the ratio of convective to diffusive mass transport), whereas a smaller bend opening ($30$~mm) led to values of $\She$ $1.3$ times larger than that for $50$~mm. Only a few studies focused on the effect of controllable parameters on the structure of the flow. \citet{liou1999fluid,liou2000non} experimentally showed that the divider thickness (the thickness of the structure between the inflow and outflow channels) mainly influenced the intensity and uniformity of turbulence.

The only reported parametric study on the effect of the bend opening ratio on hydrodynamic flow around a 180-degree sharp bend was carried out numerically by \citet{zhang2013influence}. They categorised the \twod\ flow into five regimes \cite[see figure~4 in][]{zhang2013influence}. Regime I was at very low $\Rey$, where the flow was laminar and remained attached to the walls throughout. Regime II emerged at higher $\Rey$, where flow separation appears at the bend, leading to the creation of the primary recirculation bubble. Regime III occurred at a yet higher Reynolds number when the adverse pressure gradient at the top wall was strong enough to create a secondary recirculation bubble there. Regime IV occurred when a small scale vortex structure was detected far downstream at high $\Rey$ before the flow entered regime V, which featured vortex shedding originating from the sharp bend at higher $\Rey$. They also reported the results of  some \threed\ simulations. According to their results, at $\Rey=2000$, two types of shedding structures were found in the spanwise direction reminiscent of A- and B-modes in the flow past a circular cylinder \citep{williamson1988defining,thompson1996three,brede1996secondary,henderson1997nonlinear}. These shedding structures disrupted the \twod\ vortex in the flow and tended to slow down the shedding mechanism. Though a very detailed study of the \twod\ $180$-degree bend flow has been provided by \citet{zhang2013influence}, the \threed\ stability of these flows is yet to be determined.

One approach towards understanding the \threed\ stability relies on a linear stability analysis, where the stability of infinitesimal \threed\ perturbations to a \twod\ flow is determined by obtaining the leading eigenmode(s) of the evolution operator of the linearised perturbation field. Combined with accurate numerical methods, this technique has been significantly contributing to a better understanding of separated flows in complex geometries over the past couple of decades. Relevant examples include backward-facing step  \citep{barkley2002three,blackburn2008convective}  and partially blocked channel flows \citep{griffith2008steady}.

Studies elucidating the stability and the three-dimensionality of the flow past a back\-ward-facing step include \citet{armaly1983experimental,ghia1989analysis,barkley2002three,wee2004self,griffith2007wake} and \citet{lanzerstorfer2012global}. Based on detailed experiments, \citet{ghia1989analysis} and \citet{armaly1983experimental} initially proposed that the
curvature in the main flow caused by the second recirculation bubble led to a Taylor--G\"{o}rtler instability. This type of instability occurs in flows with curved streamlines when the fluid velocity decreases radially, and the centrifugal force drives pairs of counter-rotating streamwise vortices \cite[][]{drazin2004hydrodynamic}. However, the linear stability analysis of \citet{barkley2002three} ruled out the effect of a Taylor--G\"{o}rtler-type instability because they found that the \twod\ flow remained linearly stable long after the secondary recirculation bubble appeared. Instead, they found the critical eigenmode to consist of a flat roll localised in the primary recirculation region at the step edge. \citet{hammond1998local} studied the instability properties of separation bubbles. They found that the instability mode associated to the inflection point of the dividing streamline became globally unstable as the peak backflow velocity approached about $30\%$ of the free stream value. As the size of the bubble grew, the peak velocity of the backflow increased correspondingly and the conditions for local absolute instability could be predicted.

No such scenario has been established for the onset of unsteadiness of flows in $180$-degree sharp bends. The purpose of the present paper is to find such a scenario using methods in the spirit of \citet{barkley2002three}. The specific aim is to thoroughly characterise the \threed\ stability of flow around a $180$-degree sharp bend as a function of Reynolds number, bend opening ratio, and spanwise wavenumber of the \threed\ disturbances. In turn it is expected that this study will provide insights into a more general separated confined flows, and this understanding may enable future enhancement of the efficiency of heat transport in such systems.

This paper is organised as follows. Problem formulation and numerical methods are presented in \S~\ref{sec:problem_formulation} and \S~\ref{sec:computational_methods}, respectively. In \S~\ref{sect:2D_flow} and \S~\ref{sec:LSA}, respectively, the base flow characteristics and the results of the stability analysis for a range of $\beta$ and $\Rey$ are discussed. Finally, the nature of bifurcation in the \threed\ flow is studied in \S~\ref{sect:3D_flow}.

\section{Problem formulation}\label{sec:problem_formulation}
\begin{figure}
\centering
  \includegraphics[width=0.65\textwidth,keepaspectratio]{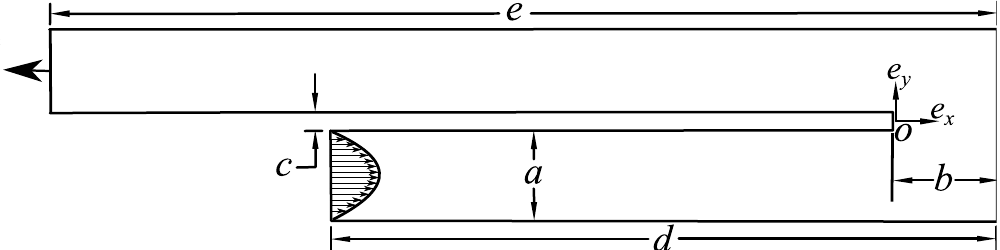}
  \caption{Flow geometry for the 180-degree sharp bend system. The fluid enters the bottom channel flowing to the right, and exits the top channel flowing leftwards.}
\label{fig:geometry2}
\end{figure}

Figure~\ref{fig:geometry2} shows the computational domain under consideration, including the geometric parameters for the problem. The channel widths in the inlet and at the bend are $a$ and $b$, respectively. The heights of the inlet and outlet channels are identical. The divider thickness is $c$, with $d$ and $e$ respectively denoting the lengths from the far wall of the bend to the inlet and outlet, respectively. The ratio of the gap $c$ to the channel height $a$ is $4\%$, while the lengths of the upstream and downstream channels are $(d-b) = 15a$ and $(e-b) = 30a$. The opening ratio of the bend is defined as $\beta = b/a$.

The fluid has a constant density $\rho$ and kinematic viscosity $\nu$. In this study, velocities are normalised by the peak inlet velocity $U_o$, lengths by inlet channel height $a$, time by $a/U_o$ and pressure by $\rho U_o^2$. The fluid motion is governed by the incompressible Navier--Stokes equations,
\begin{gather}
\pde{\vect{u}}{t} = \vect{\mathrm{N}}(\vect{u}) - \bnabla p + \frac{1}{\Rey}\nabla^2 \vect{u},\label{eq:NS_momentum}\\
\bnabla\bcdot\vect{u} = 0,\label{eq:NS_continuity}
\end{gather}
where $\vect{u}$ is the velocity field, $p$ is the pressure, the Reynolds number
\BEQ
\Rey\equiv\frac{U_o a}{\nu}{,}
\EEQ
and the non-linear advection term is calculated in convective form as $\vect{\mathrm{N}}(\vect{u})\equiv -(\vect{u}\bcdot\bnabla)\vect{u}$.

Fluid enters from the inlet, flows around the sharp bend and through the outlet channel. With the origin of a Cartesian coordinate system positioned at the mid-point of the divider surface at the inside of the bend (see figure~\ref{fig:geometry2}), boundary conditions are imposed as follows: at the inlet ($x = b-d, -1.02 \leq y \leq -0.02$), a Poiseuille velocity profile $u_x = 1 - 4(y + 0.52)^2, u_y = 0, u_z = 0$ is imposed. A no slip boundary condition ($\vect{u}=0$) is imposed at all solid walls. At the outlet ($x = b - e, 0.02 \leq y \leq 1.02$), a standard outflow boundary is enforced with a Dirichlet reference pressure ($p=0$) and a weakly enforced zero normal velocity gradient \citep{barkley2008direct}.

\section{Computational methods}\label{sec:computational_methods}
The governing equations are spatially discretised using a spectral-element method and time integrated using a third-order backward differentiation scheme \citep{karniadakis1991high}. The \twod\ Cartesian formulation of the present code has been validated and employed in several confined channel-flow problems \cite[\eg\ ][]{neild2010swirl,hussam2012enhancing,hussam2012optimal}. The method used to analyse the linear stability of perturbations is based on time integration of the linearized Navier--Stokes equations following \citet{barkley1996three} and references therein. The technique is in fact a Floquet problem for time-periodic base flows, but is applied to the steady-state base flows in the present problem for convenience, as it was already implemented and validated within our code.

Velocity and pressure fields are decomposed into a \twod\ base flow and infinitesimal fluctuating disturbance components
\begin{gather}
\vect{u} = \vect{u}_\textrm{2D} + \vect{u}^\prime,\label{eq:velocity_decomposed}\\
p = p_\textrm{2D} + p^\prime \label{eq:pressure_decomposed}.
\end{gather}
Substituting equations (\ref{eq:velocity_decomposed}) and (\ref{eq:pressure_decomposed}) into (\ref{eq:NS_momentum}, \ref{eq:NS_continuity}) and retaining terms in the first order of the perturbation field yields the linearised Navier--Stokes equations describing the evolution of infinitesimal \threed\ disturbances,
\begin{gather}
\pde{\vect{u}^\prime}{t} = -\vect{\mathrm{DN}}(\vect{u}^\prime) - \bnabla p^\prime + \frac{1}{\Rey}\nabla^2 \vect{u}^\prime,\label{eq:linearized_NS_momentum}\\
\bnabla\bcdot\vect{u}^\prime = 0,\label{eq:linearized_NS_continuity}
\end{gather}
where we calculate the linearised advection term as $\vect{\mathrm{DN}}(\vect{u}^\prime) = (\vect{u}_\textrm{2D}\bcdot\bnabla)\vect{u}^\prime + (\vect{u}^\prime\bcdot\bnabla)\vect{u}_\textrm{2D}$.

Since the base flow is invariant in the spanwise direction, we can decompose general perturbations into \Fourier\ modes with spanwise wavenumber
\BEQ
k = \frac{2 \pi}{\lambda},
\EEQ
where $\lambda$ is the wavelength in the spanwise direction. As per equation (\ref{eq:linearized_NS_momentum}, \ref{eq:linearized_NS_continuity}), the equations are linear in $\vect{u}^\prime$ and therefore \Fourier\ modes are linearly independent and coupled only with the \twod\ base flow. The absence of any spanwise component to the base flow further permits a single phase of the complex \Fourier\ mode to be considered, \ie
\BEQ\label{Ch1:Eq-var-four-exp}
    \left.
        \begin{array}{rcl}
            \vect{u}^\prime(x, y, z, t)&=&\left<\hat{u} (x, y, t) \cos (kz),\hat{v} (x, y, t) \cos (kz),\hat{w} (x, y, t) \sin (kz)\right>\\
            p^\prime(x, y, z, t)&=&\hat{p} (x, y, t) \cos (kz)
        \end{array}
    \right\}.
\EEQ
The flow stability therefore reduces to a three-parameter problem in $\Rey$, $\beta$ and $k$. Following \citet{barkley1996three} and others, spanwise phase locked perturbations of the form in equation (\ref{Ch1:Eq-var-four-exp}) remain in this form under the linearised evolution equations
 (\ref{eq:linearized_NS_continuity})-(\ref{eq:linearized_NS_momentum}): for a $\Rey$, $\beta$ and spanwise wavenumber $k$, the \threed/three-component perturbation field $\vect{u}^\prime(x,y,z,t)$ then reduces to a \twod/three-component field
\BEQ
\vect{\hat{u}}(x,y,t)=\left<\hat{u}(x,y,t), \hat{v}(x,y,t), \hat{w}(x,y,t)\right>,
\EEQ
which is computed on the same \twod\ domain as the base flow. The perturbation $z$-velocity features a sine function rather than a cosine as only $z$-derivatives of this term (yielding a cosine) interact with other terms under (\ref{eq:linearized_NS_continuity})-(\ref{eq:linearized_NS_momentum}).

By defining $\mathscr{A}(\tau)$ to represent the linear evolution operator for time integration (via equations~(\ref{eq:linearized_NS_momentum}) and~(\ref{eq:linearized_NS_continuity})) of a perturbation field comprising a single phase-locked spanwise \Fourier\ mode $\vect{\hat{u}}$ over time interval $\tau$, \ie
\BEQ
\vect{\hat{u}}(t+\tau) = \mathscr{A}(\tau)\vect{\hat{u}},
\EEQ
an eigenvalue problem may then be constructed as
\BEQ
\label{eq:LSA-eig-prob}
\mathscr{A}(\tau)\vect{\hat{u}}_{k} = \mu_{k}\vect{\hat{u}}_{k},
\EEQ
having complex eigenvalues $\mu_{k}$ and eigenvectors $\vect{\hat{u}}_k$. Eigenvalues $\mu_k$ are Floquet multipliers that relate to the eigenmode's exponential growth rate $\sigma$ and angular frequency $\omega$ through
\BEQ
\label{eq:LSA-Floq-mult-defn}
\mu \equiv e^{(\sigma + \mathrm{i}\omega)\tau},
\EEQ
where the subscripts have been omitted for clarity. As the base flows are time-invariant in this study, the usual time period is replaced by an arbitrary time interval for $\tau$. An appealing feature of this technique is that solutions to the eigenvalue problem (\ref{eq:LSA-eig-prob}) may be obtained using iterative methods involving time-integration of the linearised perturbation field via (\ref{eq:linearized_NS_momentum})-(\ref{eq:linearized_NS_continuity}), which avoids the substantial cost of explicitly constructing the very large operator $\mathscr{A}(\tau)$.

Stability is dictated by the leading eigenmode (\ie\ $\mu = \mu_k$ having largest $\left|\mu_k\right|$). Neutral stability corresponds to $\abs{\mu}=1$, while $\left|\mu\right| > 1$ and $\left|\mu\right| < 1$ describe unstable and stable flows, respectively. The bifurcation may be either synchronous ($\omega=0$) or oscillatory ($\omega\neq 0$). The smallest Reynolds number for which any spanwise wavenumber $k$ yields $\abs{\mu}=1$ is the critical Reynolds number for the onset of instability.

The following steps are taken in order to solve this problem numerically. The time invariant base flow at a given $\Rey$ and $\beta$ is obtained by solving the \twod\ Navier--Stokes equations (\ref{eq:NS_momentum})-(\ref{eq:NS_continuity}) and stored. Subsequently, random initial perturbation fields are constructed for one or more spanwise wavenumbers $k$, and an implicitly restarted Arnoldi method in conjunction with time integration of the linearised Navier--Stokes equations (\ref{eq:linearized_NS_momentum})-(\ref{eq:linearized_NS_continuity}) is used to determine the leading eigenmodes governing stability. The ARPACK \cite[][]{ARPACK1998} implementation of the implicitly restarted Arnoldi method is used, and the present formulation has been validated and employed across \cite{SheardFitzgeraldRyan2009, Sheard2011, VoMontaboneSheard2014, VoMontaboneSheard2015}.

\subsection{Test of base flow structure computations}
\begin{figure}
\centering
  \includegraphics[width=0.6\textwidth,keepaspectratio]{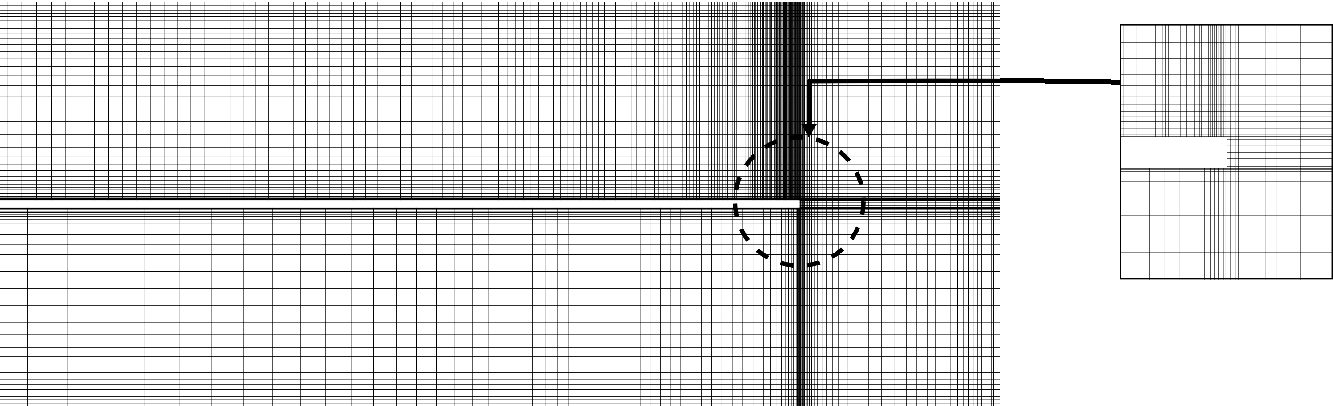}
  \caption{Details of the mesh around turning part area with polynomial order $N=3$.}
\label{fig:mesh}
\end{figure}

The flow past a 180-degree sharp bend is a deceptively difficult problem to fully resolve, especially at large Reynolds number due to the sensitivity of the solution to the mesh structure mainly near the sharp bend. This section describes the tests used to validate the numerical algorithm, and to select appropriate meshes and element order. For the spatial resolution study, we varied element polynomial degree from $N=4$ to $N=8$ of a mesh based on domain length parameters from the mesh domain. For consistency with the domain size study, the mesh employed in this study models a 180-degree sharp bend with opening ratio $\beta = 1$ and $\Rey=500$. In this regime, the flow is steady, with two recirculation bubbles. Figure~\ref{fig:mesh} shows the detail of the mesh with $N=3$. The mesh is structured and refined in the vicinity of the sharp bend as well as in the downstream channel in order to capture the detailed structure of the flow that passes around the bend.

\begin{table}
\begin{tabular}{>{\centering\arraybackslash} >{\centering\arraybackslash}p{0.7cm} >{\centering\arraybackslash}p{0.7cm} >{\centering\arraybackslash}p{1.15cm} >{\centering\arraybackslash}p{1.2cm} >{\centering\arraybackslash}p{1.2cm} >{\centering\arraybackslash}p{1.5cm} >{\centering\arraybackslash}p{1.5cm} >{\centering\arraybackslash}p{1.8cm} >{\centering\arraybackslash}p{1.8cm}}
$N$ & A & B & C & D & $L_\mathrm{R_1}$ & $L_\mathrm{R_2}$ & $\%L_\mathrm{R_1}$ & $\%L_\mathrm{R_2}$ \\[6pt]
  4   & 0   & -4.8057 & -3.7077 & -9.7726 & 4.80573 & 6.06489 & 0.0307 & 0.0797 \\
  5   & 0   & -4.8072 & -3.7069 & -9.7765 & 4.80716 & 6.06955 & 0.0009 & 0.0029 \\
  6   & 0   & -4.8073 & -3.7071 & -9.7766 & 4.80730 & 6.06962 & 0.0019 & 0.0018 \\
  7   & 0   & -4.8073 & -3.7070 & -9.7766 & 4.80725 & 6.06966 & 0.0009 & 0.0010 \\
  8   & 0   & -4.8072 & -3.7069 & -9.7767 & 4.80720 & 6.06972 & --- & --- \\
\end{tabular}
\caption{Dependence of recirculation length on polynomial order. Parameter $N$ indicates the independent polynomial order of the base flow. Two separation points (A and C) and two reattachment points (B and D) as indicated in figure~\ref{fig:recirculation_points} computed on the mesh at $\Rey=500$ and $\beta=1$ are given. $L_\mathrm{R_1}$ and $L_\mathrm{R_2}$ represent the recirculation length of the first and the second bubble, respectively. Errors on bubble lengths at each $N$ relative to the highest $N$ are also provided.} \label{tab:grs_recirculation_length}
\end{table}
\begin{figure}
\centering
  \includegraphics[width=0.8\textwidth,keepaspectratio]{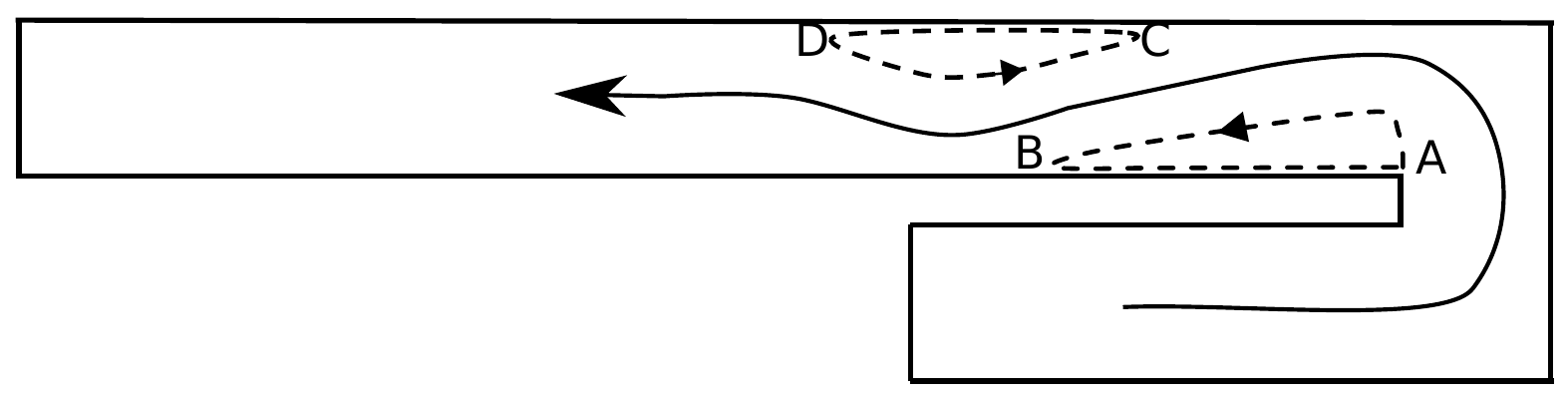}
  \caption{Sketch of separation and re-attachment points defining the locations of all recirculations. The unbroken arrow represents the direction of the main bulk flow as it navigates the bend.}
\label{fig:recirculation_points}
\end{figure}

To demonstrate the accuracy of computing recirculation length in the base flow, table~\ref{tab:grs_recirculation_length} shows the relative error of several measured quantities as a function of polynomial order. From computation of error on the recirculation length (Table \ref{tab:grs_recirculation_length}), it was found that the polynomial order $N=5$ provides a good accuracy to run the base flow computations. To examine the effect of downstream channel length on the solutions, a convergence study was conducted on the lengths of the first recirculation bubble on the bottom wall of the downstream channel and the secondary recirculation bubble on the top wall. The results shown in Table \ref{tab:domain_study} demonstrate that an outlet length of $10$ results in an error in the determination of the primary recirculation bubble length of approximately $0.007\%$, and approximately $4.5\%$ for the secondary bubble. The relatively large error in the size of the secondary bubble is caused by its proximity to the outlet. Outlet lengths of $20$ to $100$ are required to achieve at least $5$ significant figures of accuracy. Hence, an outlet length of $30$ is considered adequately long to be used throughout this study. \citet{barton1997entrance} and \citet{cruchaga1998study} studied the entrance effect for backward-facing step flow with expansion ratio of 2 and found that the inlet length of $10h$ and $2h$ (where $h$ is the step height), respectively, return slightly different numerical solutions compared to that with zero inlet length. In this study, the inlet length is $15$ which is adequately long for the velocity flow to be fully developed before reaching the bend.

\begin{table}
\centering
\begin{tabular}{>{\centering\arraybackslash} >{\centering\arraybackslash}p{2.7cm} >{\centering\arraybackslash}p{2cm} >{\centering\arraybackslash}p{2cm} >{\centering\arraybackslash}p{2cm} >{\centering\arraybackslash}p{2cm} >{\centering\arraybackslash}p{2cm}}
  $\mathrm{Outlet ~length}~(e-b)$ & $L_\mathrm{R_1}$ & $L_\mathrm{R_2}$ & $\%L_\mathrm{R_1}$ & $\%L_\mathrm{R_2}$ \\[6pt]
  10                & 4.80485494           & 5.79491686           & 0.00723577          & 4.482661 \\
  20                & 4.80452028           & 6.06684499           & 0.00027037          & 0.000483 \\
  30                & 4.80452020           & 6.06685749           & 0.00026858          & 0.000277 \\
  40                & 4.80452009           & 6.06685810           & 0.00026644          & 0.000267 \\
  50                & 4.80451921           & 6.06685923           & 0.00024807          & 0.000248 \\
  60                & 4.80451788           & 6.06686092           & 0.00022029          & 0.000220 \\
  70                & 4.80451875           & 6.06685982           & 0.00023848          & 0.000239 \\
  80                & 4.80451505           & 6.06686448           & 0.00016141          & 0.000161 \\
  90                & 4.80450991           & 6.06687100           & 0.00005442          & 0.000054 \\
  100               & 4.80450729           & 6.06687431           & ---                 & --- \\
\end{tabular}
\caption{Dependence of the length of the primary and secondary recirculation bubbles ($L_{\mathrm{R_1}}$ and $L_{\mathrm{R_2}}$, respectively) on outlet channel length $(e-b)$. Percent differences between bubble lengths at each $e-b$ relative to the longest-outlet case $e-b=100$ are also provided. Outlet lengths of $e-b=20$ and higher capture the bubble lengths to a precision of at least $5$ significant figures.} \label{tab:domain_study}
\end{table}

\subsection{Test of eigenvalue computations}
The precision of eigenvalue $\mu$ and eigenmode $\vect{\hat{u}}$ produced by subspace iteration were quantified by the residual
\BEQ
r = \parallel \mathscr{A}\vect{\hat{u}} - \mu \vect{\hat{u}}\parallel{,}
\EEQ
where $\parallel \cdot \parallel$ was the standard vector norm and where eigenmodes were normalised ($\parallel\vect{\hat{u}}\parallel = 1$). The linear stability analysis technique relied on an iterative process to obtain the leading eigenvalues and eigemodes of the system. The process ceased when $r< 10^{-7}$ was achieved. Nevertheless, the eigenmodes are also resolution-dependent. Table~\ref{tab:grs_eigenvalue} reports the accuracy of the eigenvalue computations as a function of element polynomial degree $N$. The leading eigenvalue for $\Rey=500$, $\beta=1$ and $k=6.4$ is real and linearly stable. It is found that the eigenvalue converged to an error of merely $0.0003180\%$ at $N=5$, which is employed hereafter.

\begin{table}
\centering
\begin{tabular}{ccc}

$N$ & $\left|\mu_\textrm{max}\right|$ & Relative error \\[6pt]
  4   & 0.9941048 & 0.0143794\% \\
  5   & 0.9942509 & 0.0003180\% \\
  6   & 0.9942458 & 0.0001927\% \\
  7   & 0.9942469 & 0.0000837\% \\
  8   & 0.9942477 & - \\

\end{tabular}
\caption{Dependence of leading eigenvalues on polynomial order. Parameter $N$ indicates the independent polynomial order of the base flow. Leading eigenvalues computed on the mesh at $\Rey=500$, $\beta=1$ and spanwise wavenumber $k=6.4$ are provided. The relative error is to the highest polynomial order case ($N=8$). Given eigenvalues are real.} \label{tab:grs_eigenvalue}
\end{table}

\subsection{Code validation}
Finally, the present model is validated against published \twod\ simulations providing the position and length of the recirculation bubble. In a viscous flow (featuring boundary-layers adjacent to no-slip surfaces), the point of reattachment can be precisely measured by finding the location where the wall shear stress $\tau_\textrm{wall}=-\mu\partial u/\partial y$ is zero.

\begin{figure}
\centering
  \includegraphics[width=0.6\columnwidth,keepaspectratio]{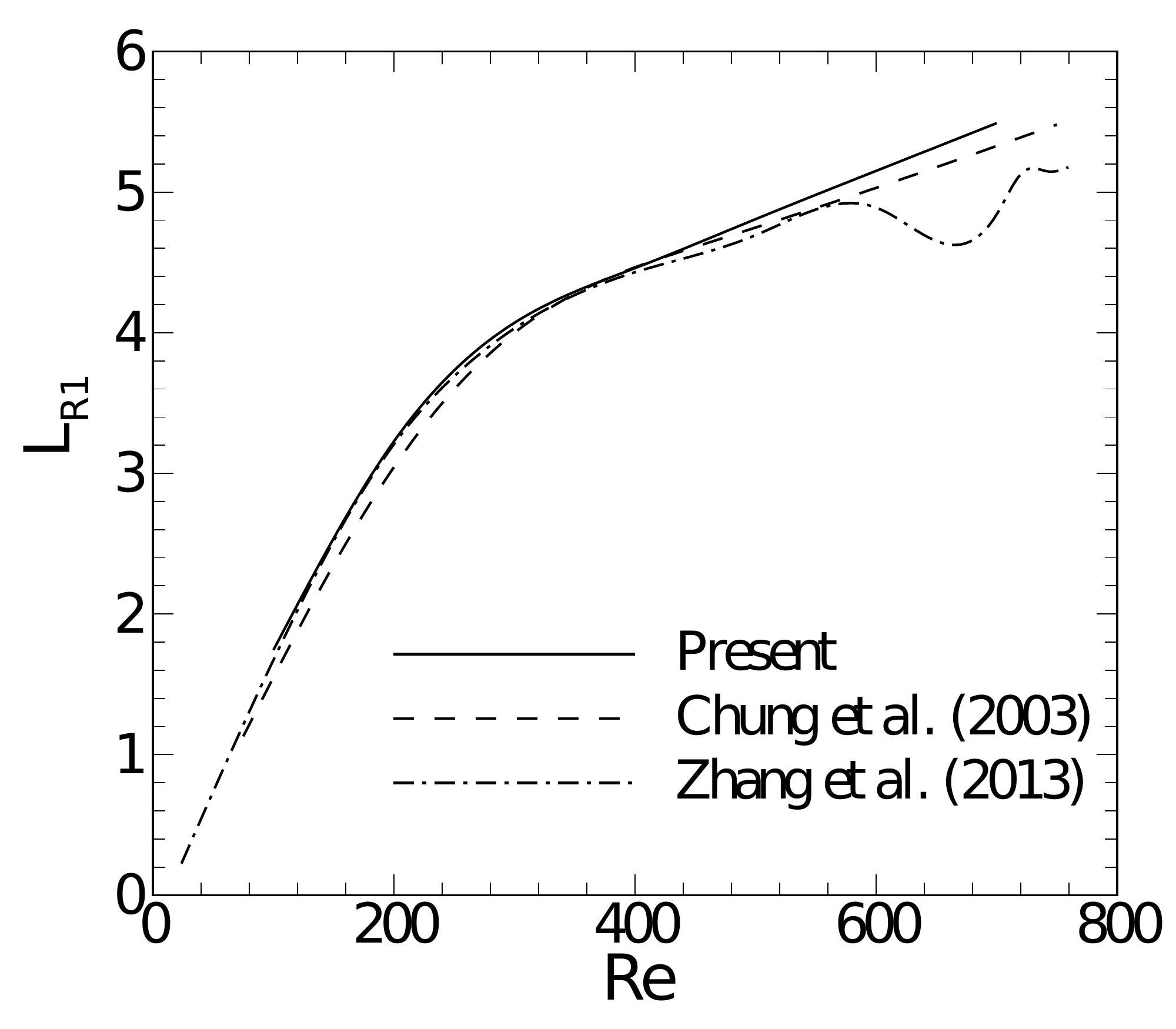}
  \caption{Length of the first recirculation bubble ($L_\mathrm{R_1}$) against Reynolds number ($\Rey$) for $\beta=1$, comparing the present results to those of \citet{zhang2013influence}  and \citet{chung2003unsteady}.}
\label{fig:recirculation_validation}
\end{figure}

Figure~\ref{fig:recirculation_validation} shows a comparison between the recirculation length of the first bubble ($L_\mathrm{R_1}$) in a flow with $\beta=1$ as a function of Reynolds number between the present and previously reported results digitised from figures in \citet{zhang2013influence} and \citet{chung2003unsteady}. The coefficient of determination, $R^2$, between the present data and those of these previous studies differ by just $2.2\%$ and $0.2\%$, respectively. The comparison displays a strong agreement between the studies, with each curve increasing rapidly and linearly at Reynolds numbers $\Rey\lesssim 200$, before transitioning to a more gradual linear regime of further bubble elongation beyond $\Rey\approx 300$. This regime terminates with the onset of unsteady flow. The sudden drop in the data from \citet{zhang2013influence}  at $\Rey\approx 600$ coincides with the onset of unsteady flow in that study. The present computations return a steady-state \cite[in agreement with][]{chung2003unsteady} up to $\Rey\approx 700$.

\section{\Twod\ base flows}\label{sect:2D_flow}

\subsection{Flow regimes}

\begin{figure}
\centering
\begin{tabular}{l}
(a) $\Rey=10$   \\
{\includegraphics[width=0.8\textwidth,keepaspectratio]{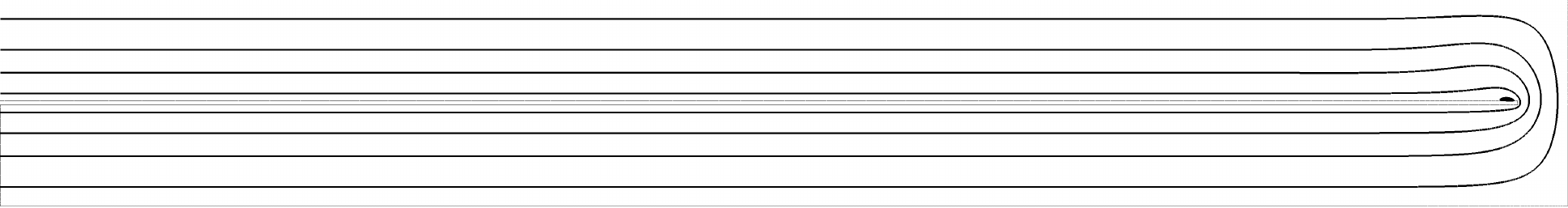}} \\
(b) $\Rey=200$ \\
{\includegraphics[width=0.8\textwidth,keepaspectratio]{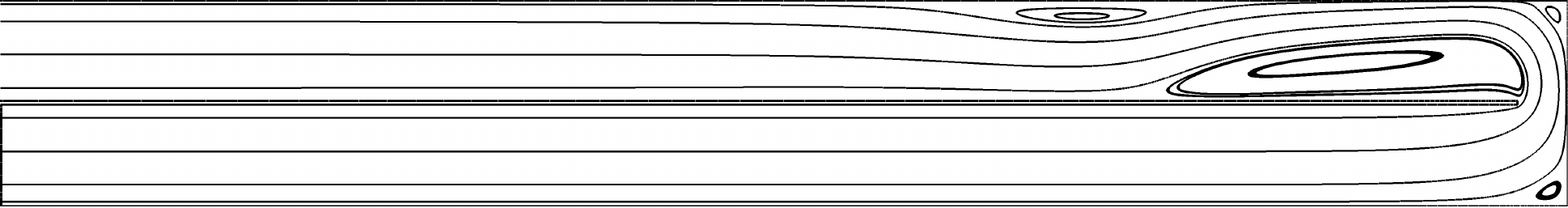}} \\
{(c) $\Rey=600$}  \\
{\includegraphics[width=0.8\textwidth,keepaspectratio]{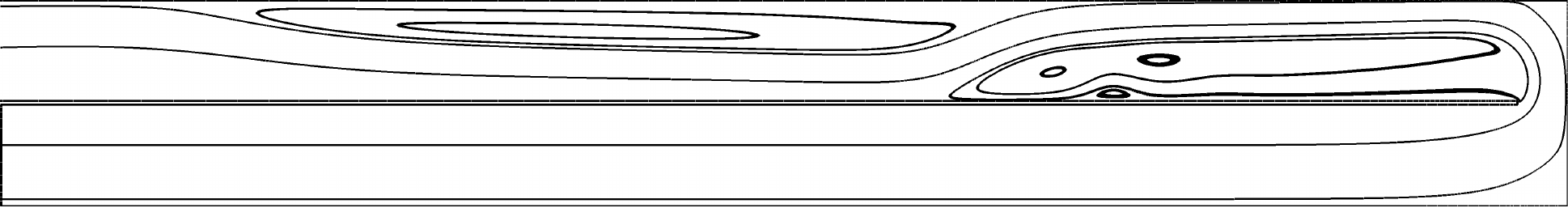}} \\
{(d) $\Rey=800$}  \\
{\includegraphics[width=0.8\textwidth,keepaspectratio]{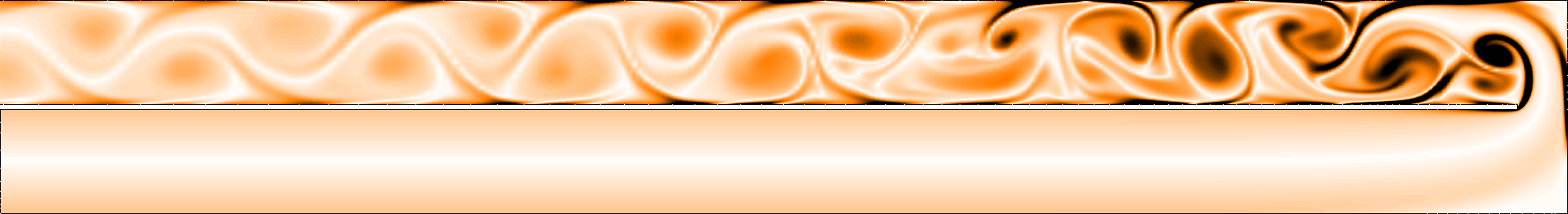}} \\
\end{tabular}
\caption{Streamlines of steady \twod\ base flow at (a) $\Rey=10$, (b) $\Rey=200$, (c) $\Rey=600$ and (d) flooded contours of vorticity magnitude at $\Rey=800$ for $\beta=0.5$. In (d), white represents zero vorticity (no rotation), and darker shading denotes arbitrarily higher vorticity magnitude levels.}
\label{fig:Base_flow}
\end{figure}

\begin{figure}
\centering
  \includegraphics[width=0.8\columnwidth,keepaspectratio]{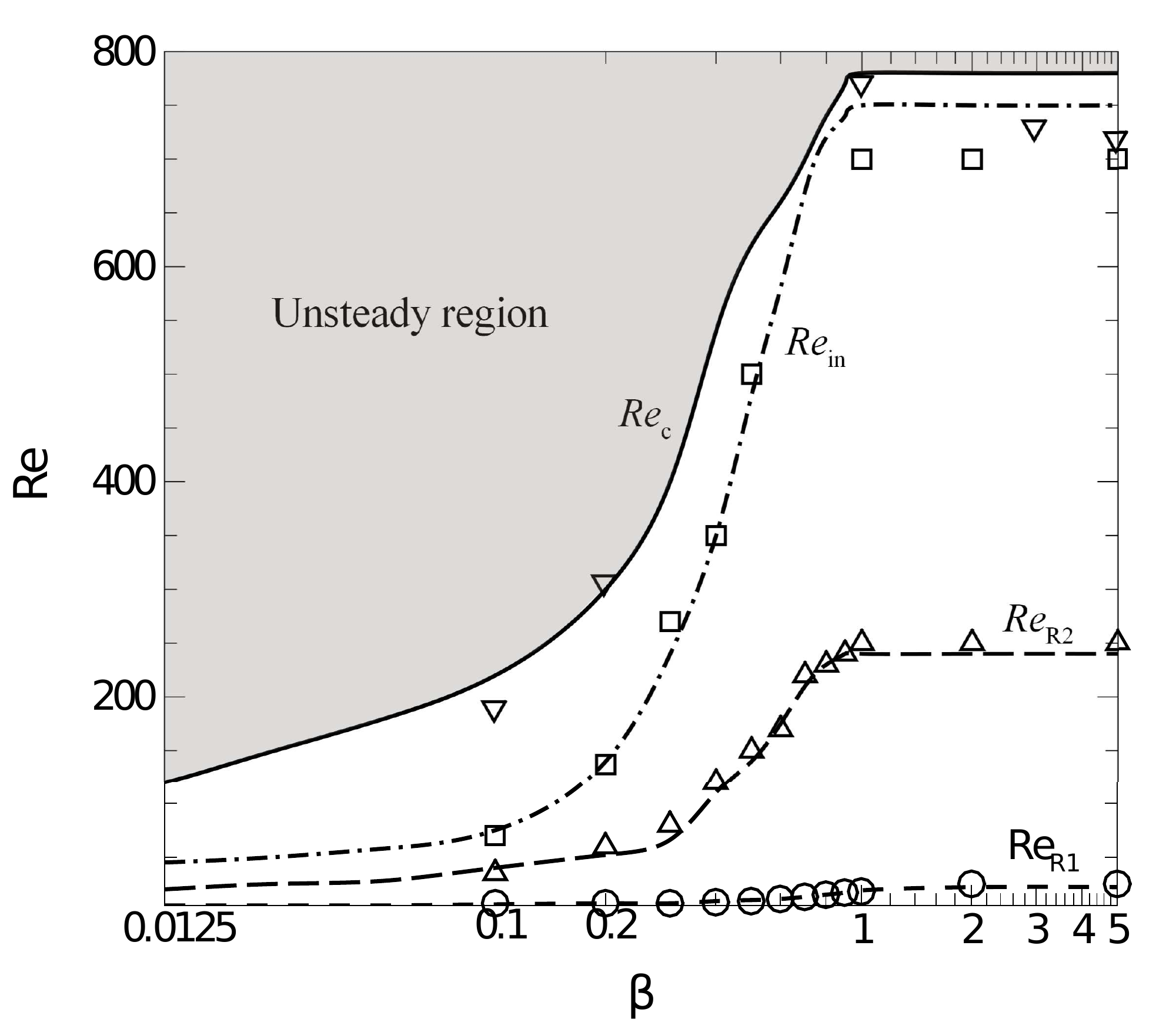}
  \caption{Reynolds number where primary recirculation bubble $\Rey_\mathrm{R_1}$, secondary recirculation bubble $\Rey_\mathrm{R_2}$ and inside recirculation bubble $\Rey_\textrm{in}$ start to appear in the \twod\ flow. $\Rey_\mathrm{c}$ is the transition from steady to unsteady. The lines are from current study and the symbols are results from \citet{zhang2013influence}. $\bigcirc$, $\bigtriangleup$, $\Box$ and $\bigtriangledown$ represent $\Rey_\mathrm{R_1}$, $\Rey_\mathrm{R_2}$, $\Rey_\textrm{in}$ and $\Rey_\mathrm{c}$, respectively.}
\label{fig:validation_Rec}
\end{figure}

In this section, we focus on the behaviour of the \twod\ base flow, especially around the sharp bend and along the downstream channel. For the range of opening ratios $\beta$ studied, four regimes are identified (figure \ref{fig:Base_flow}). The first regime exhibits only a single recirculation bubble immediately behind the sharp bend.  The second regime sees the emergence of a bubble at the opposite wall slightly downstream of the first bubble. The third regime reveals the appearance of a small counter-rotating recirculation bubble between the primary recirculation bubble and the bottom wall.  Finally, the fourth regime marks the development of an unsteady \twod\ flow.  The Reynolds numbers at the onset of each regime are denoted by $\Rey_{\mathrm{R_1}}$, $\Rey_{\mathrm{R_2}}$, $\Rey_\textrm{in}$ and $\Rey_{\mathrm{c}}$, respectively as shown in figure~\ref{fig:validation_Rec}. The results of the current study agree with those of  \citet{zhang2013influence} but small discrepancies are found for $\Rey_\textrm{in}$ at $\beta\geq1$, which are attributed to the different nodes density in the meshes at high $\Rey$ in both studies. The present study also extends the lower end of the range of $\beta$ from $\beta=0.1$ \citep{zhang2013influence} to $\beta=0.0125$. The constriction at the bend at smaller $\beta$ leads to high velocities and shear in that region.

For clarity, we shall highlight the main features of the regimes shown in figure \ref{fig:validation_Rec}, but a more detailed description can be found at \citet{zhang2013influence}. The onset of all regimes is delayed consistently to larger $\Rey$ at larger $\beta$ over $\beta \lesssim 1$, and it remains almost constant when $\beta \gtrsim 1$. It is worth noting that the behaviour of the flow is different between $\beta < 0.2$, $0.2 < \beta < 1$ and $\beta \gtrsim 1$. For $\beta \lesssim 0.2$; a flow resembling jet flow is created through the narrow bend orifice, causing the flow to accelerate transversely and hit the top wall before deflecting towards the streamwise direction. This causes the onset of each regime to occur at low $\Rey$. By contrast, at $0.2 \lesssim \beta \lesssim 1$, the width of the opening of the bend is sufficient for the flow to turn smoothly to the streamwise direction around the bend; hence, $\beta$ influences the onset of each regime. However, at $\beta \gtrsim 1$, the onsets do not vary significantly because a recirculation bubble develops at the outer wall of the bend, which confines the turning flow, such that the true breadth of the bend opening is no longer apparent. Thus the flow behaves similarly to that of $\beta=1$ and the onset $\Rey$ for the flow regimes remain almost constant for $\beta \gtrsim 1$.

At very low $\Rey$, the flow in the downstream and upstream channel is almost symmetrical with respect to $y=0$. However, as $\Rey$ increases, the flow streamlines at the bottom wall move upward until an inflexion point becomes visible behind the edge of the sharp bend. Consequently, the flow separation occurs and a recirculation bubble is formed. In the range of $\beta$ studied, the primary recirculation bubble appears at very small $\Rey$ because of a strong adverse pressure gradient behind the bend, as can be seen in figure~\ref{fig:Base_flow}(a). Theoretically a sharp edge always causes separation to a flow, even at very low $\Rey \rightarrow 0$ \citep{taneda1979visualization,moffatt1985magnetostatic}. The finite $\Rey_\mathrm{R_1}$ captured in this study is a numerical artefact of finite spatial resolution at the sharp bend corner. The finite discretisation obscures the bubble at very low $\Rey$.

As $\Rey$ increases, the secondary recirculation bubble appears when the flow streamlines in the bulk flow above the primary recirculation bubble move away from the top wall. This results in a strong adverse pressure gradient at the wall, which causes another separation to occur at $\Rey_{R_2}$ (figure~\ref{fig:Base_flow}(b)). Under the same circumstances, an inner counter-rotating recirculation bubble as seen in figure~\ref{fig:Base_flow}(c), is formed between the primary recirculation bubble and the bottom wall when the backflow in the primary recirculation bubble moves away from the wall.

\begin{figure}
\centering
  \includegraphics[width=0.65\columnwidth,keepaspectratio]{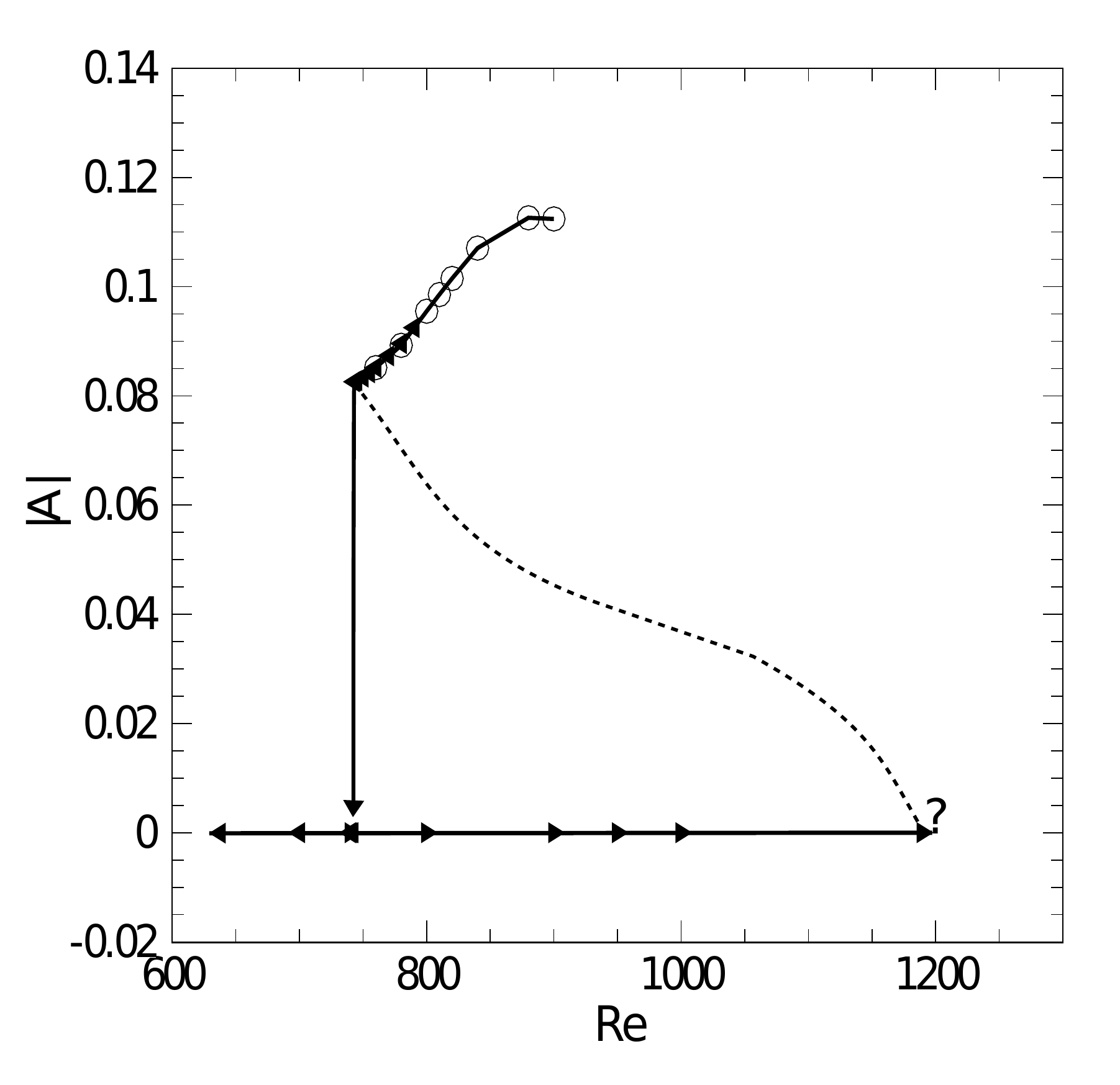}
  \caption{Hysteretic behavior described by the fluctuation of the integral of velocity magnitude throughout the domain as a function of $\Rey$ at $\beta=1$, reducing $\Rey$ from an unsteady flow ($\blacktriangleleft$) and increasing $\Rey$ from a steady flow ($\blacktriangleright$) give different $\Rey_\mathrm{c}$. $|A|=0$ indicates steady flow solution.}
\label{fig:l2norm_hysteresis}
\end{figure}

\citet{zhang2013influence} found an additional regime between $\Rey_\textrm{in}$ and $\Rey_\mathrm{c}$, where they reported a small scale vortices structure far downstream of the channel; this state is not observed in the current study. $\Rey_\mathrm{c}$ in figure~\ref{fig:validation_Rec} indicates an unsteady flow where large eddies or vortices are shed downstream from the sharp corner of the bend as depicted in figure~\ref{fig:Base_flow}(d). $\Rey_\mathrm{c}$ is found to be in agreement with those of \citet{zhang2013influence} study, and it is important to note that both studies started simulations from rest to obtain this $\Rey_\mathrm{c}$. In a further analysis, hysteretic behaviour has been observed when the simulation is started from different initial conditions. Considering the unsteady flow as a departure from the base steady flow at the same Reynolds number, its amplitude $\left|A\right|$ is measured as the $\mathscr{L}^2$ norm (the integral of the magnitude of velocity over the computational domain) of the difference between velocity fields in these states. Figure~\ref{fig:l2norm_hysteresis} shows the variations of $\left|A\right|$ and therefore regimes of unsteady flow where $\left|A\right| > 0$ when Reynolds number is incrementally varied for $\beta=1$. Two distinct onsets of \twod\ unsteadiness are found by initiating the simulation from three different initial conditions which are (i) flow at rest, (ii) a snapshot of unsteady flow computed at a slightly lower Reynolds number, and (iii) the steady flow solution obtained at a slightly lower Reynolds number. It is evident from the figure that the simulations starting from the flow at rest and from an unsteady flow yield the same value of $\Rey_\mathrm{c} = 742$, whereas the simulations starting from a steady flow become unsteady at a higher $\Rey$, $\Rey_\mathrm{c}\approx 1150$.

\begin{figure}
\centering
\begin{tabular}{l}
(a)  \\
{\includegraphics[width=0.8\textwidth,keepaspectratio]{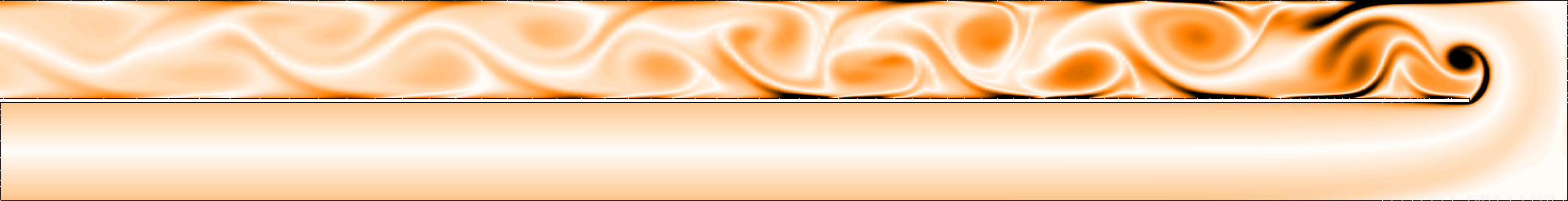}} \\
(b)  \\
{\includegraphics[width=0.8\textwidth,keepaspectratio]{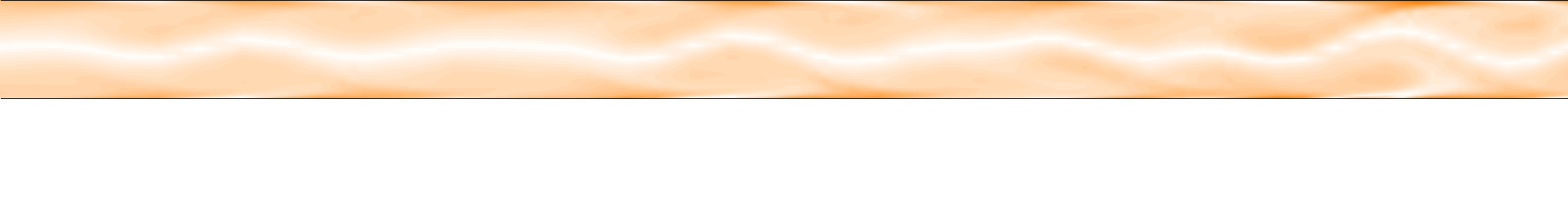}} \\
(c)  \\
{\includegraphics[width=0.8\textwidth,keepaspectratio]{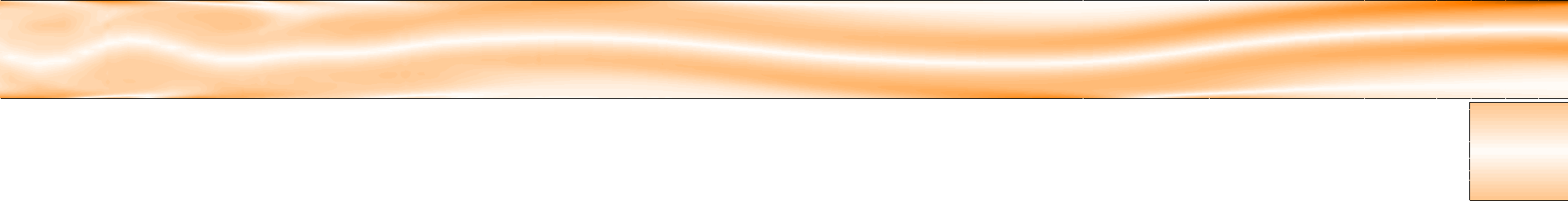}} \\
(d)  \\
{\includegraphics[width=0.8\textwidth,keepaspectratio]{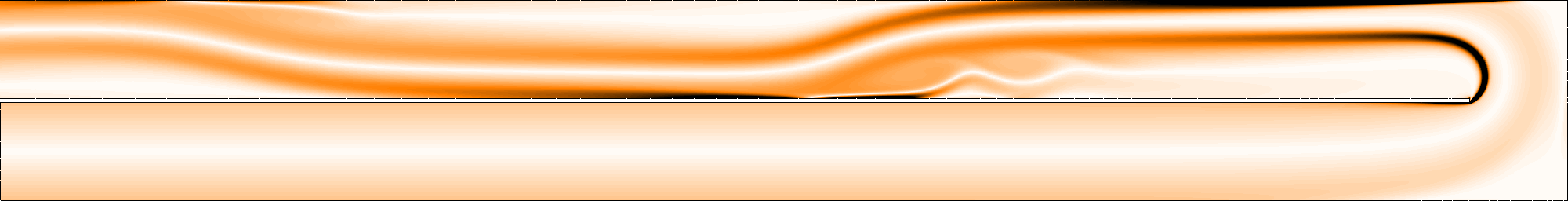}} \\
\end{tabular}
\caption{Flooded contours of vorticity magnitude demonstrating unsteady saturated flows for (a) $\Rey=800$ at $\beta=1$ (which was initiated from rest), and (b)-(d) $\Rey=1152$ at $\beta=1$ (which was initiated from a saturated steady-state flow solution at a lower Reynolds number). Contour levels are as per figure~\ref{fig:Base_flow}(d). (b), (c) and (d) show domain segments $-45\leq x \leq -29$, $-30\leq x \leq -14$ and $-15\leq x \leq 1$, respectively.}
\label{fig:hysteresis_example}
\end{figure}

The location where shedding initiates at the onset depends on initial conditions too. When a simulation is initiated from a flow at rest, the vortex shedding can be seen to emerge from the sharp corner of the bend as illustrated in figure~\ref{fig:hysteresis_example}(a). It is likely that a large-amplitude perturbations caused by the impulsive initiation of the flow are sufficient to provoke a shedding from the bend that bypasses the downstream destabilisation, and beyond $\Rey_\mathrm{c} \approx 742$ (at least in these simulations) is self-sustaining. A similar observation was made when reducing $\Rey$ from an unsteady flow: a large sudden decrease could revert the unsteady flow to steady state at a Reynolds number at which the unsteady state could be preserved via a gradual decrement in Reynolds number.

When $\Rey$ is increased gradually along the steady-flow branch, the flow remains steady up to $\Rey\simeq1150$. Figure~\ref{fig:hysteresis_example}(b)-(d) depicts the saturated flow at $\Rey=1152$. This final state exhibits a wavy disturbance extending $25$ diameters downstream of the bend. In this case, unsteadiness first manifested in the shear layers behind the secondary recirculation bubble in the form of small eddies, but over time the flow in this region more proximate to the bend re-stabilised and the unsteady region retreated to its ultimate position further downstream. This has some resemblance to the regime described by \citet{zhang2013influence} before the flow becomes unsteady in their study. It is likely that they found this regime at lower $\Rey$ due to the high sensitivity to mesh resolution of this feature. Since the flow is already unstable far downstream of the bend, noise tended to be amplified and created small vortical structures. In testing this hypothesis, we found that by increasing resolution of the mesh in our study, the onset of unsteadiness could be delayed significantly by increasing $\Rey$ gradually. It is also plausible that the length of the outlet channel may influence the upper limit of the steady-state branch, though this was not tested. Finally, it is noted that when comparing the flow pattern and the region of the flow producing the unsteady flow features between figure~\ref{fig:hysteresis_example}(a) and figure~\ref{fig:hysteresis_example}(b)-(d) that the two depicted unsteady branches are different. It will be shown in \S~\ref{sec:LSA} that the \twod\ flows are unstable to \threed\ perturbations at Reynolds numbers below those of this hysteretic zone, so it will not be characterised further here.

\subsection{Bubble separation points (steady flows)}
\begin{figure}
\centering
  \includegraphics[width=0.8\columnwidth,keepaspectratio]{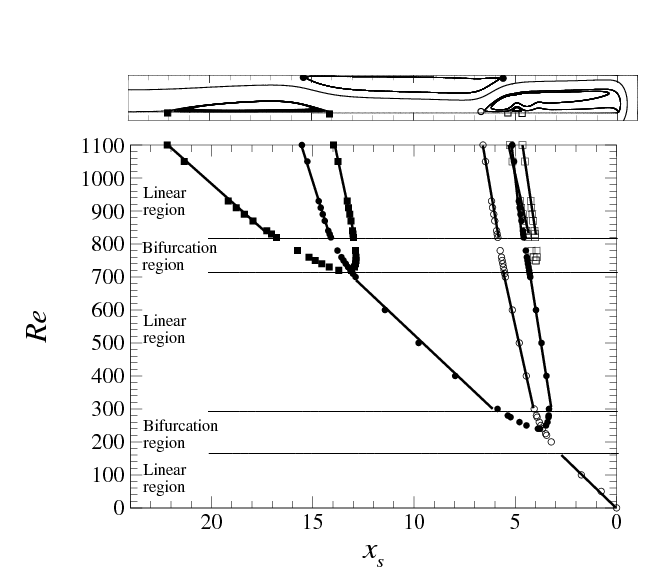}
  \caption{A plot showing the Reynolds-number-dependence of the locations of separation points (measured by their distances from the bend - left of the origin), $x_s$, for the base flows as functions of Reynolds number. Open circles represent the stagnation points for the primary recirculation bubbles. Solid circles denote the stagnation points for the secondary recirculation bubbles which forms at $\Rey \approx 240$ and $x_s \approx 3.77$. Solid and open squares represent the stagnation points for third bubble and inner recirculation bubble, respectively, which are formed at higher $\Rey$. The top frame shows the streamlines of the base flow and the separation points of primary, secondary, inner and third recirculation bubbles at $\Rey = 1100$ in the outlet channel.}
\label{fig:separating_points}
\end{figure}

Figure~\ref{fig:separating_points} shows the Reynolds number dependence of the separation and reattachment points (expressed by their distance downstream of the bend, $x_s$) on both the bottom and top walls of the channel for $\beta=1$. The empty circle symbol indicates the limit of the primary recirculation bubble behind the sharp corner. At $\Rey=240$, a secondary recirculation bubble appears on the top wall at $x_s = 3.77$. At $\Rey\approx 700$, two additional recirculation bubbles appear at the bottom wall; one is a small bubble in the primary recirculation bubble and the other is near the reattachment point of the secondary recirculation bubble.

From figure~\ref{fig:separating_points}, we notice that there are regions where the location of the separation and reattachment points are almost linear functions of $\Rey$. It can clearly be seen in figure~\ref{fig:separating_points} that when the bifurcation region around $\Rey=300$ and 750 is being excluded, the locations of both points for all bubbles behave almost linearly with $\Rey$. When a new recirculation bubble appears, the growth of the recirculation bubble just upstream of it is affected and so is the variation with Reynolds of its reattachment point. The same behaviour was also found in the backward-facing step flow by \citet{erturk2008numerical}.

\begin{figure}
\centering
  \includegraphics[width=0.6\columnwidth,keepaspectratio]{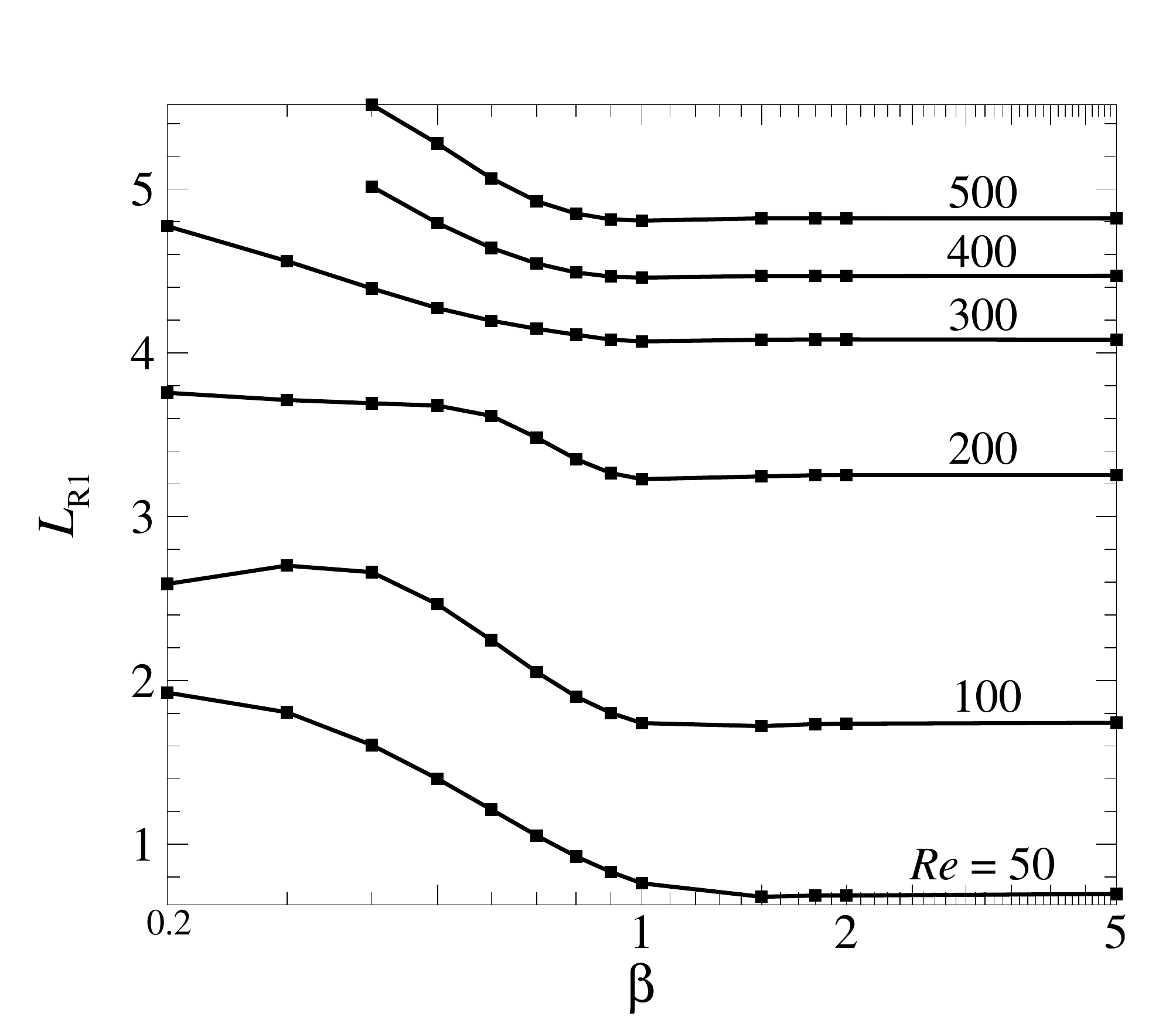}
  \caption{A plot of the length of the primary recirculation bubble as a function of $\beta$ for Reynolds numbers as labelled.}
\label{fig:Lr_vs_beta}
\end{figure}

The effect of $\beta$ on the size of the primary recirculation bubble is illustrated in figure~\ref{fig:Lr_vs_beta}. At small values of $\beta$, the main bulk flow is accelerated by the small jet opening, making the primary recirculation bubble elongated in the streamwise direction. This causes the bubble to be bigger than that at larger $\beta$. $L_\mathrm{R1}$ decreases as $\beta$ increases, but as $\beta>1$ and $\Rey\gtrsim200$, the size of the bubble increases due to the effect of the recirculation bubble at the far end of the bend wall.

\begin{figure}
\centering
\begin{tabular}{l}
(a) \hspace{62ex}$\beta_\textrm{eff}=\beta$\\
{\includegraphics[width=0.76\textwidth,keepaspectratio]{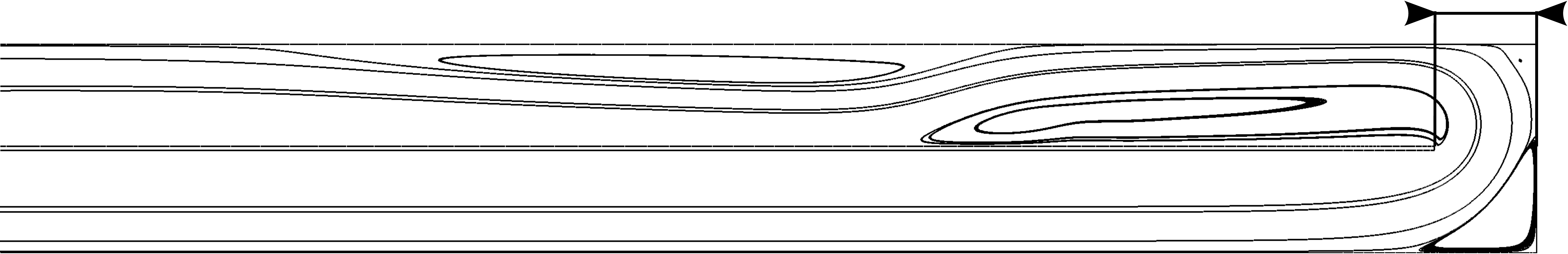}}		 \\
(b) \hspace{62ex}$\beta_\textrm{eff}<\beta$\\
{\includegraphics[width=0.8\textwidth,keepaspectratio]{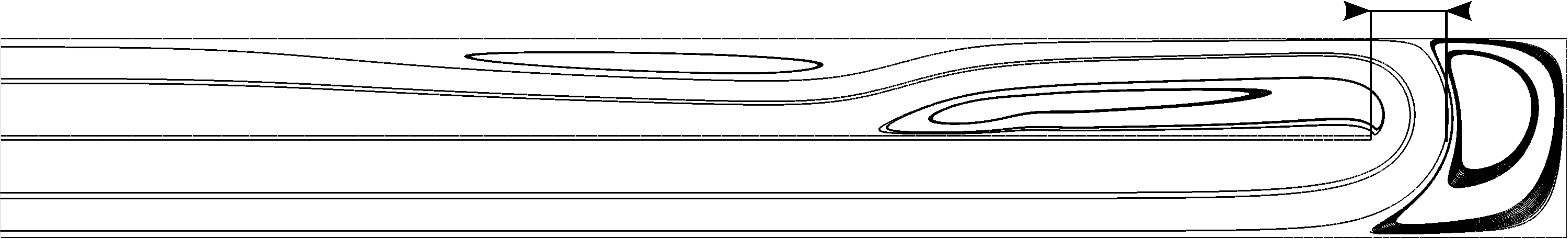}} 	 \\
(c) \\
{\includegraphics[width=0.7\textwidth,keepaspectratio]{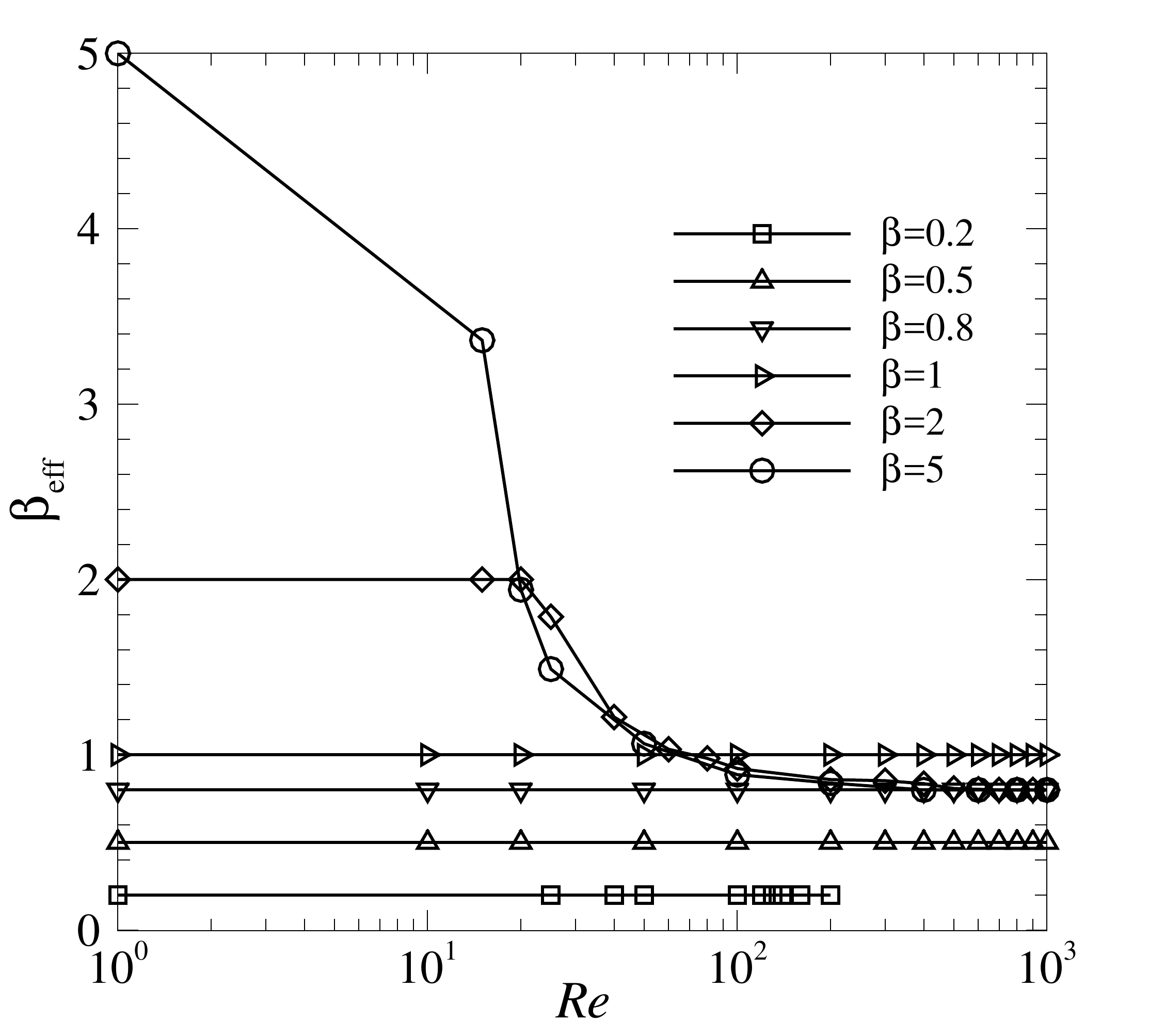}} 	 \\
\end{tabular}
\caption{Streamlines of steady \twod\ flow for $\Rey=600$, (a) $\beta=1$ and (b) $\beta=2$. A big recirculation bubble appears at the far end of the wall when $\beta>1$ causing the effective bend opening ratio to be lesser than the actual opening. $\beta_{\mathrm{eff}}(\Rey)$ is plotted in (c) for several value of $\beta$.}
\label{fig:end_wall_recirculation}
\end{figure}

Since for $\beta>1$, part of the flow in the bend is trapped in a closed eddy outside of the through-flow taking the bend, \citet{zhang2013influence} defined an effective opening ratio $\beta_{\mathrm{eff}}$ based on the horizontal thickness of the flow effectively turning from inlet to outlet at $y=0$ --- that is, the distance from the inner vertical wall at the bend to the dividing streamline separating the turning flow from the closed recirculation. The same definition is used in the present study. For $\beta \leq 1$, $\beta_\textrm{eff}=\beta$ (\eg\ see figure~\ref{fig:end_wall_recirculation}(a)), but for $\beta>1$, starting from $\Rey\approx 30$, $\beta_\textrm{eff}$ is smaller than $\beta$ (\eg\ see figure~\ref{fig:end_wall_recirculation}(b)). At $\Rey > 200$, $\beta_\textrm{eff}$ saturated close to $0.7$. This degradation of $\beta_\textrm{eff}$ to values below $\beta$, and its saturation behaviour at large $\Rey$, can be observed in figure~\ref{fig:end_wall_recirculation}(c). For $\beta=2$ in figure~\ref{fig:end_wall_recirculation}(b), the effective $\beta$ is found to be $\beta_\textrm{eff}\approx0.8$ which is slightly higher than what was found by \citet{zhang2013influence}.

\section{Linear stability}\label{sec:LSA}
\subsection{Growth rates and marginal stability}\label{subsect:stab_curve}
\begin{figure}
\centering
    \begin{tabular}{ll}
    (a) & (b) \\
    \multicolumn{1}{c}{\includegraphics[width=0.48\textwidth,keepaspectratio]{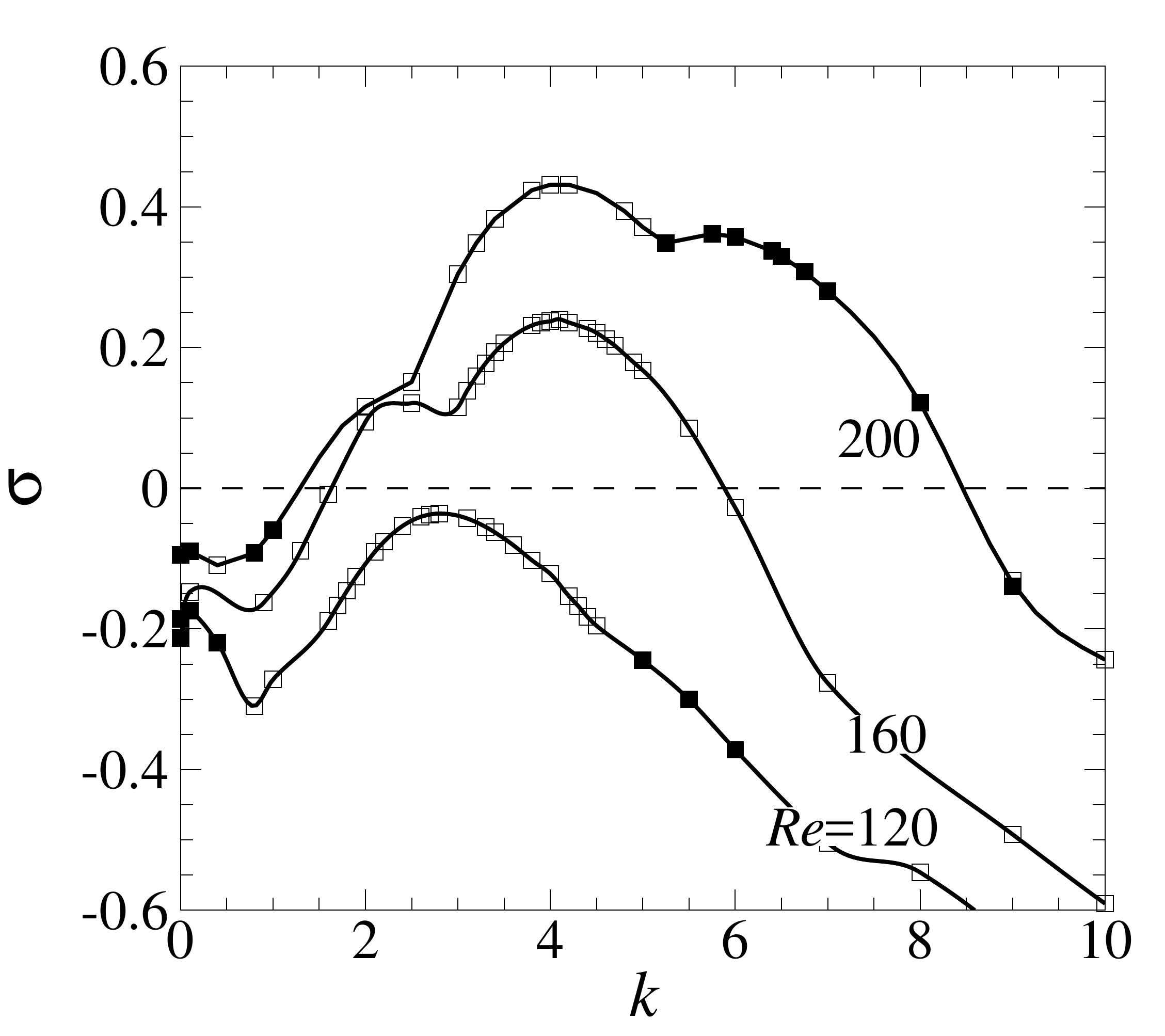}} &
    \multicolumn{1}{c}{\includegraphics[width=0.48\textwidth,keepaspectratio]{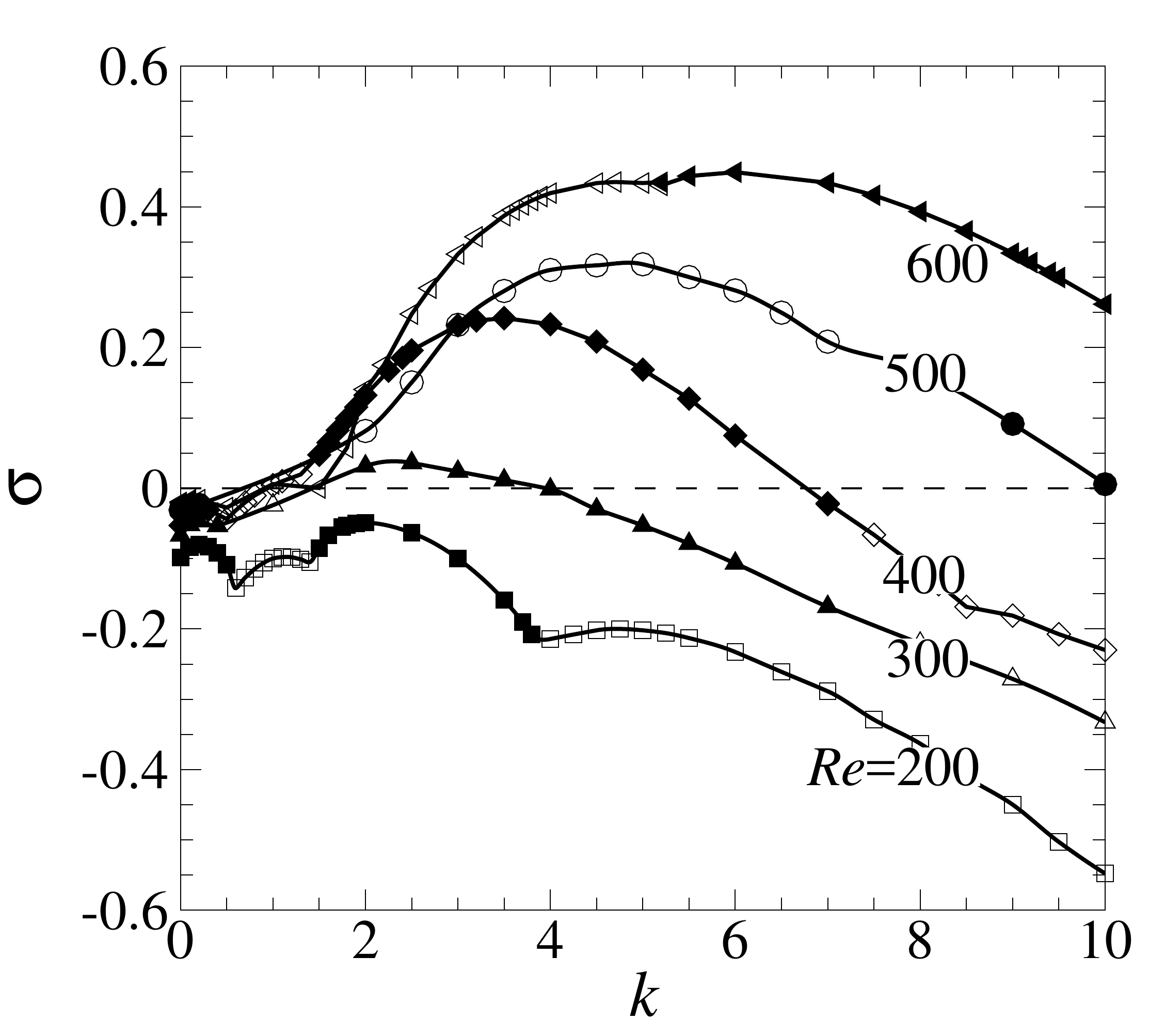}} \\
    (c) & (d) \\
    \multicolumn{1}{c}{\includegraphics[width=0.48\textwidth,keepaspectratio]{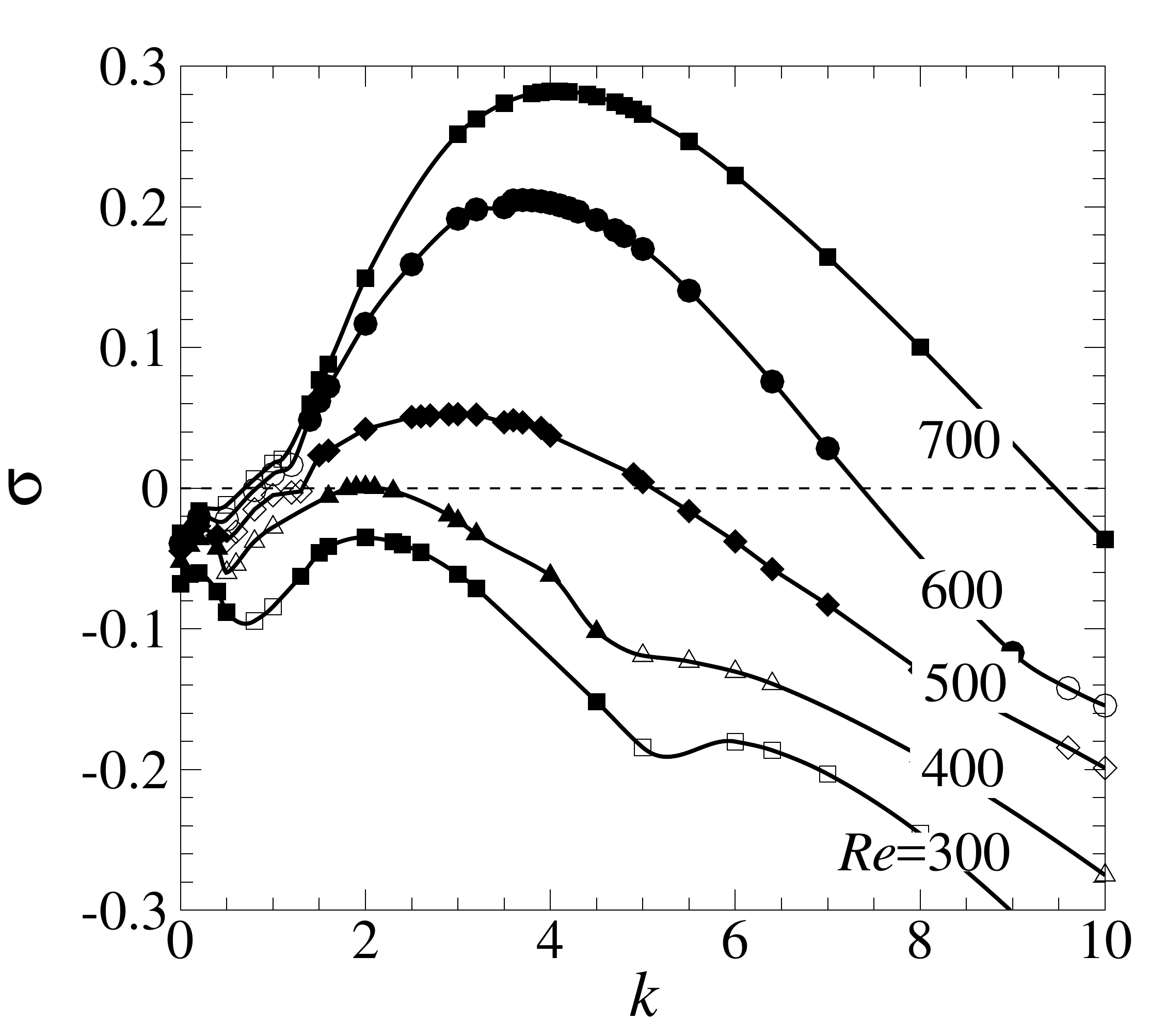}} &
    \multicolumn{1}{c}{\includegraphics[width=0.48\textwidth,keepaspectratio]{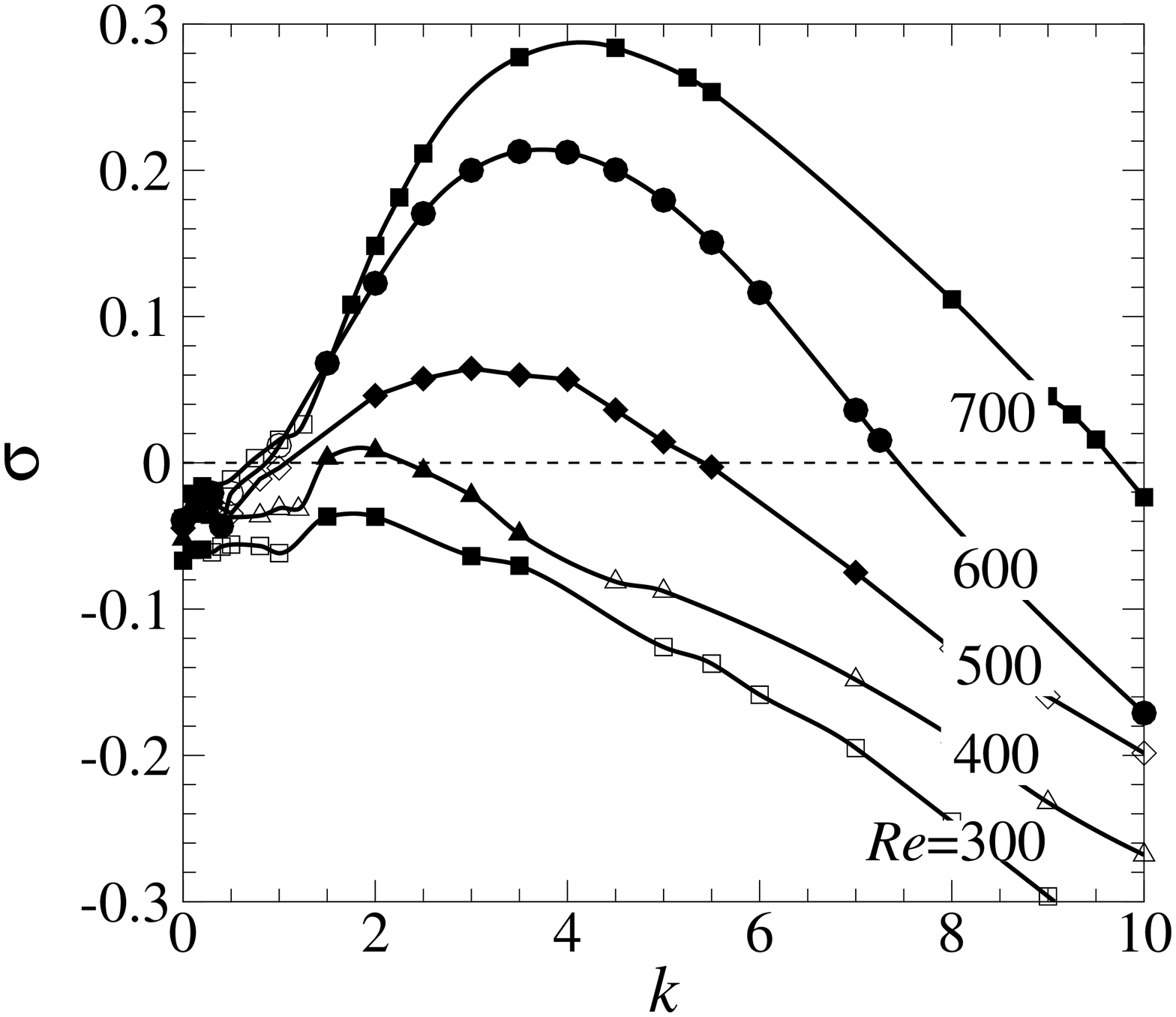}} \\
    \end{tabular}
  \caption{Growth rates of leading eigenmodes as a function of spanwise wavenumber $k$ for (a) $\beta=0.2$ and $\Rey\leq200$, (b) $\beta=0.5$ and $\Rey\leq600$, (c) $\beta=1$ and $\Rey\leq700$, and (d) $\beta=2$ and $\Rey\leq700$ . Solid symbols represent real leading eigenvalues, while hollow symbols represent complex-conjugate pairs of non-real leading eigenvalues. Solid lines connect all dominant leading eigenvalues from several branches of the same Reynolds number.}
\label{fig:lsa_curve}
\end{figure}

This subsection analyses the dependence of the perturbation growth on Reynolds number $\Rey$, spanwise wavenumber $k$ and opening bend ratio $\beta$. Figure~\ref{fig:lsa_curve} shows the predicted growth rates as a function of the Reynolds number and spanwise wavenumber $k$ for $\beta=0.2$, 0.5, 1 and 2. The primary linear spanwise instability is obtained via polynomial interpolation to determine the lowest Reynolds number that first produces $\sigma=0$, and the wavenumber at which this occurs. Both are reported in table \ref{tab:critical}. For $\beta>0.2$, at very low wavenumber $k \lesssim 0.3$, a local maximum is observed, but the corresponding mode is always stable. Between this local maximum and the primary maximum, a small range of $k$ produces leading non-real eigenvalues. Larger wavenumbers that those shown in figure~\ref{fig:lsa_curve} were also studied and found decay increasingly fast at large $k$. This trend of the growth rate as a function of $\Rey$ and $k$ shows a good resemblance with those of backward-facing step flow \citep{barkley2002three}. However, for all $\beta$, the critical $\Rey$ for the flow to become \threed\ are found to be much lower compared to the flow in backward-facing step \citep{barkley2002three,armaly1983experimental} and partially blocked channel \citep{griffith2007wake}. This finding is not surprising, as the \twod\ flow around sharp bend becomes unsteady at much lower $\Rey$ compared to those geometries.
\begin{table}
\centering
\begin{tabular}{ccc}

{$\beta$} & {$\Rey_\mathrm{c}$} & {$k_c$} \\[6pt]
  0.2 & 125 & 3 \\
  0.5 & 278 & 2.05 \\
  1   & 397 & 1.98 \\
  2   & 387 & 1.93 \\
\end{tabular}
\caption{Critical Reynolds number and corresponding spanwise wavenumber at the onset of instability for $\beta = 0.2$, $0.5$, $1$ and $2$.} \label{tab:critical}
\end{table}

Figure~\ref{fig:lsa_curve} shows that $\beta=0.2$ always has an oscillatory leading mode for the range of $\Rey$ investigated.
Meanwhile, $\beta=0.5$ has a synchronous leading mode at $\Rey \leq 400$, before an oscillatory mode becomes dominant at higher $\Rey$. On the other hand, $\beta>1$ always has a synchronous leading mode. The transition from oscillatory to synchronous behaviour as $\beta$ increases is similar to what was observed by \citet{lanzerstorfer2012global} in backward facing step flow, where they found that the the flow with very large step height is destabilised by an oscillatory mode, and this changes from oscillatory to synchronous if the step height is further decreased. To clarify the shift of the leading mode from synchronous to oscillatory, several of the leading eigenvalues have been computed for $\beta=0.5$ at each wavenumber and three different Reynolds numbers.

\begin{figure}
\centering
\begin{tabular}{ll}
(a)   & (b)  \\
\multicolumn{1}{c}{\includegraphics[width=0.48\textwidth,keepaspectratio]{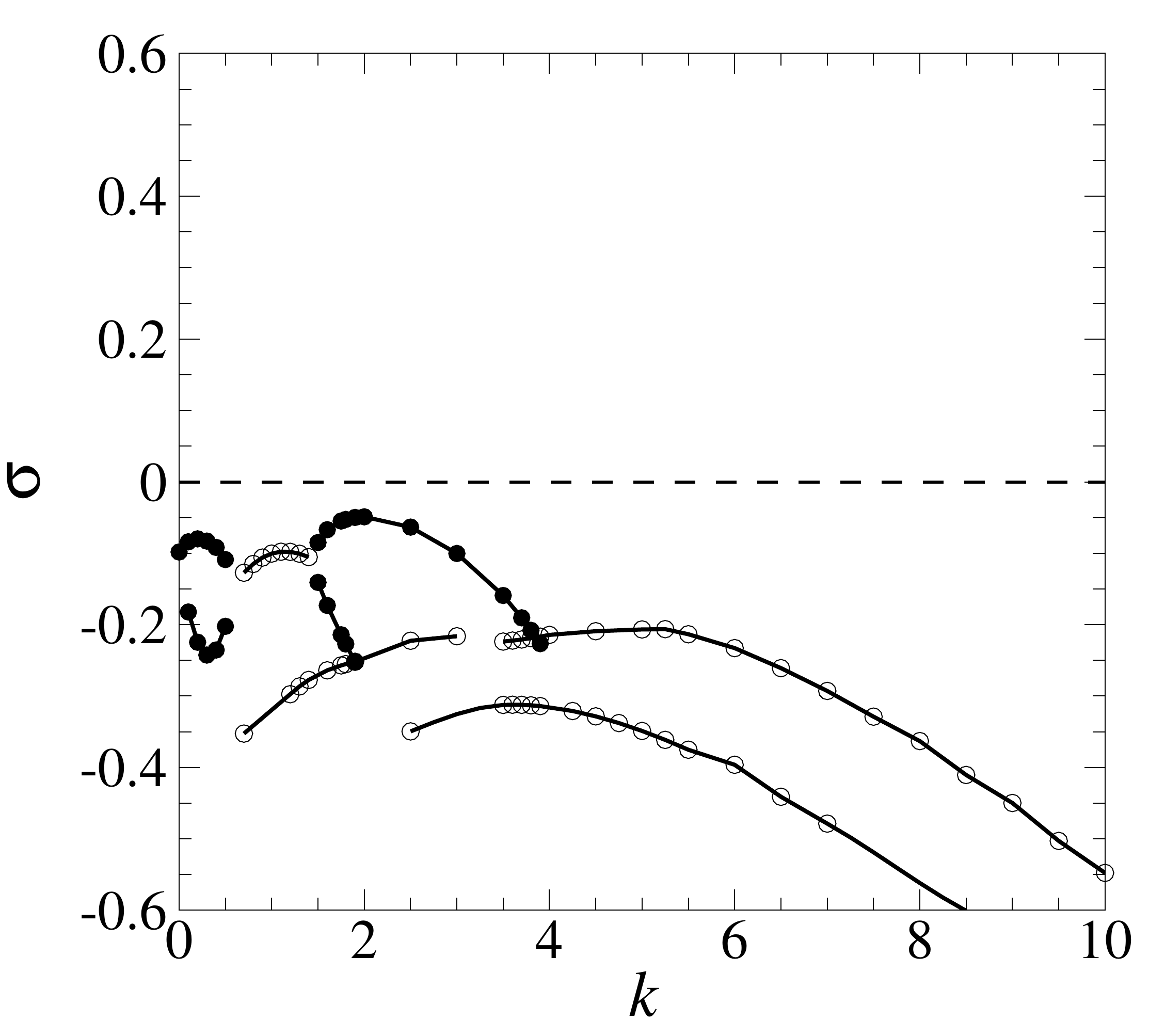}} &
\multicolumn{1}{c}{\includegraphics[width=0.48\textwidth,keepaspectratio]{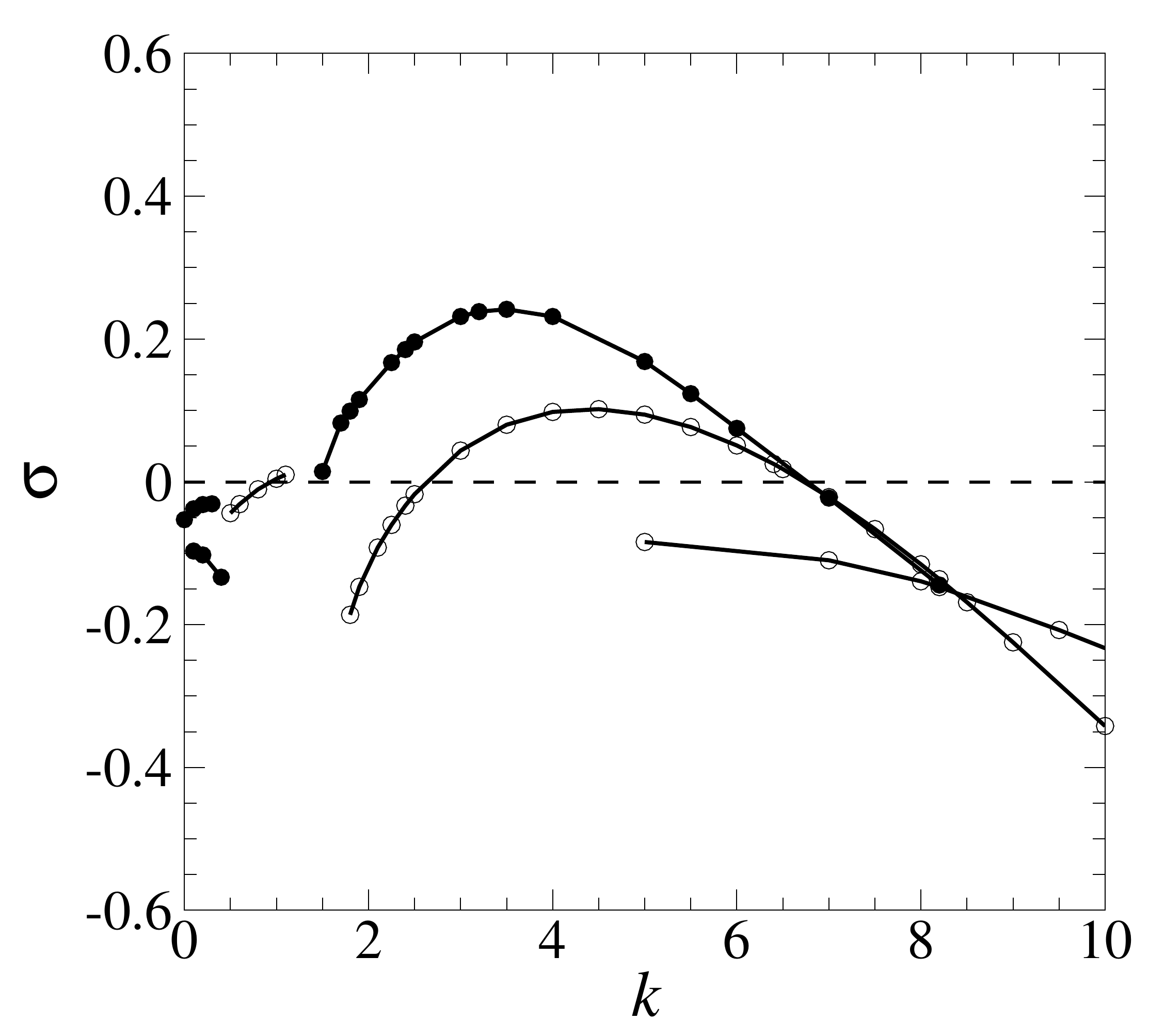}} \\
\multicolumn{2}{l}{\hspace{100pt}(c)}   \\
\multicolumn{2}{c}{\includegraphics[width=0.48\textwidth,keepaspectratio]{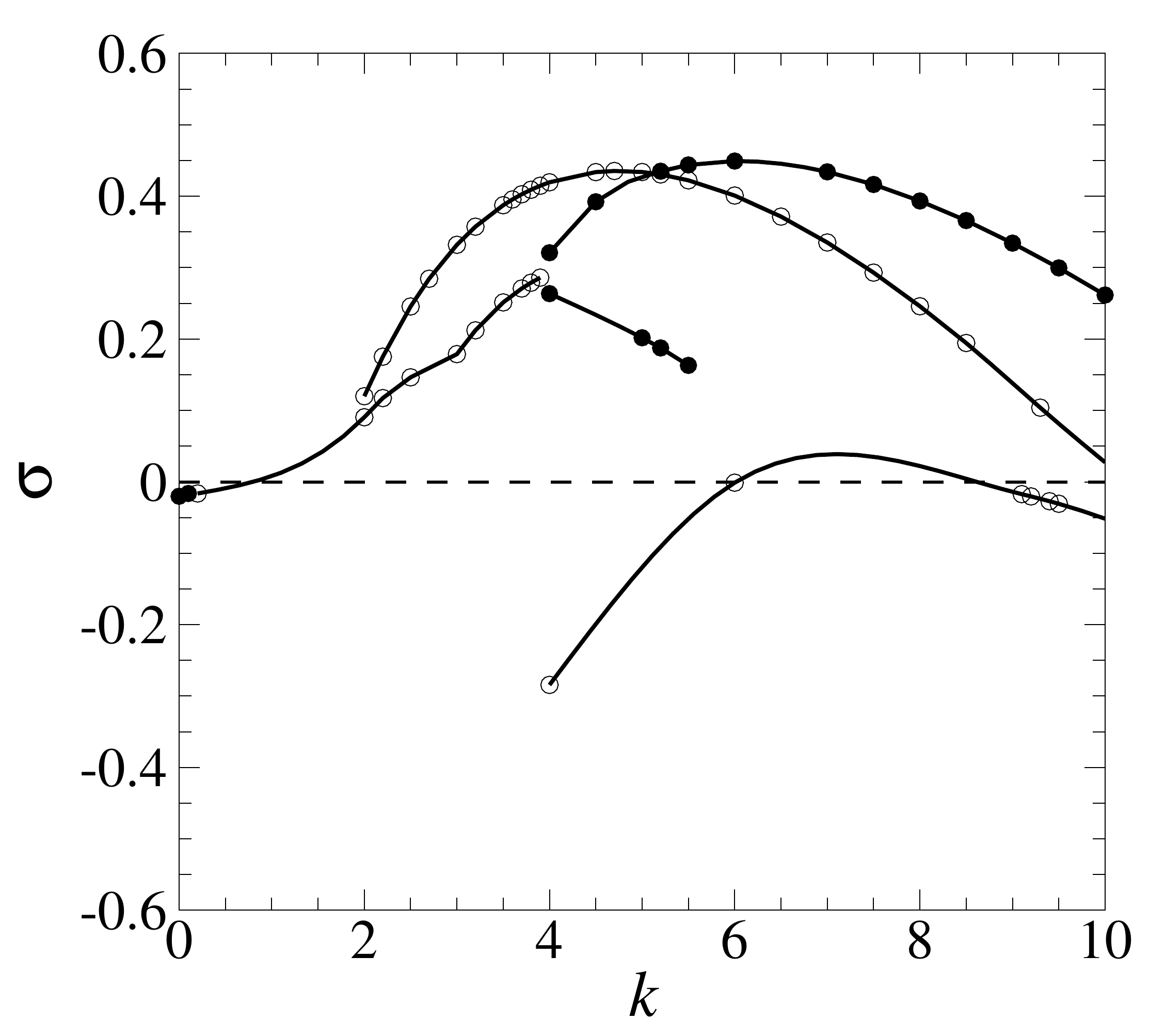}} \\
\end{tabular}
\caption{Growth rates of leading eigenmodes plotted against spanwise wavenumber at $\beta=0.5$ for (a) $\Rey=200$, (b) $\Rey=400$ and (c) $\Rey=600$. Solid symbols represent real leading eigenvalues, meanwhile hollow symbols represent complex-conjugate pairs of leading non-real eigenvalues.}
\label{fig:Three_leading_eigenvalues}
\end{figure}

The results are shown in figure~\ref{fig:Three_leading_eigenvalues}. The curves closely resemble those for the flow over a backward-facing step \citep{barkley2002three}, which consists of two branches of real eigenvalues at low wavenumber coalescing into a single branch of non-real eigenvalue as $k$ increases. In \cite{barkley2002three}'s case, however, the primary leading eigenvalues appear at higher wavenumbers, and (as with the present study for $\beta>0.5$) the real branch at high wavenumber is the first to become unstable. All branches shift to higher $\sigma$ as $\Rey$ increases. However, as $\Rey$ increases further, it can be seen that the leading oscillatory mode is more stable than the real one (figure~\ref{fig:Three_leading_eigenvalues}(b)) before it becomes dominant at $\Rey\approx600$ as shown in figure~\ref{fig:Three_leading_eigenvalues}(c). A similar observation was made by \citet{natarajan1993instability} in flow past spheres and disks, by \citet{tomboulides2000numerical} in the weak turbulent flow past a sphere, and by \citet{johnson1999flow} in a numerical and experimental study on flow past a sphere up to $\Rey_d=300$.

\subsection{Structure of the eigenvalue spectra}\label{subsect:structure}
\begin{figure}
\centering
\begin{tabular}{ll}
\parbox{0.45\textwidth}{
(a)\\
\includegraphics[width=0.45\textwidth,keepaspectratio]{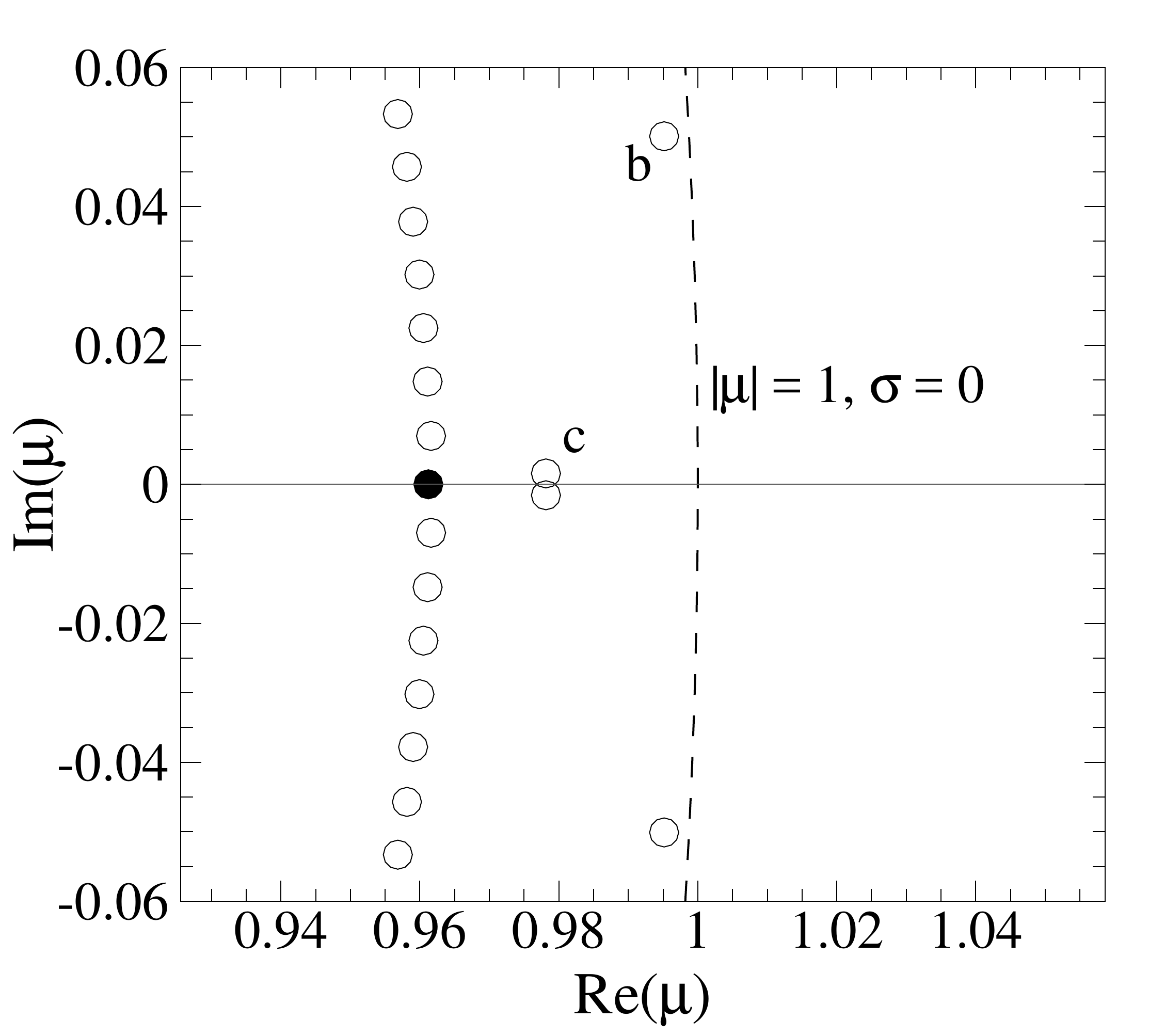}
} &
\parbox{0.5\textwidth}{
(b)\\
\includegraphics[width=0.5\textwidth,keepaspectratio]{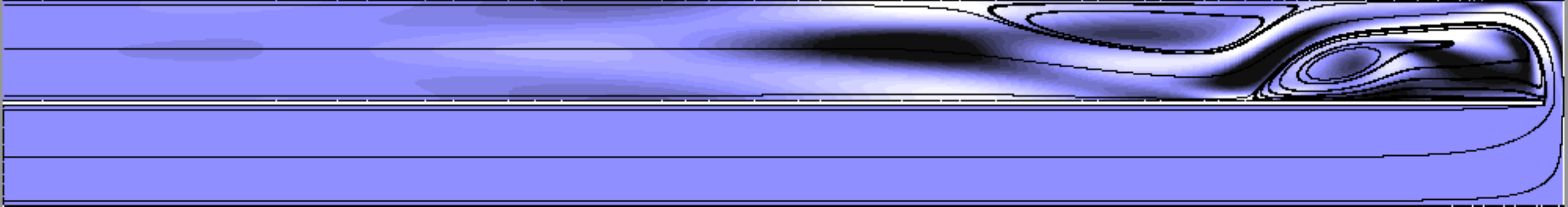}\\
(c) \\
\includegraphics[width=0.5\textwidth,keepaspectratio]{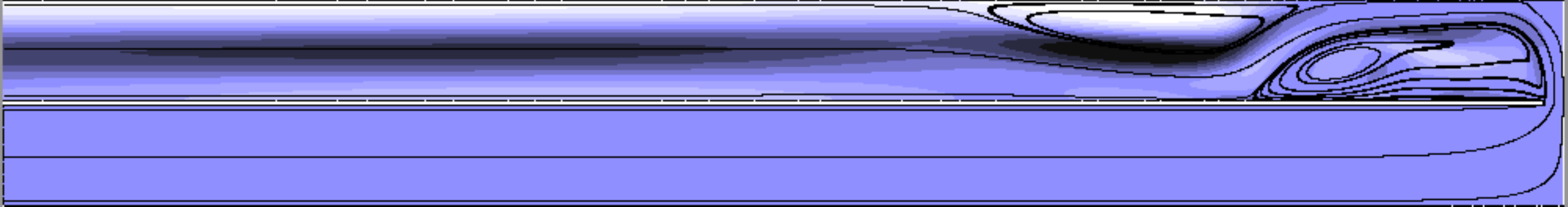}
}\\
\parbox{0.45\textwidth}{
(d)\\
\includegraphics[width=0.45\textwidth,keepaspectratio]{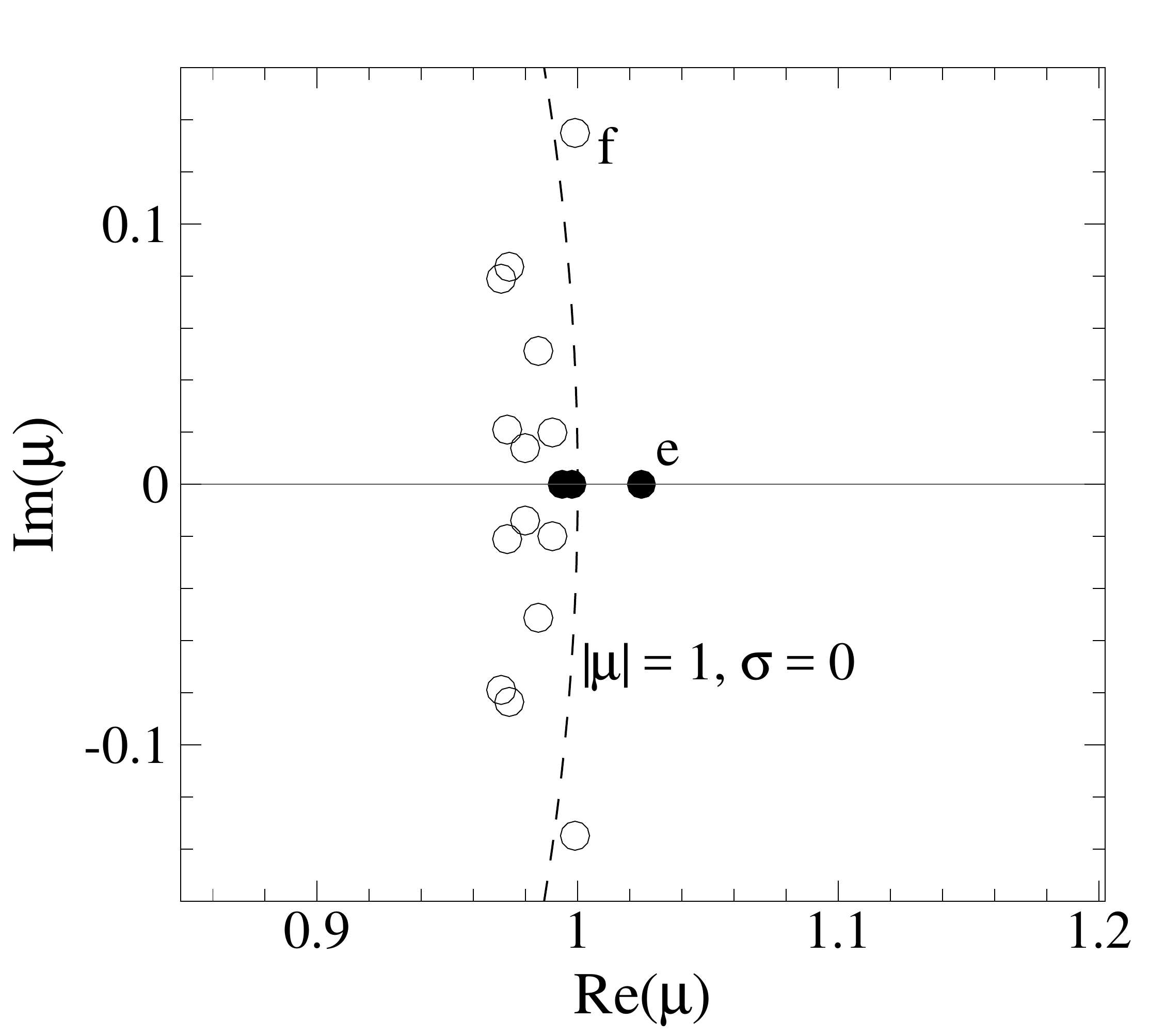}
} &
\parbox{0.5\textwidth}{
(e)\\
\includegraphics[width=0.5\textwidth,keepaspectratio]{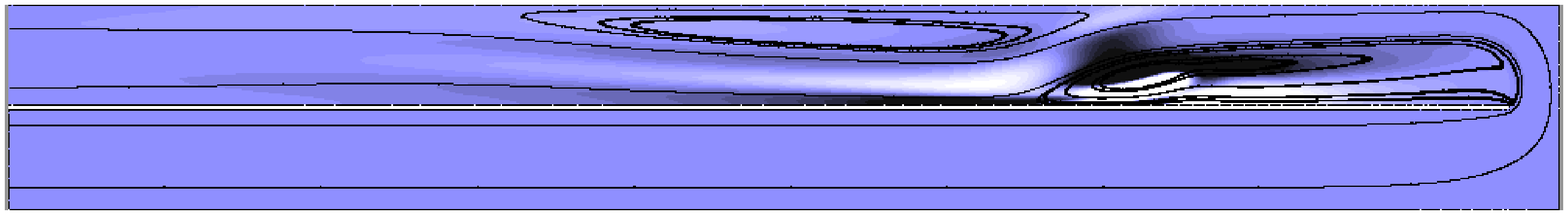}\\
(f) \\
\includegraphics[width=0.5\textwidth,keepaspectratio]{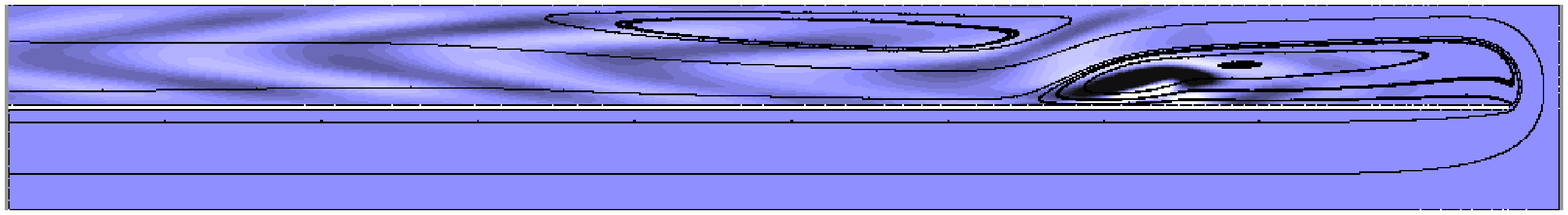}
}\\
\parbox{0.45\textwidth}{
(g)\\
\includegraphics[width=0.45\textwidth,keepaspectratio]{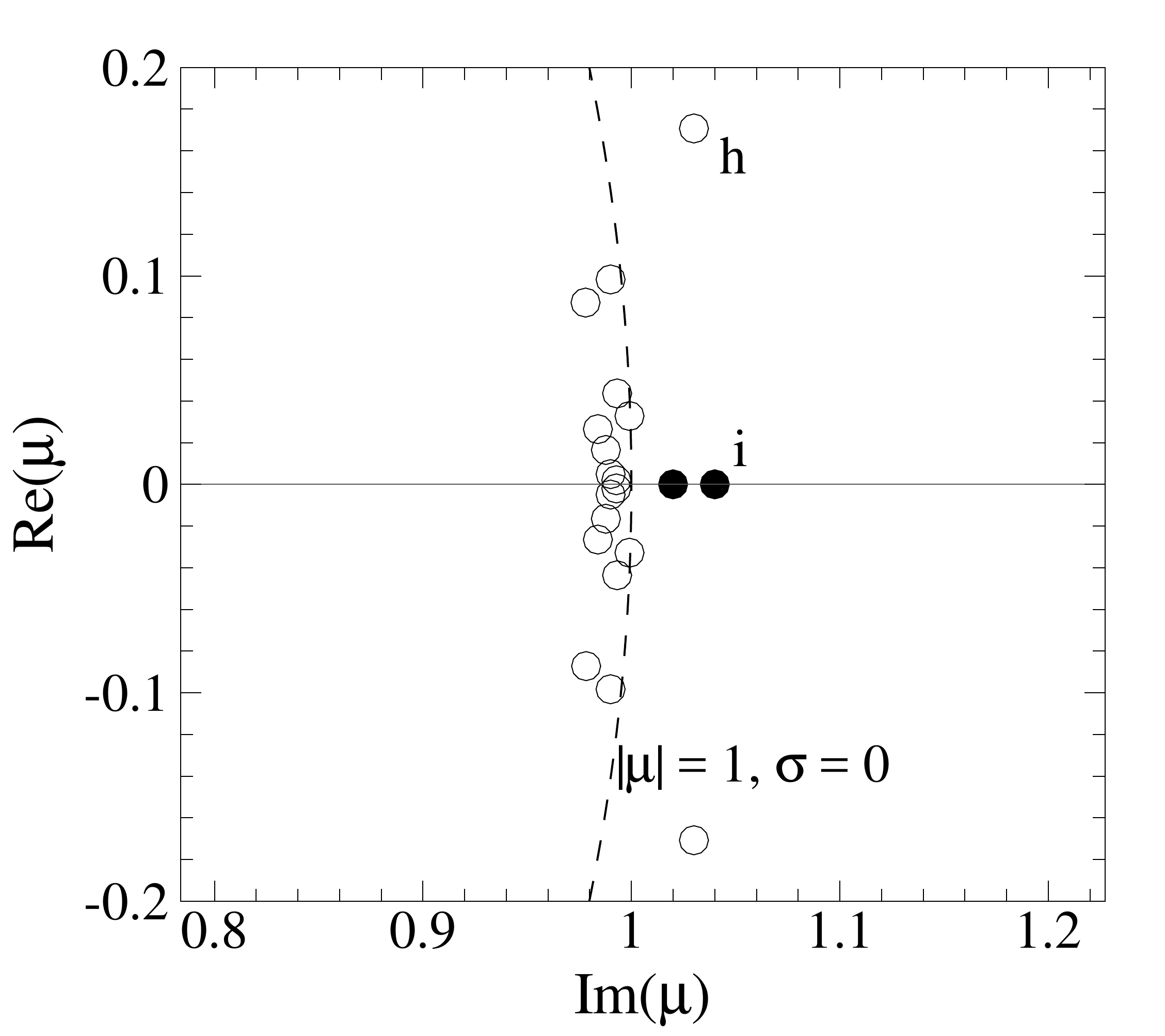}
} &
\parbox{0.5\textwidth}{
(h)\\
\includegraphics[width=0.5\textwidth,keepaspectratio]{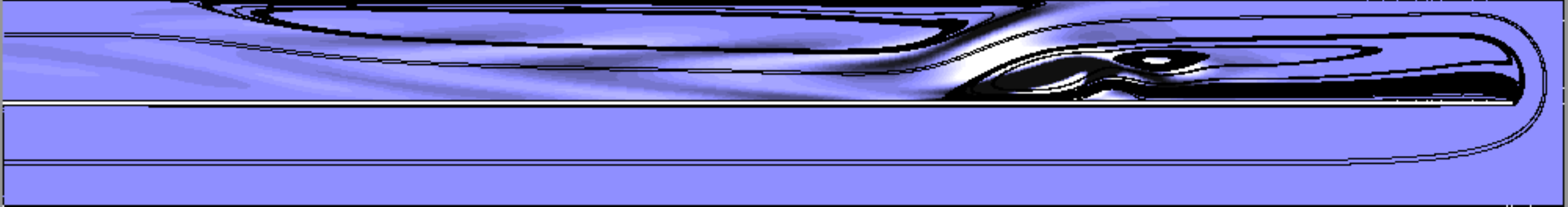}\\
(i) \\
\includegraphics[width=0.5\textwidth,keepaspectratio]{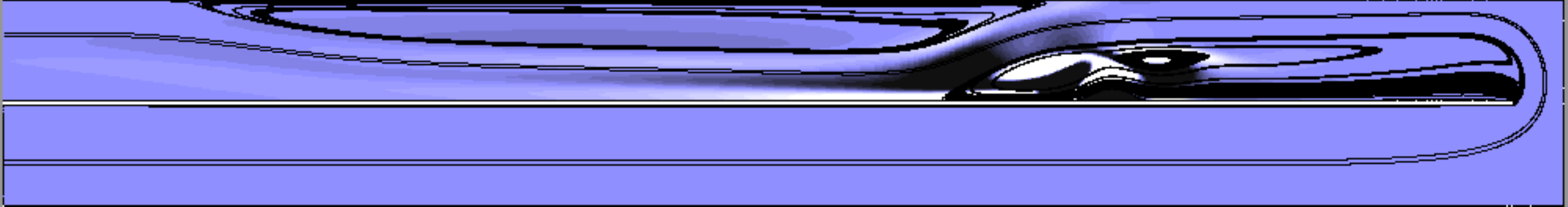}
}\\
\end{tabular}
\caption{Eigenvalue spectra for (a) $\beta=0.2$, $\Rey=120$, $k=2.8$, (d) $\beta=0.5$, $\Rey=400$, $k=3.5$, and (g) $\beta=0.5$, $\Rey=600$ and $k=4.5$. (b, c, e, f, h, i) visualise the real part of the complex eigenvector fields via spanwise vorticity on the plane at $z=0$. (b), (e) and (h) show the respective leading eigenvalues, while (c), (f) and (i) show the corresponding field for the second most dominant eigenmode. In these vorticity plots, zero vorticity is shown by the mid-level shading, while darker and lighter shading respectively show negative and positive shading.}
\label{fig:Spectra}
\end{figure}
Figure~\ref{fig:Spectra}(a, d, g) show the eigenvalue spectra for three different cases. Each has a different type of leading eigenmode, and the dashed line indicates the onset of instability ($\sigma=0$). The real part of the eigenmodes are visualised via plots of spanwise vorticity on the plane $z=0$ in figure~\ref{fig:Spectra}(b, c, e, f, h, i). Figure~\ref{fig:Spectra}(a) depicts the eigenvalue spectrum for $\beta=0.2$, $\Rey=120$ and $k=2.8$, which is near to the onset of instability. Two complex-conjugate pairs of non-real eigenvalues are the fastest-growing eigenvalues in the spectrum. The leading pair (\eg\ figure~\ref{fig:Spectra}(b)) exhibits a strong growth rate, with strong perturbation structure in the primary recirculation bubble, while the second pair (\eg\ figure~\ref{fig:Spectra}(c)) has perturbation structure mainly localised in the bulk flow near the secondary recirculation bubble.

In contrast to $\beta=0.2$, larger $\beta$ tend to favour synchronous leading modes. $\beta=0.5$ is particularly interesting because this is the point a transition from synchronous ($\Rey=400$ in figure~\ref{fig:Spectra}(d)) to oscillatory ($\Rey=600$ in figure~\ref{fig:Spectra}(g)) leading mode is seen. The leading eigenvalue for $\Rey=400$ is synchronous, while the second-largest eigenmode is oscillatory. However, at higher $\Rey$, the oscillatory mode has higher growth rate compared to the synchronous mode, as can be seen in figure~\ref{fig:Spectra}(g). The perturbation fields associated with these modes are strongest in the same vicinity, which is in the primary recirculation bubble near to the reattachment point. From the contours represented in figures~\ref{fig:Spectra}(h) and~(i), we can see that the contours in figure \ref{fig:Spectra}(b), \ref{fig:Spectra}(e) and \ref{fig:Spectra}(i) have a similar synchronous eigenmode structure; meanwhile figures \ref{fig:Spectra}(c), \ref{fig:Spectra}(f) and \ref{fig:Spectra}(h) have a consistent oscillatory eigenmode structure. The oscillatory mode structure is distinguished from the synchronous mode structure by the presence of an array of chevron-shaped vorticity structures following the core flow downstream from the aft end of the primary recirculation bubble.

\subsection{Dependence on $\beta$ and analogy with related flows}\label{subsect:dependence_beta}
From the known influence of the expansion ratio to the \threed\ characteristics of backward-facing step \citep{barkley2002three,lanzerstorfer2012global} and opening ratio in partially blocked channel \citep{griffith2007wake} and to the \twod\ characteristics of 180-degree sharp bend \citep{zhang2013influence}, we shall expect that the \threed\ flow in a 180-degree sharp bend flow is also dependant on the same physical parameter (here $\beta$).

\begin{figure}
\centering
\begin{tabular}{ll}
(a) & (b) \\
\multicolumn{1}{c}{\includegraphics[width=0.48\textwidth,keepaspectratio]{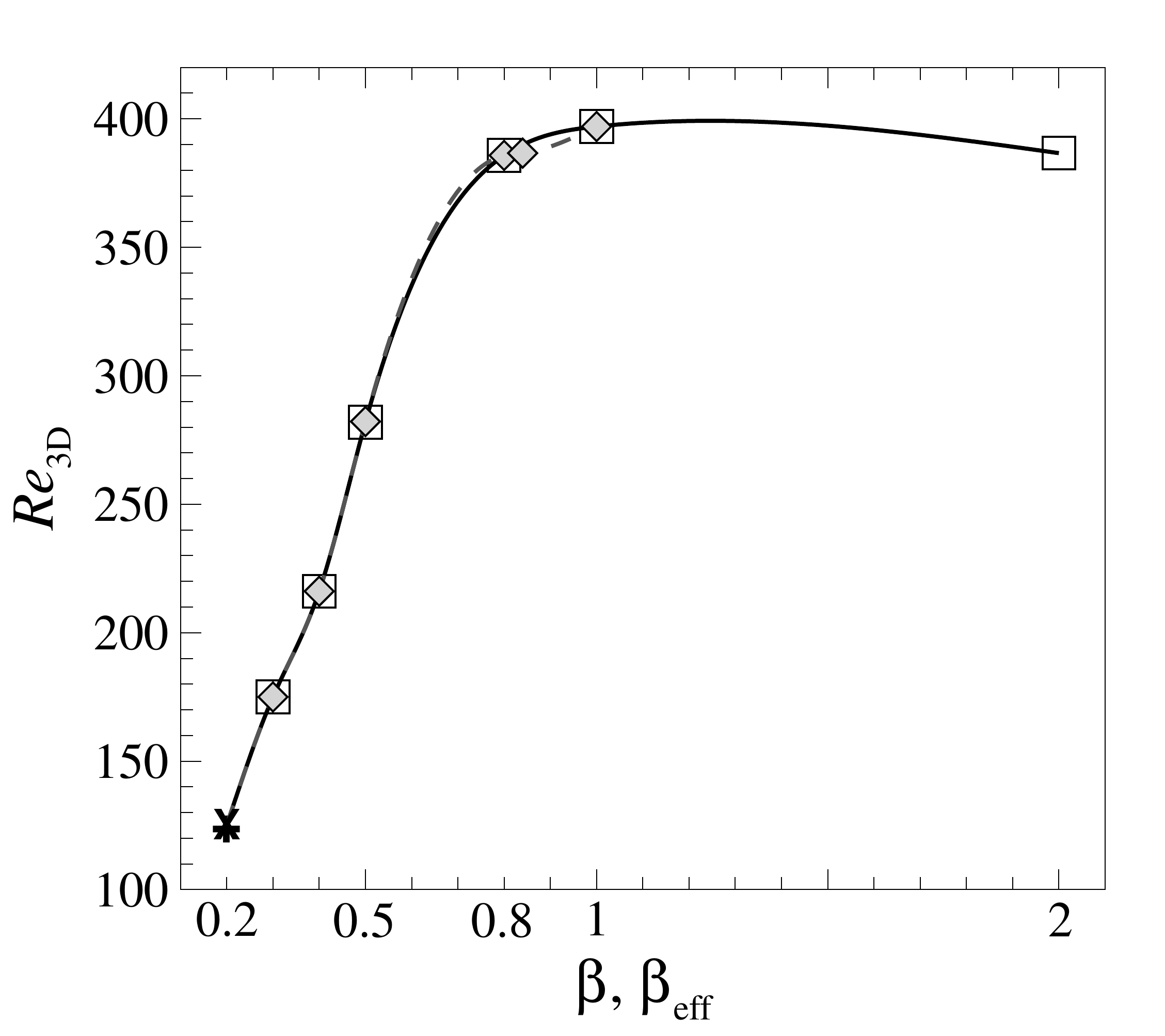}} &
\multicolumn{1}{c}{\includegraphics[width=0.48\textwidth,keepaspectratio]{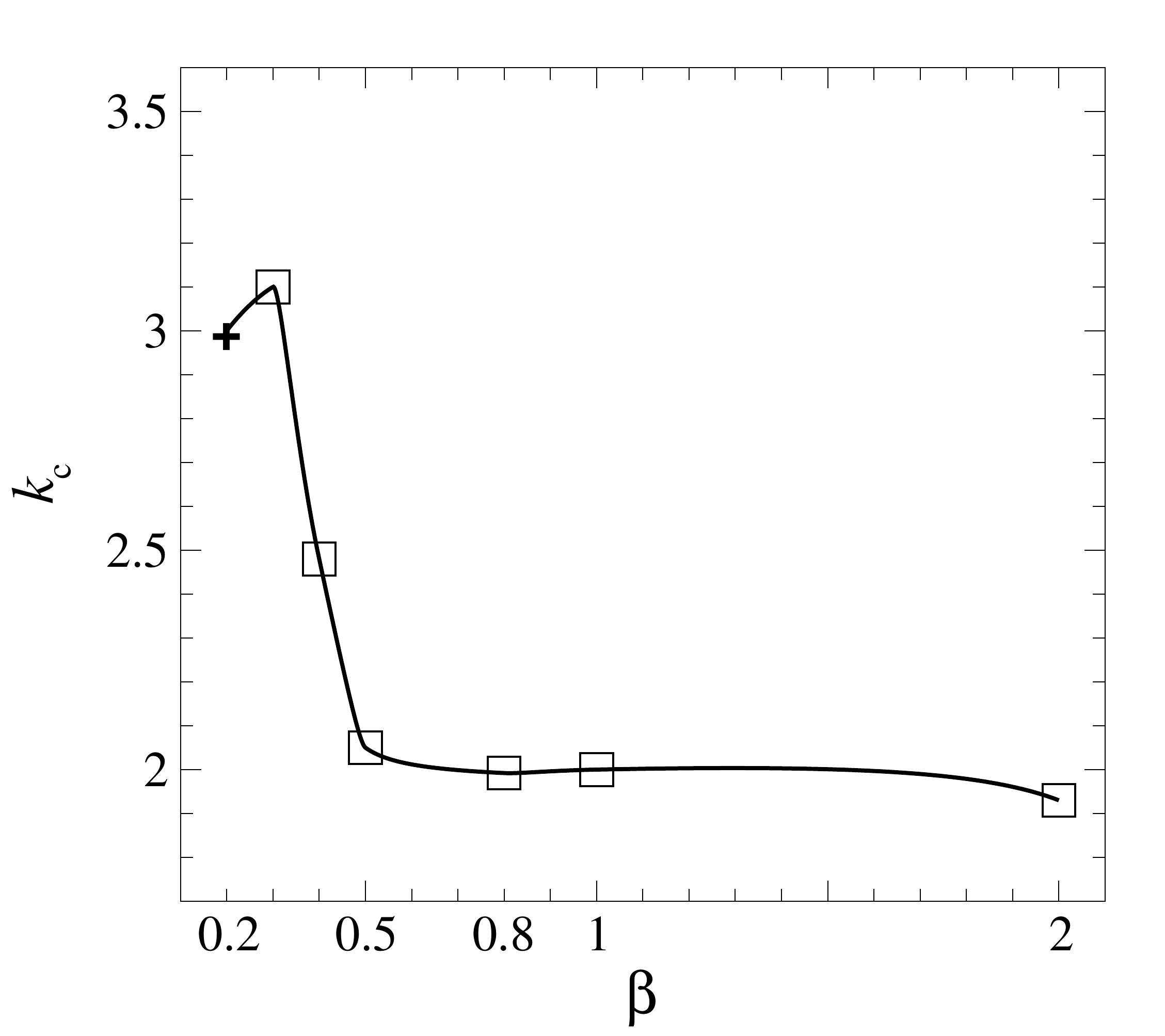}} \\
\end{tabular}
\caption{Critical (a) Reynolds number and (b) wavenumber as a function of opening ratio. In (a) the critical Reynolds number is also plotted against effective opening ratio $\beta_\textrm{eff}$. For $\beta$ data, synchronous modes and oscillatory modes are represented respectively by ``$\square$'' and ``$+$''. For $\beta_\textrm{eff}$ data, synchronous and oscillatory modes are respectively represented by ``$\diamond$'' and ``$\times$''.}
\label{fig:critical_value_vs_beta}
\end{figure}
Figure~\ref{fig:critical_value_vs_beta} illustrates the effect of $\beta$ on the $\Rey$ and $k$ at the onset of \threed\ instability. In the range of $\beta$ studied, only the case $\beta=0.2$ becomes \threed\ through the onset of an oscillatory mode; while all other cases transition through the onset of a non-oscillatory mode. Apparently, $\Rey_\textrm{3D}$ increases steadily as the bend opening becomes larger until $\beta\approx 1$. $\Rey_\mathrm{3D}$ at $\beta=2$ is lower than $\beta=1$ because of the appearance of the recirculation bubble at the far end of the bend wall that limits the width of the bulk flow in the bend, causing a reduction in effective value of $\beta$. If we look closely, the value of $\Rey_\textrm{3D}$ at $\beta=2$ is almost the same as that of $\beta=0.8$, which testifies to the importance of $\beta_\textrm{eff}$ in determining the stability of the flow. In figure~\ref{fig:critical_value_vs_beta}(a), $\Rey_\textrm{3D}$ is also plotted against $\beta_\textrm{eff}$ to illustrate this behaviour.

The eigenvector fields at the onset of instability for all synchronous modes ($\beta\gtrsim 0.3$) have perturbation structure located in the recirculation bubble, similar to what was found in the backward-facing step flow \citep{barkley2002three,alam2000direct}, partially blocked channel flow \citep{griffith2007wake} and separation flow \citep{hammond1998local}. \citet{barkley2002three} concluded that the size and shape of the bubble directly affected the instability. Conversely, \citet{hammond1998local} and \citet{alam2000direct} confirmed in their studies that the onset of local absolute instability depended on the backflow of the bubble. Interestingly, $\beta=0.8$ and $\beta=2$ have comparable bubble size, peak backflow velocity and location of the peak backflow velocity (with discrepancies of only $4.3\%$, $1.0\%$ and $0.4\%$, respectively). This supports the view that for $\beta>1$, the stability of the flow is characterised by $\beta_\textrm{eff}$, in both of these cases $\beta_\textrm{eff}=0.8$.

The dependence of $k_c$ on $\beta$ is also illustrated in figure~\ref{fig:critical_value_vs_beta}. As $\beta$ increases from 0.3 to 0.5, the dominant wavelength of the instability increases from $\lambda \approx 2\pi/3$ to $\approx\pi$. This is perhaps due to the transition from jet-like flow around the bend to a broader turning flow. $k_c$ at $\beta=0.2$ does not follow the trend observed between $\beta=0.3$ and 0.5, due to the different mechanism. For $\beta\geq 0.5$, the critical wavenumber exhibits little dependence on $\beta$.

\begin{figure}
\centering
  \includegraphics[width=0.6\columnwidth,keepaspectratio]{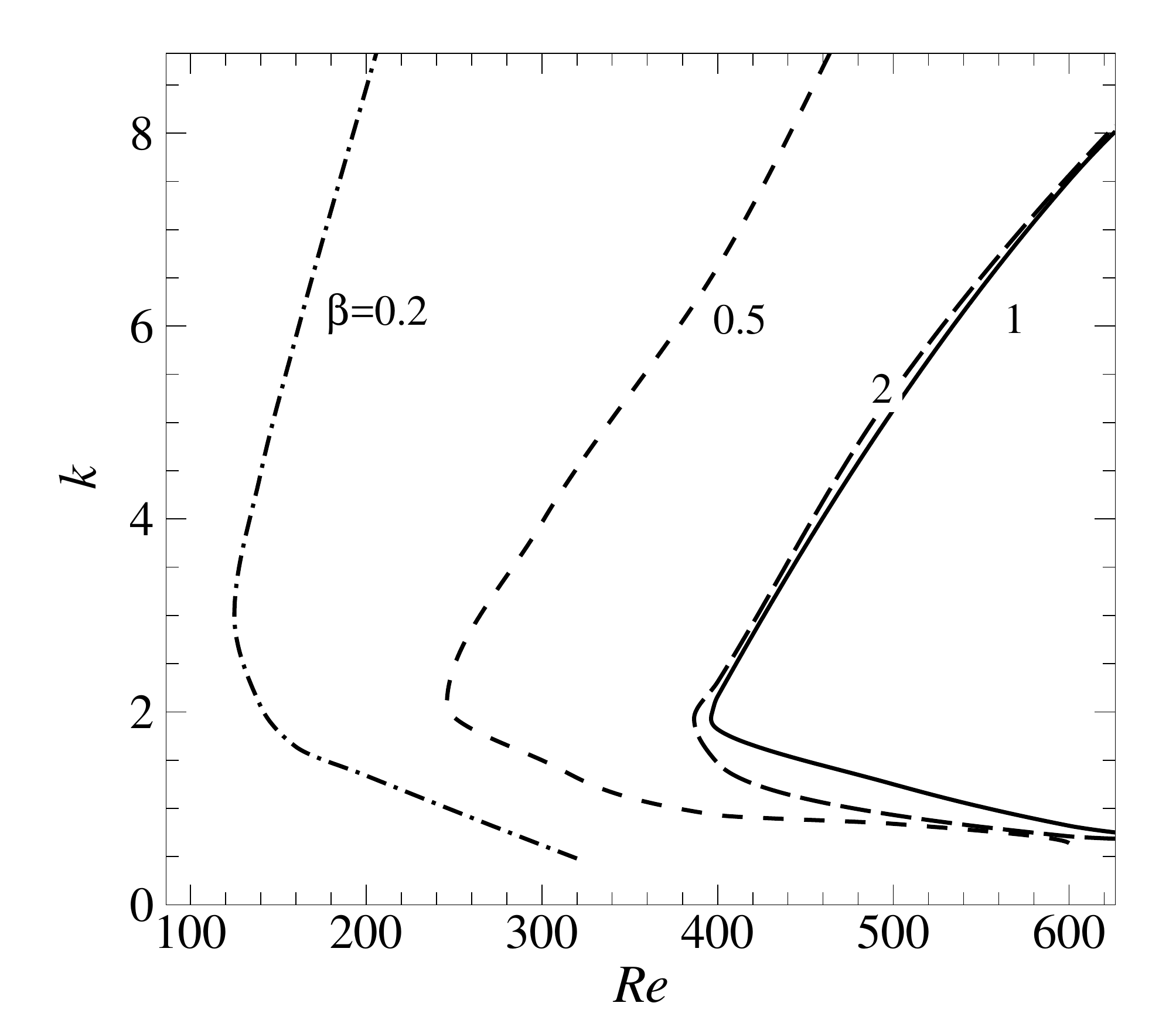}
  \caption{Marginal stability curves for sharp 180-degree bend flow with $\beta=0.2$, 0.5, 1, and 2. Regions on the right of the curves represent flow conditions that are linearly unstable to \threed\ perturbations for that particular $\beta$.}
\label{fig:stability_curve}
\end{figure}
\begin{figure}
\centering
\begin{tabular}{ll}
(a) & (b) \\
\multicolumn{1}{c}{\includegraphics[width=0.48\textwidth,keepaspectratio]{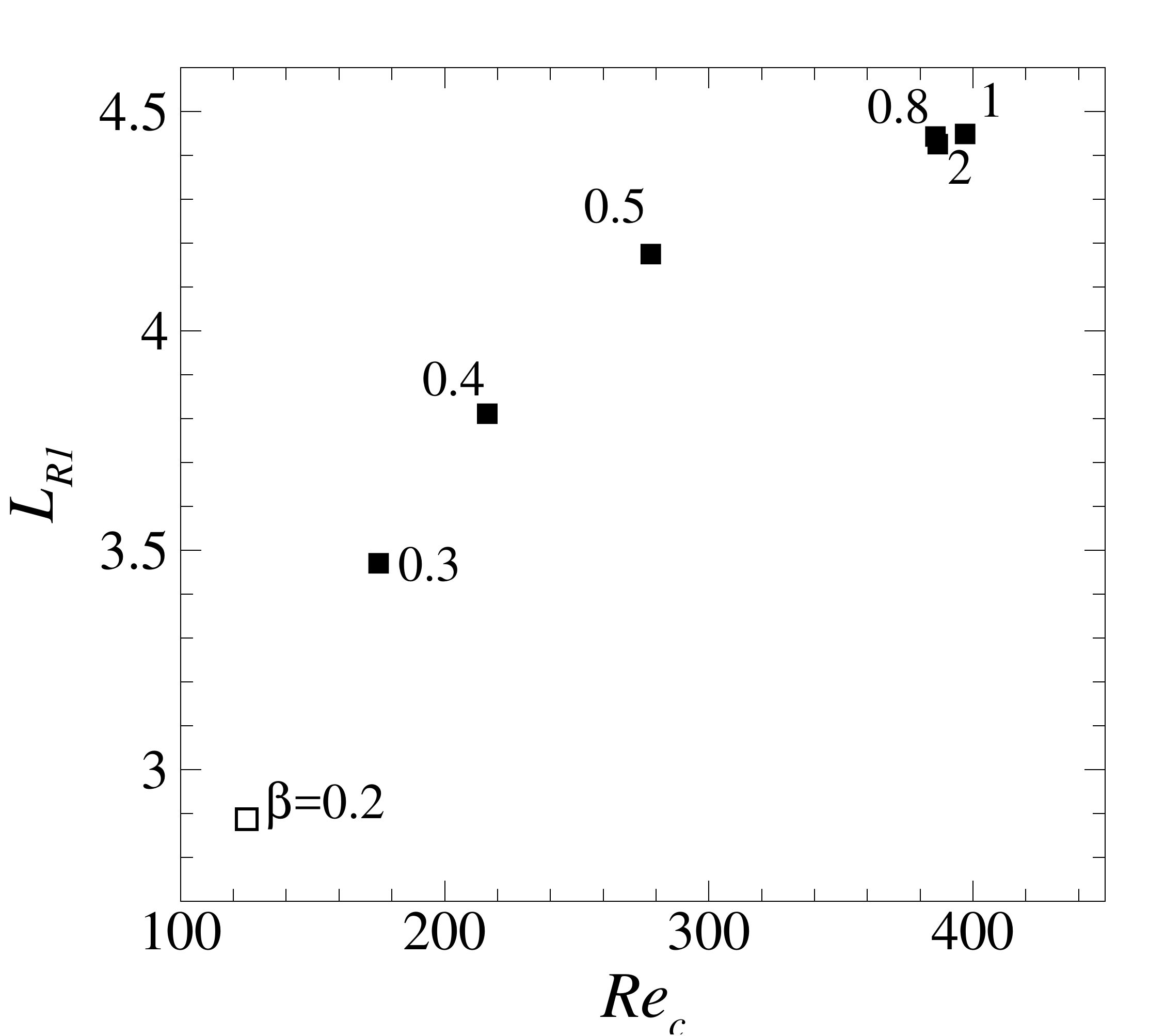}} &
\multicolumn{1}{c}{\includegraphics[width=0.48\textwidth,keepaspectratio]{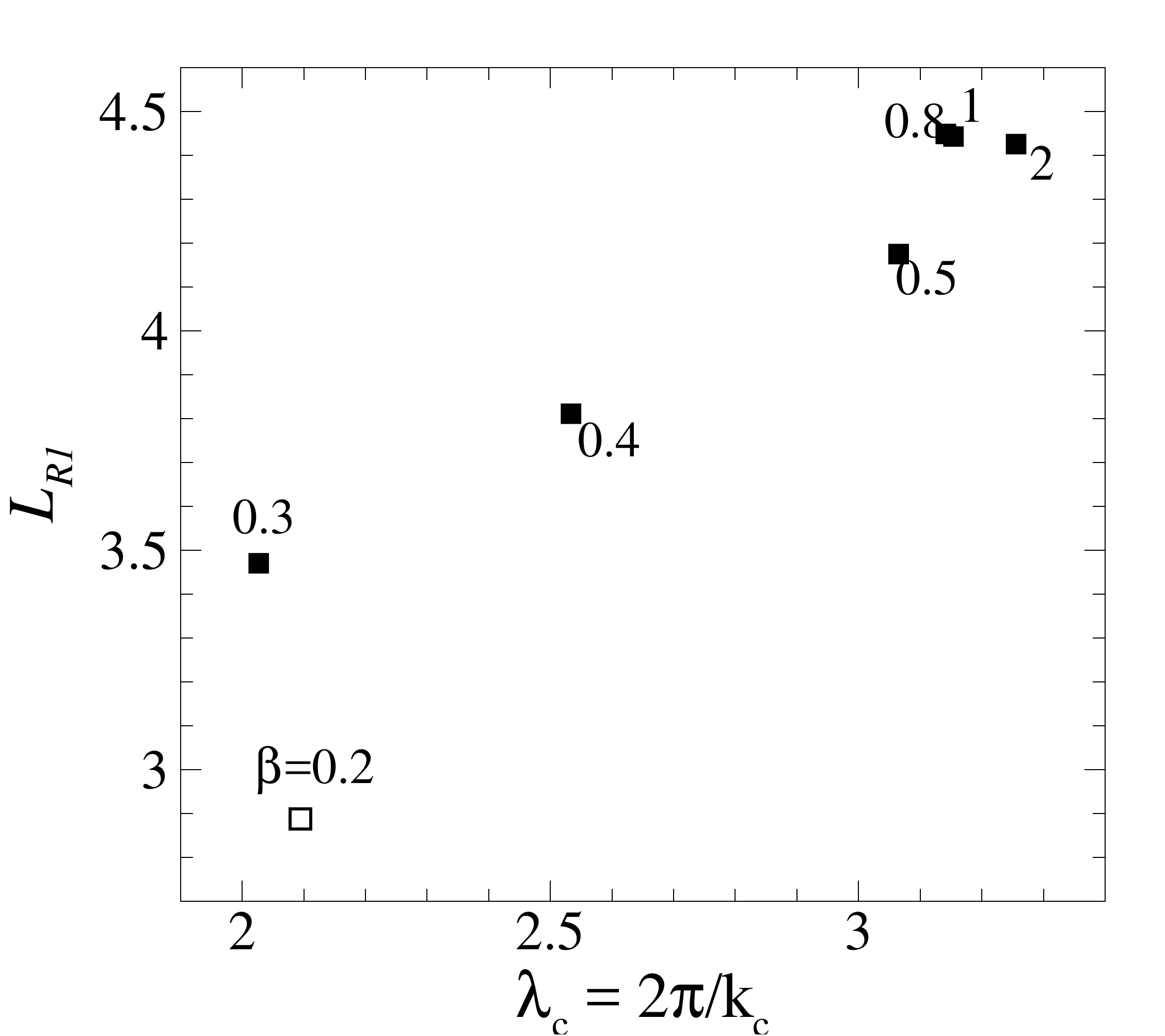}} \\
\end{tabular}
\caption{Length of primary recirculation bubble as a function of (a) critical Reynolds number and (b) critical wavelength.}
\label{fig:Lr_vs_kc}
\end{figure}
Figure~\ref{fig:stability_curve} shows the marginal stability curves at several $\beta$. The marginal curves are obtained by interpolating $\sigma$($k,\Rey$) to zero growth rate for each $\Rey$ from figure~\ref{fig:lsa_curve}. For $\beta \leq 1$, with increasing $\beta$ the neutral stability curve shifts to the right as expected as the flow with wider bend opening ratio is more stable than those with smaller opening ratios. Beyond $\beta\approx 1$, the stability curve recesses slightly towards lower Reynolds numbers, occupying the region between the marginal stability curves for $\beta=0.5$ and~$1$. This is explained by the decrease in $\beta_\textrm{eff}$ with $\beta$ for $\beta>1$ (see \S~\ref{sect:2D_flow}). The stability curves also show a decrease in dominant wavenumber with increasing $\beta$. This is because the instability is scaled with bubble size as discussed by \citet{barkley1996three}. The length of the primary recirculation bubble as a function of critical Reynolds number and critical wavelength is depicted in figure~\ref{fig:Lr_vs_kc}. It can be seen from figure~\ref{fig:Lr_vs_kc}(a) that the flow becomes unstable at higher $\Rey$ at bigger $\beta$. For the synchronous modes ($0.3\lesssim\beta\lesssim 1$), the critical wavelength increases as the size of the primary recirculation bubble increases.

\begin{figure}
\centering
  \includegraphics[width=0.6\columnwidth,keepaspectratio]{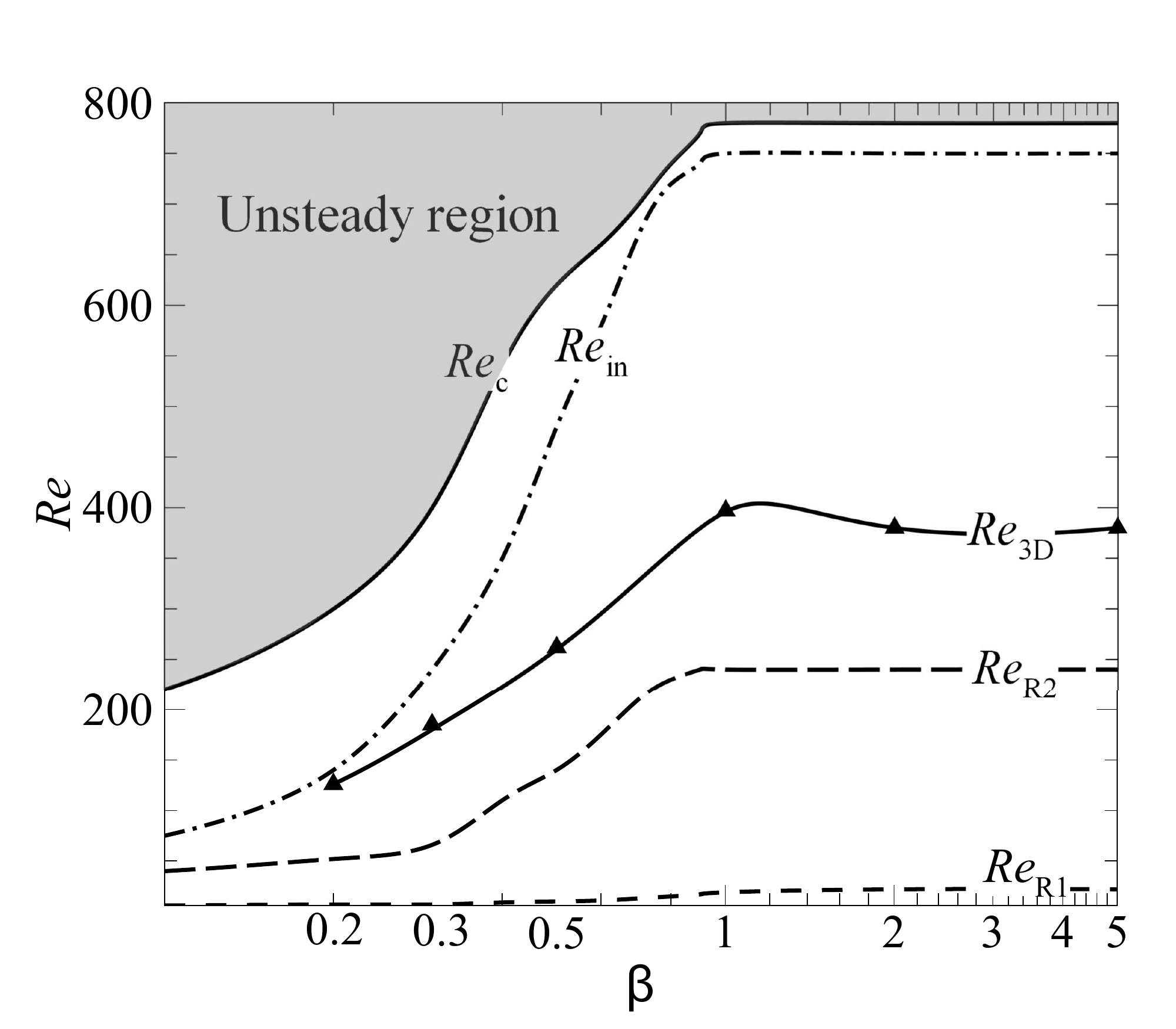}
  \caption{Parameter space for flow regimes and leading peak eigenvalues of \threed\ stability. The onset of \threed\ instability is represented by $\triangle$ symbols. Lines are included for guidance, and the shaded region shows the parameter values exhibiting unsteady \twod\ solutions.}
\label{fig:regime+eigen}
\end{figure}
In order to consider the \threed\ stability of these flows in the context of the underlying \twod\ flows, figure~\ref{fig:regime+eigen} summarizes the ($\Rey,\beta$) parameter space explored. Within this range, instability to \threed\ perturbations always occurs in the regime where both primary and secondary recirculation bubbles exist. This agrees with \citet{armaly1983experimental}, who found for the backward-facing step that three-dimensionality appeared in the flow after the secondary recirculation bubble had formed. In our study, $\Rey_\textrm{3D}$ increases monotonically with $\beta$ before slightly reducing and becoming independent on $\beta$ as $\beta$ exceeds unity. The critical eigenvalue for $\beta=0.2$ is found to be non-real, while it is real for $\beta=0.5$, $1$ and $2$. Across all considered $\Rey$ studied at $\beta=0.2$, the dominant eigenvalues are non-real. On the other hand, for $\beta\gtrsim 1$, the dominant eigenvalues are consistently real. A transition from real to non-real (from solid to hollow symbol) can be seen at $\beta=0.5$ in figure~\ref{fig:regime+eigen} which occurs near the regime where the inside recirculation appears in the primary recirculation bubble.

\subsection{Instability mode structure}\label{subsect:mechanism}
In this section, the mechanisms by which the \threed\ infinitesimal perturbations are amplified are addressed. The obvious mechanism seen in the \twod\ flow is Kelvin--Helmholtz instability. \citet{zhang2013influence} found that in regime~\rom{4}, the shear layers around both bubbles are subject to it, and lead to unsteadiness. However, as this study demonstrates, absolute \threed\ instabilities are found at $\Rey \ll \Rey_\mathrm{c}$ involving different mechanisms.

As mentioned earlier, the structure of instability affecting the primary recirculation bubble bears a strong similarity to both the flow over a backward-facing step and in a partially blocked channel. \citet{ghia1989analysis} suggested that the appearance of the secondary bubble introduced a concave curvature in the streamlines of the bulk flow, thus inducing Taylor--G\"{o}rtler instability. However, this scenario has been ruled out \citep{barkley2002three,griffith2007wake} because instability arises neither in the secondary recirculation bubble nor in the main bulk flow between the primary and secondary recirculation bubble zones. The leading instability mode from our analysis is also localised in a different location; though the second-leading lower-wavenumber eigenmode is found to be located in these regions.

\begin{figure}
\centering
\begin{tabular}{l@{}l@{}}
(a) $\beta=0.2$, $\Rey=123$ and $k=2.8$ & \\
{\begin{tabular}{l}
(i)\\
{\includegraphics[width=0.40\textwidth,keepaspectratio]{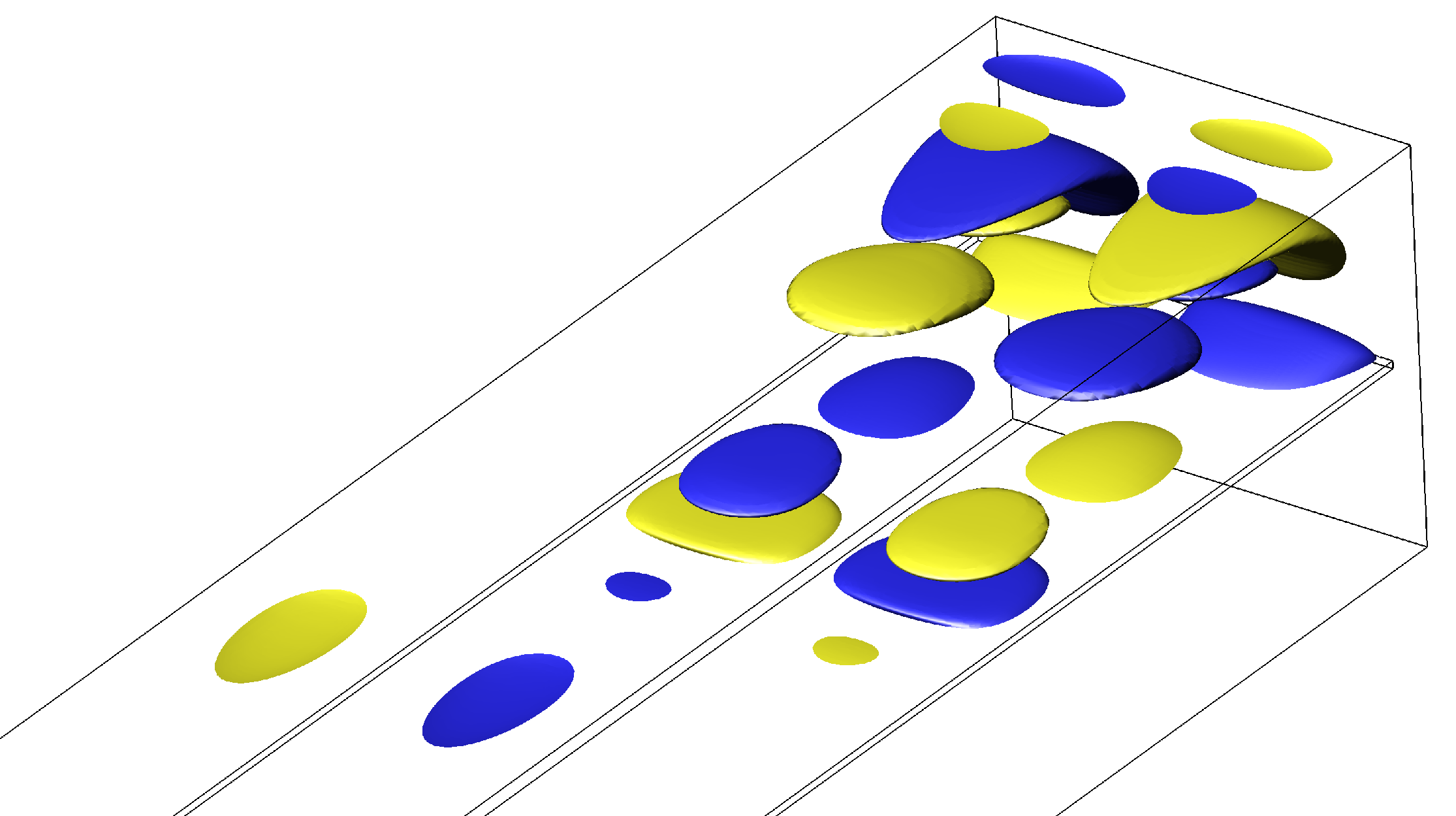}}
\end{tabular}} & {\begin{tabular}{l}
(ii)\\
{\includegraphics[width=0.45\textwidth,keepaspectratio]{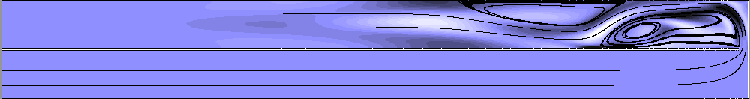}} \\
(iii)\\
{\includegraphics[width=0.45\textwidth,keepaspectratio]{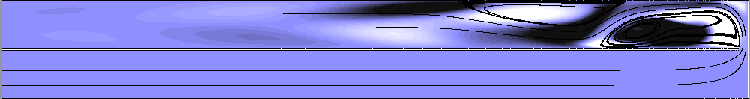}} \\
\end{tabular}}\\
(b) $\beta=0.5$, $\Rey=278$ and $k=2$ & \\
{\begin{tabular}{l}
(i)\\
{\includegraphics[width=0.40\textwidth,keepaspectratio]{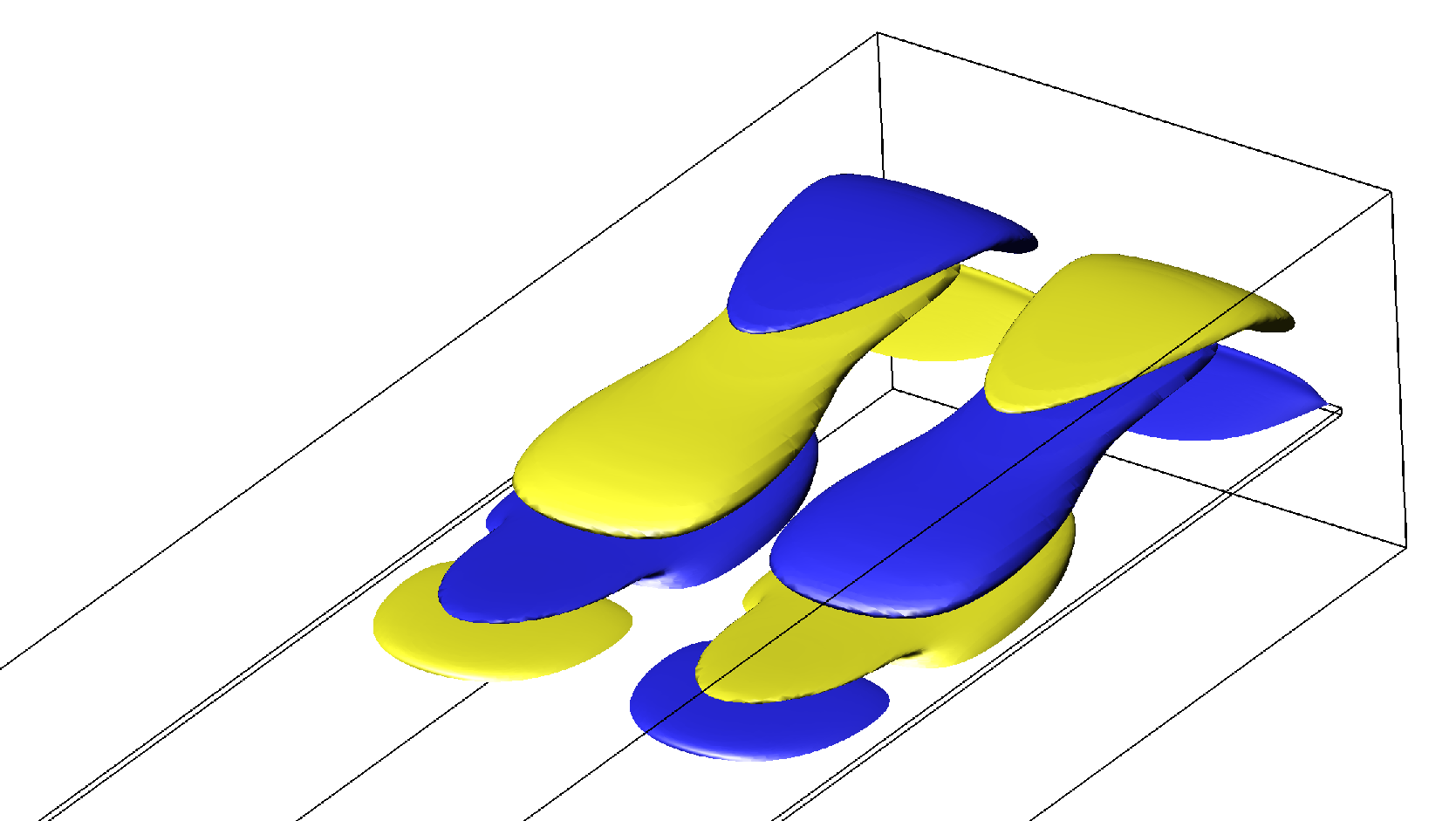}}
\end{tabular}} & {\begin{tabular}{l}
(ii)\\
{\includegraphics[width=0.45\textwidth,keepaspectratio]{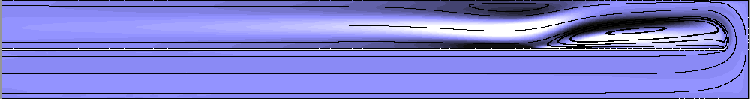}} \\
(iii)\\
{\includegraphics[width=0.45\textwidth,keepaspectratio]{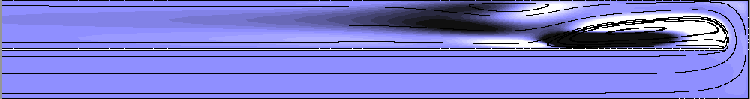}} \\
\end{tabular}}\\
(c) $\beta=1$, $\Rey=397$ and $k=2$ & \\
{\begin{tabular}{l}
(i)\\
{\includegraphics[width=0.40\textwidth,keepaspectratio]{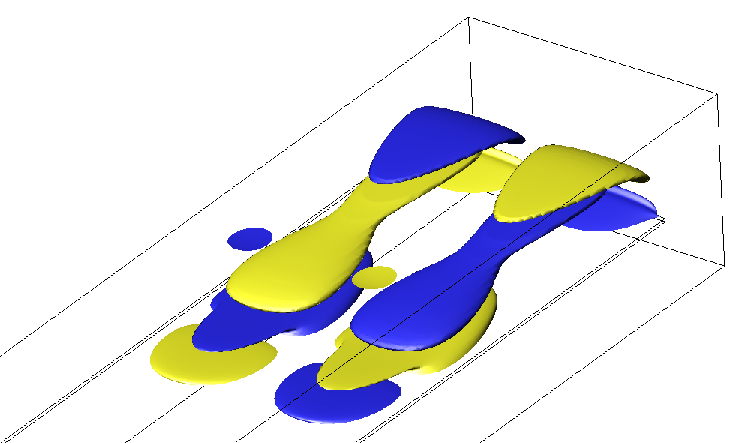}}
\end{tabular}} & {\begin{tabular}{l}
(ii)\\
{\includegraphics[width=0.45\textwidth,keepaspectratio]{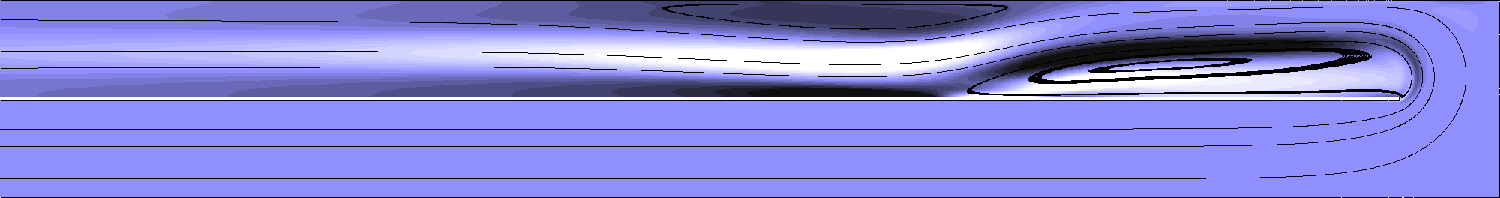}} \\
(iii)\\
{\includegraphics[width=0.45\textwidth,keepaspectratio]{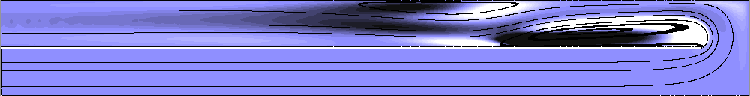}} \\
\end{tabular}}\\
(d) $\beta=2$, $\Rey=380$ and $k=2$ & \\
{\begin{tabular}{l}
(i)\\
{\includegraphics[width=0.40\textwidth,keepaspectratio]{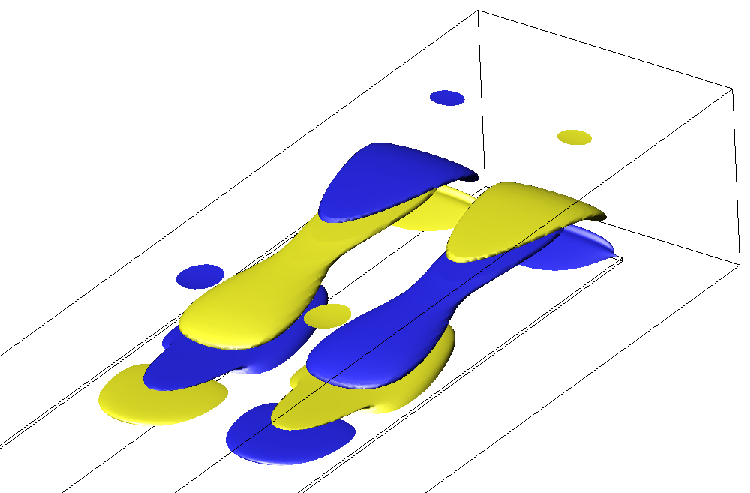}}
\end{tabular}} & {\begin{tabular}{l}
(ii)\\
{\includegraphics[width=0.45\textwidth,keepaspectratio]{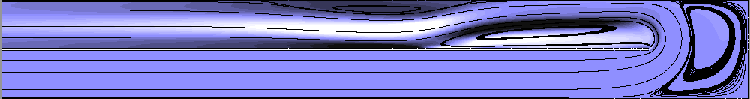}} \\
(iii)\\
{\includegraphics[width=0.45\textwidth,keepaspectratio]{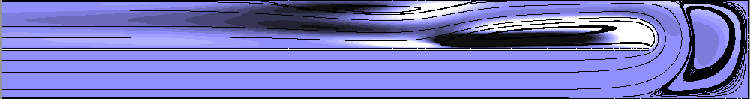}} \\
\end{tabular}}\\
\end{tabular}
\caption{Visualisation of the real part of leading eigenmodes at (a) $\beta=0.2$, (b) $\beta=0.5$, (c) $\beta=1$ and (d) $\beta=2$, consisting of (i) a \threed\ visualisation of the streamwise ($x$-component) of vorticity, (ii) spanwise ($z$-component) vorticity and (iii) spanwise velocity contours overlaid with the base flow streamlines. Dark-mid-light shading denotes negative-zero-positive levels, respectively, while (ii) and (iii) are plotted on the plane $z=0$.}
\label{fig:eigenmodes}
\end{figure}
Figure~\ref{fig:eigenmodes} visualises the real part of the leading eigenmode at gap ratios spanning $0.2 \leq \beta \leq 2$. Three-dimensionality appears at the reattachment and  separating points of the primary recirculation bubble as shown by the spanwise velocity component in figure~\ref{fig:eigenmodes}(b)(iii), (c)(iii) and (d)(iii). The perturbation is strongest within the closed streamlines of the primary recirculation bubble (ref.\ the isosurface plots in figure~\ref{fig:eigenmodes}), with further perturbation structure also present in the secondary bubble and propagating downstream into the core jet flow.

Spanwise perturbation vorticity contour plots shown in figure~\ref{fig:eigenmodes} exhibit perturbation vorticity structures that resemble those arising from an elliptic instability, namely a pair of counter-rotating vortices inside the recirculation bubble. A similar interaction of counter-rotating vortices in \twod\ elliptic streamlines was seen in \citet{thompson2001physical}. \citet{leweke1998cooperative} suggested that when two counter-rotating vortices balance each other, the radial component of the strain field leads the disturbance to grow exponentially.


\citet{lanzerstorfer2012global} observed that the combination of the flow deceleration near the reattachment point, a lift up process on both sides of the bulk flow between the primary and the secondary recirculation bubbles, and an amplification due to streamline convergence near and in the separated flow regions is the cause of the flow instability in the flow over backward-facing step with expansion ratio of $0.5$. The same observation was made by \citet{wee2004self} where they found that the backward-facing step flow was locally absolutely unstable near the middle of the primary recirculation bubble. The backflow was found to be high and the shear layer was sufficiently thick to support an absolutely unstable mode; hence, an absolute mode was more likely to originate in the middle of the bubble. By contrast, \citet{marquillie2003onset} studied a flow behind a bump and observed that the structural changes near the reattachment point of the primary recirculation bubble behind the bump triggered an abrupt local transition from convective to absolute instability. It is observed that for the flow around a 180-degree sharp bend, as $\Rey$ increases, the location of the peak backflow in the primary recirculation bubble shifted towards the reattachment point. The close gap between the peak backflow and the reattachment point means that the flow is strongly decelerated upon approaching the reattachment point. Interestingly, the peak backflow at the onset of instability in this study for $0.3\lesssim \beta \lesssim 2$ is consistently located about $0.22L_R$ from the reattachment point which is also the same location of the peak perturbation spanwise velocity from the linear stability analysis (figure~\ref{fig:eigenmodes} biii, ciii, diii). This suggests that as in the other geometries mentioned, the instability in the $180$-degree sharp bend for $\beta\geq0.3$ is localised near the peak of back flow intensity.

The oscillatory critical mode at $\beta=0.2$ exhibits strong spanwise velocities at the upstream end of the primary recirculation bubble and quite strong values around the intense vortex in the bubble. This is strongly consistent with a mechanism involving a centrifugal instability around the intense vortex. A similar mode was seen by \citet{lanzerstorfer2012global} in the flow over a backward-facing step with a very small opening. There, the perturbations were found to be a spanwise travelling wave that displaces the jet and the intense vortex periodically.


The structure of the mode that destabilizes the flow at $\beta=0.5$, $\Rey=278$ and $k=2$ is shown in figure~\ref{fig:eigenmodes}(b) in the isosurface plots of streamwise vorticity, spanwise vorticity and $w$ velocity contours. The isosurface consists of positive (light) and negative (dark) vorticity contours located almost entirely in the primary recirculation bubble, near the separation and reattachment points. The spanwise vorticity contour plot shows that there is a pair of counter-rotating vortices in the primary recirculation bubble which resemble the flow in a partially blocked channel \citep{griffith2007wake}. The spanwise velocity contours are also qualitatively similar to those of the unstable mode in the flow over a backward-facing step \citep{barkley2002three} where a ``flat roll'' (\ie\ a roll in a horizontal plane about a vertical axis) mode structure exists in the bubble near the reattachment point. Across the opening ratios studied, the same mode structure has been found for all real primary leading eigenmodes indicated by the solid symbols in figure~\ref{fig:regime+eigen}.

The same type of plots describing the dominant eigenmodes for $\beta=0.2$, $\Rey=123$ and $k=2.8$ are shown in figure~\ref{fig:eigenmodes}(a). The mode appears to grow in the primary recirculation bubble near the separation point and upstream of the intense vortex close to the reattachment point. Both of these modes have strong $x$-component of perturbation vorticity in the upstream part of the primary recirculation bubble which is where \citet{zhang2013influence} found secondary instability in their \threed\ simulation of an unsteady flow at $\Rey=2000$ and $\beta=1$ with spanwise periodic domain of length $2$ units (this is equivalent to the case of wavenumber $k=\pi$ in the present notations).

\begin{figure}
\centering
\begin{tabular}{l}
(a) Streamwise vorticity  \\
{\includegraphics[width=0.6\textwidth,keepaspectratio]{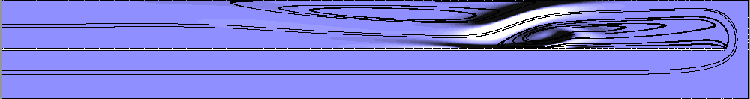}} \\
(b) $u$-velocity  \\
{\includegraphics[width=0.6\textwidth,keepaspectratio]{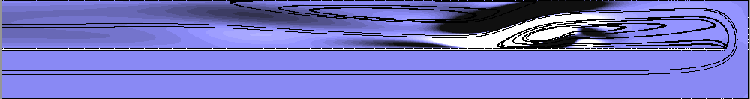}} \\
(c) $v$-velocity \\
{\includegraphics[width=0.6\textwidth,keepaspectratio]{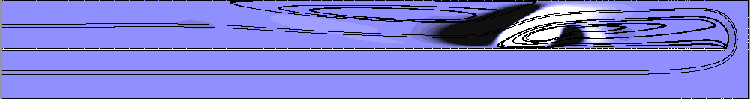}} \\
(d) $w$-velocity  \\
{\includegraphics[width=0.6\textwidth,keepaspectratio]{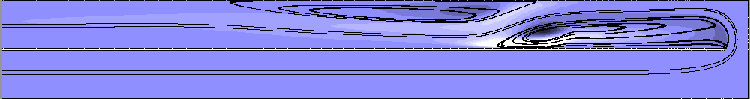}} \\
\end{tabular}
\caption{Structure of the real part of the leading eigenmode at $\beta=0.5$, $\Rey=400$ and $k=0.2$ depicted on the plane $z=0$: flooded contours of (a) streamwise vorticity and (b) $u$, (c) $v$ and (d) $w$-velocity overlaid with the base flow streamlines. For clarity, only the vicinity of the bend is shown. Dark-mid-light shading denotes negative-zero-positive levels, respectively.}
\label{fig:uvw_b1_r400_k0.2}
\end{figure}
The structure of the most unstable eigenmode at very small wavenumber is shown in figure~\ref{fig:uvw_b1_r400_k0.2} as a plot of spanwise vorticity and $(u,v,w)$ velocity contours. The structure consist of spanwise vortices in the main bulk flow located between the primary and secondary recirculation bubbles. The spanwise velocity contour in figure~\ref{fig:uvw_b1_r400_k0.2}(d) clearly shows that the bifurcating mode is located in the main bulk flow near the closed streamlines of both bubbles.

\subsection{\Twod\ instability}\label{subsect:2D_instability}
\begin{figure}
\centering
\includegraphics[width=0.6\textwidth]{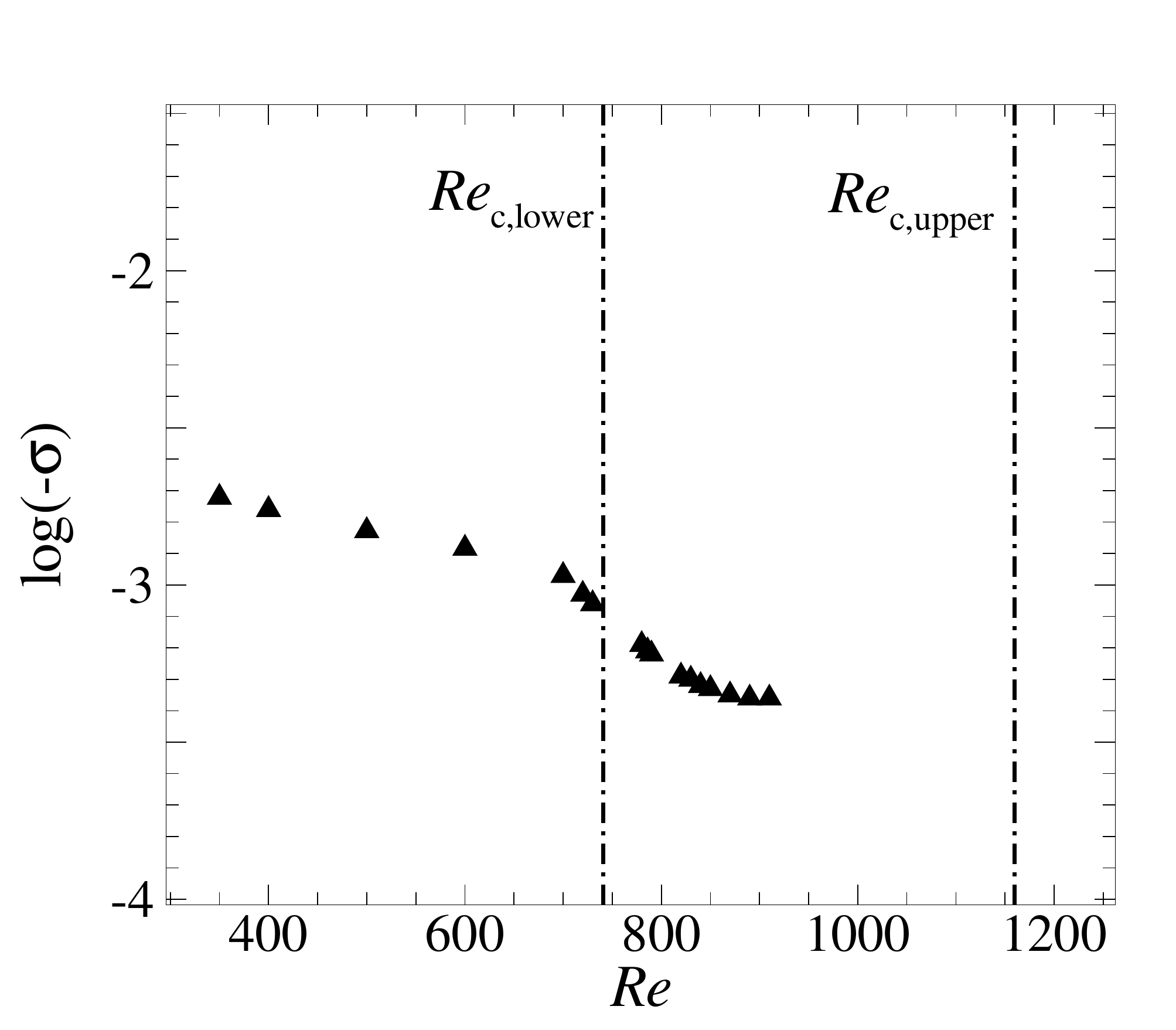} \\
\caption{Leading \twod\ ($k=0$) eigenvalues for  $\beta=1$. Due to the hysteresis, there are two $\Rey_c$ for this flow, $\Rey_{\textrm{c},\textrm{lower}}$  is the lowest $\Rey$ the flow can become unsteady, meanwhile $\Rey_{\textrm{c},\textrm{upper}}$ is the highest $\Rey$ the flow remains steady. The leading eigenvalues remain to have finite value of log$(-\sigma)$ as $\Rey\rightarrow\Rey_\mathrm{c}$. Note that the flow will become linearly neutrally stable when $\sigma\rightarrow0^-$, $\log(-\sigma)\rightarrow -\infty$.}
\label{fig:2D_eigen}
\end{figure}
\S~\ref{subsect:stab_curve} describes how for $\Rey < \Rey_\mathrm{c}$ at any $\beta$, the flow is stable to \twod\ infinitesimal perturbations ($k=0$). The linear stability analysis found that all eigenvalues for $k=0$ are real and have negative growth rate. The \twod\ flow around a sharp 180-degree bend becomes unsteady at small $\Rey$ depending on $\beta$, which is almost twice the critical value for the onset of \threed\ instability. From \twod\ simulations, \citet{zhang2013influence} observed that \twod\ instability starts in the shear layer between the two steady recirculation bubbles and sheds throughout the whole width of the channel. This agrees with the mechanism found for the \twod\ leading eigenmode. As $k\rightarrow 0$, a leading real eigenmode splits into two complex-conjugate pairs. Figure~\ref{fig:2D_eigen} is plotted similarly to figure~13 in \citet{barkley2002three}; $\log(-\sigma)$ will approach $-\infty$ as $\sigma\rightarrow 0$ from below. However, the data here shows that the eigenmode growth rate remains negative as the critical Reynolds number is approached, lacking any evidence of a departure towards $-\infty$. 
While it was not possible due to compute time limitations to obtain stability data closer to $\Rey_{c,\mathrm{upper}}$, this nevertheless suggests that the transition to unsteadiness of the steady-state \twod\ solution branch is not due to a global linear instability. This is supported by our earlier observation that $\Rey_{c,\mathrm{upper}}$ was resolution-sensitive. Similar observations to that shown in figure~\ref{fig:2D_eigen} were made for flow behind a backward-facing step \citep{barkley2002three} and through a partially blocked channel \citep{griffith2007wake}. The data exhibits a kink at $\Rey\approx 700$, were a sudden steepening in gradient is observed, before the data reverts to a nearly horizontal trend to higher Reynolds numbers. \cite{barkley2002three} observed a similar behaviour for the backward-facing step flow (in that case occurring at $\Rey\approx 1250$), and found by inspection of secondary eigenvalues that the kink occurred due to an ``avoided crossing'' of the two eigenmode branches.

\section{Non-linear analysis of the bifurcation to \threed\ state}\label{sect:3D_flow}
In this section, \threed\ direct numerical simulation (DNS) is performed to assess the linear stability analysis predictions and to understand the nature of the bifurcation arising from the predicted linear instability modes. The \threed\ algorithm exploits the spanwise homogeneity of the geometry, combining the \twod\ spectral-element discretisation in the $x$-$y$ plane with a \Fourier\ spectral method in the out-of-plane $z$-direction \cite[for more details see][]{RyanButlerSheard2012, SheardFitzgeraldRyan2009}. The span of the domain in $z$ may be specified, and periodic boundary conditions are naturally enforced in the $z$-direction.

Tests were conducted to determine the dependence of computed \threed\ solutions on the number of Fourier modes $N_f$ included in the simulations. In these tests, the spanwise wavenumber was selected to match a predicted linear instability mode above the critical Reynolds number, and a superposition of the \twod\ base flow and \threed\ eigenvector field of the predicted linear instability was used as an initial condition. The \threed\ flow was then evolved in time until it saturated, at which point measurements of the domain integral of $|w|$ and a point-measurement of $w$-velocity were taken. Results are shown in table~\ref{tab:DNS-Fourier-resolution}, demonstrating that the solution having $8$ Fourier modes is converged to within at least $2$ and $3$ significant figures to the result obtained with $16$ modes. This is deemed sufficient to capture the non-linear growth behaviour and the saturated state of the mode, so $8$ Fourier modes is employed hereafter.
\begin{table}
\centering\begin{tabular}{ccccc}
$N_f$ & $\int_\Omega\left|w\right|\,\mathrm{d}\Omega$ & $\epsilon_1$ & Point $w$-velocity & $\epsilon_2$ \\[6pt]
$2$   & $0.343542421$               & $2.01\%$     & $-0.044052007$ & $42.4\%$ \\
$4$   & $0.350601512$               & $1.91\%$     & $-0.030932087$ & $7.52\%$ \\
$8$   & $0.344016047$               & $1.11\%$     & $-0.028768317$ & $0.04\%$ \\
$16$  & $0.347868534$               & ---          & $-0.028780194$ & ---
\end{tabular}
\caption{Convergence of the saturated \threed\ DNS solution with number of Fourier modes included in the simulation ($N_f$) for a test case having $\Rey = 400$, $\beta=0.8$ and $k=2.0$. Percent differences between successive $\int_\Omega\left|w\right|\,\mathrm{d}\Omega$ and point $w$-velocity measurements are $\epsilon_1$ and $\epsilon_2$, respectively.}
\label{tab:DNS-Fourier-resolution}
\end{table}

\subsection{Non-linear evolution of the unstable modes}
\begin{figure}
\centering
    \begin{tabular}{ll}
    (a) & (b) \\
    \multicolumn{1}{c}{\includegraphics[width=0.48\textwidth,keepaspectratio]{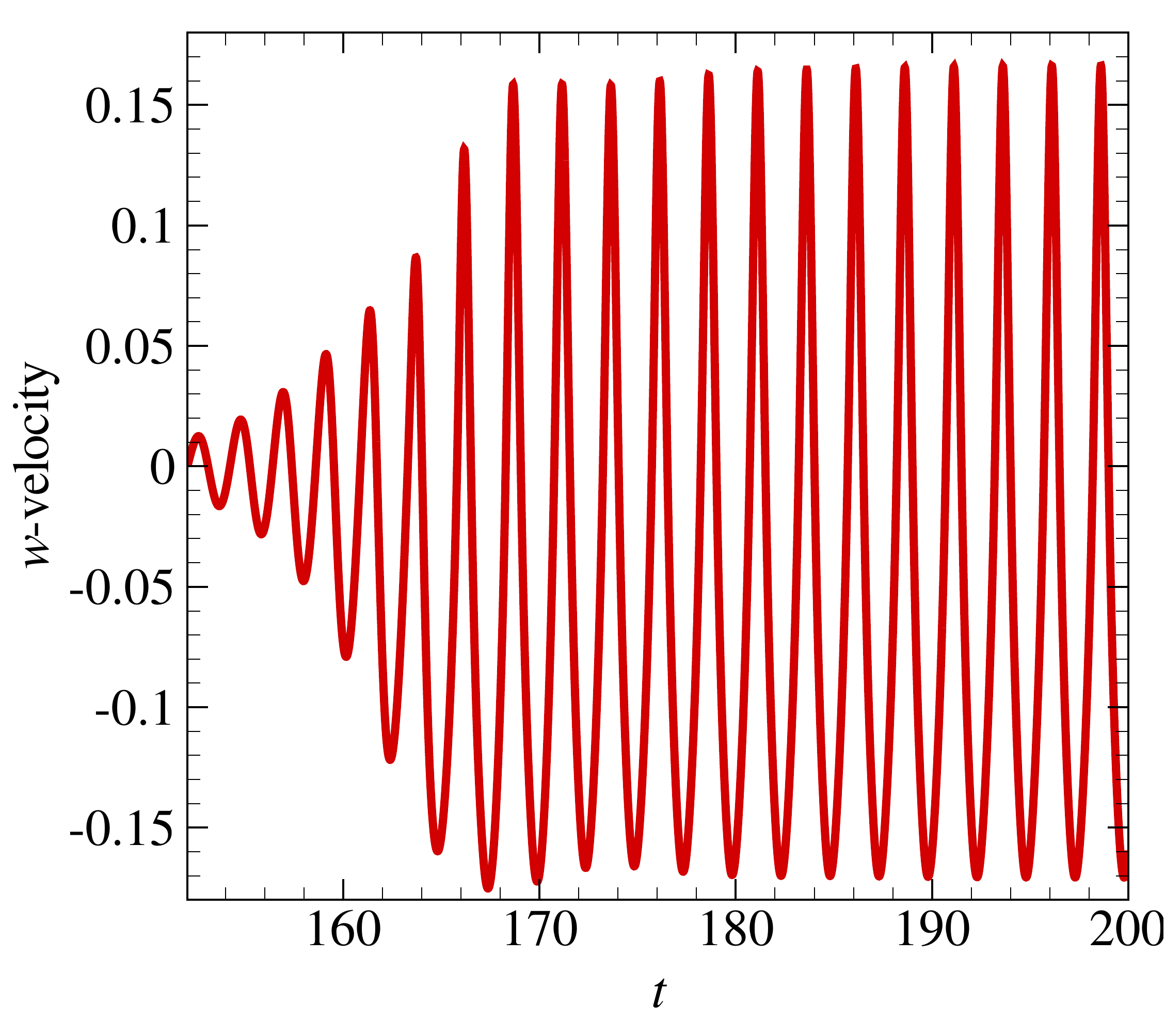}} &
    \multicolumn{1}{c}{\includegraphics[width=0.48\textwidth,keepaspectratio]{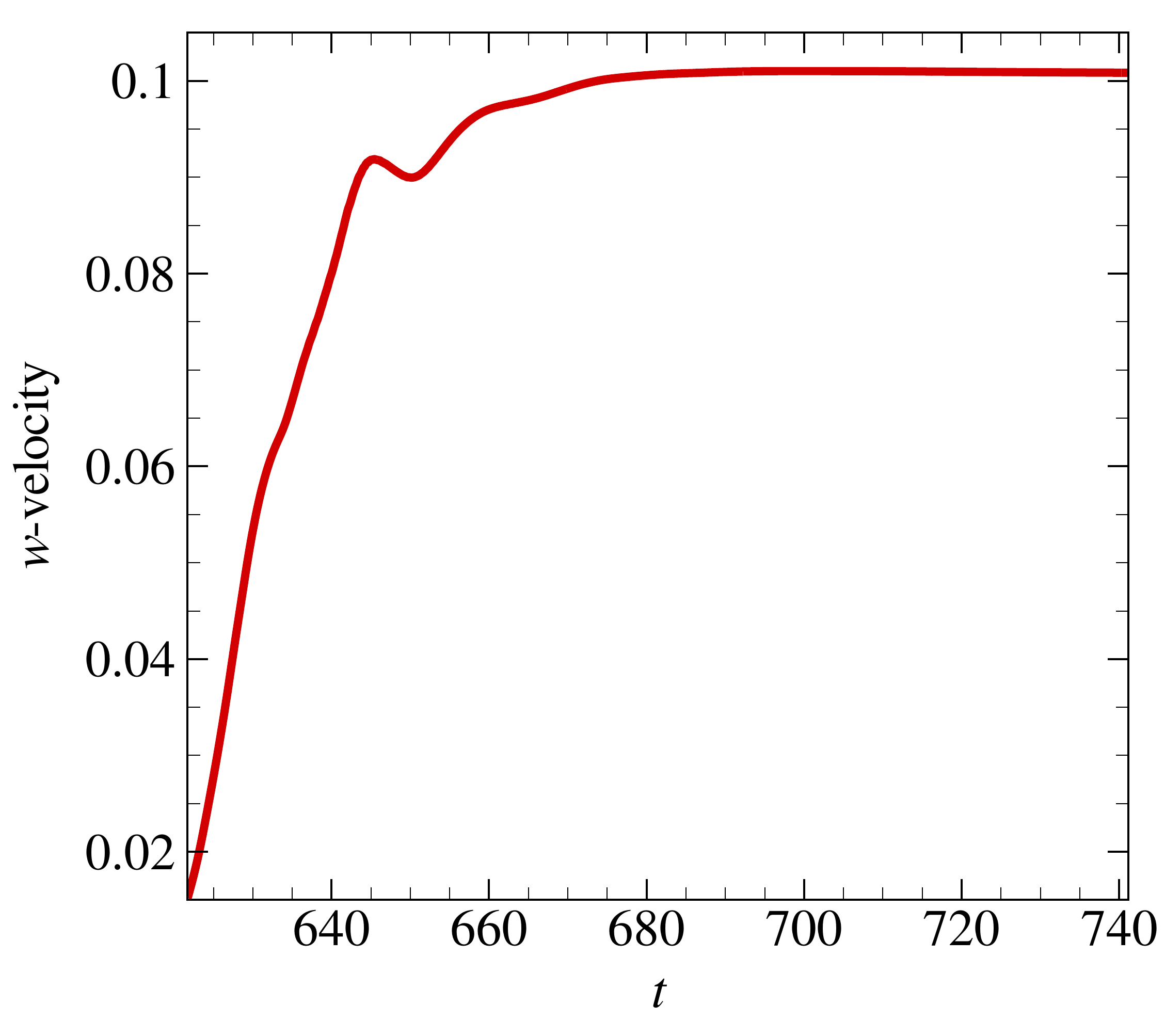}} \\
    \end{tabular}
  \caption{Time histories of $w$-velocity measured at a local point $(x,y,z)=(-2,0.52,1)$. Because $w$ is zero in the underlying \twod\ base flow, non-zero $w$ is an indicator for \threed\ flow development. In (a) $\beta=0.2$, $\Rey=160$ and the \threed\ spanwise domain wavenumber is $k=4$. In (b) $\beta=1$, $\Rey=600$ and $k=4.5$.}
\label{fig:2D_to_3D}
\end{figure}
\begin{table}
\centering
\begin{tabular}{cccccc}
$\beta$ &   $\Rey/\Rey_{\mathrm{3D}}$  &   $k$ &   $\sigma$(LSA) & $\sigma$(3D DNS) & Percentage difference \\[6pt]
  0.2   &   1.12     &   3   &   0.06648 &   0.06667 & 0.27$\%$    \\
  0.4   &   1.16     &   3.2 &   0.09877 &   0.09909   & 0.3$\%$    \\
  0.5   &   1.22     &   2   &   0.03131 &   0.03133   & 0.011$\%$   \\
  0.8   &   1.03     &   2   &   0.01454 &   0.01455   & 0.043$\%$   \\
  2     &   1.55     &   4.5 &   0.20032 &   0.20102 &   0.35262$\%$ \\
\end{tabular}
\caption{Comparison between the growth rates calculated from linear stability analysis (LSA) and \threed\ DNS simulations.} \label{tab:LSA_vs_3D}
\end{table}
\begin{figure}
\centering
    \begin{tabular}{ll}
    (a) $\beta=0.2, \Rey=160, k=4$\\
    (i) & (ii) \\
    \multicolumn{1}{c}{\includegraphics[width=0.48\textwidth,keepaspectratio]{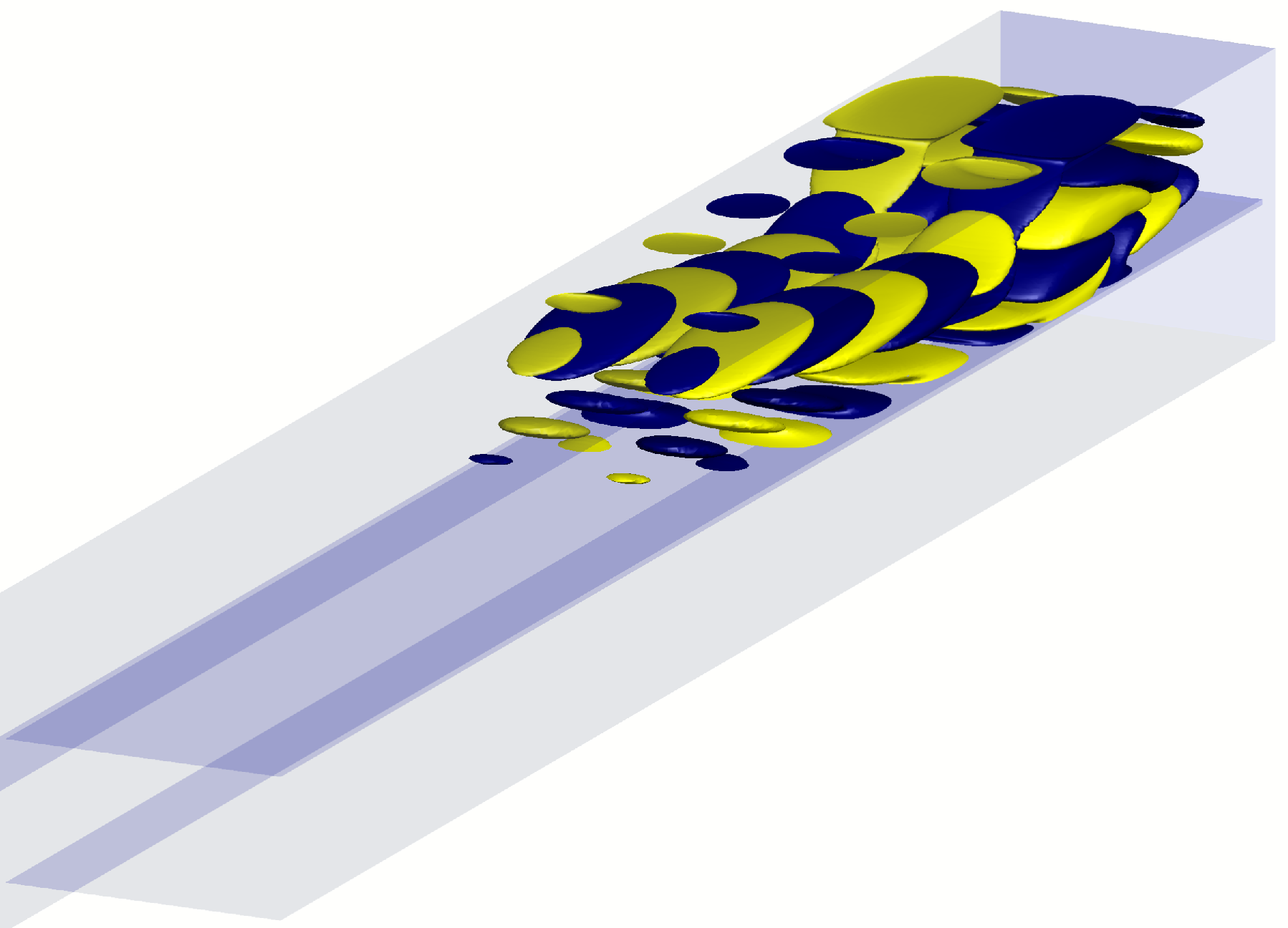}} &
    \multicolumn{1}{c}{\includegraphics[width=0.48\textwidth,keepaspectratio]{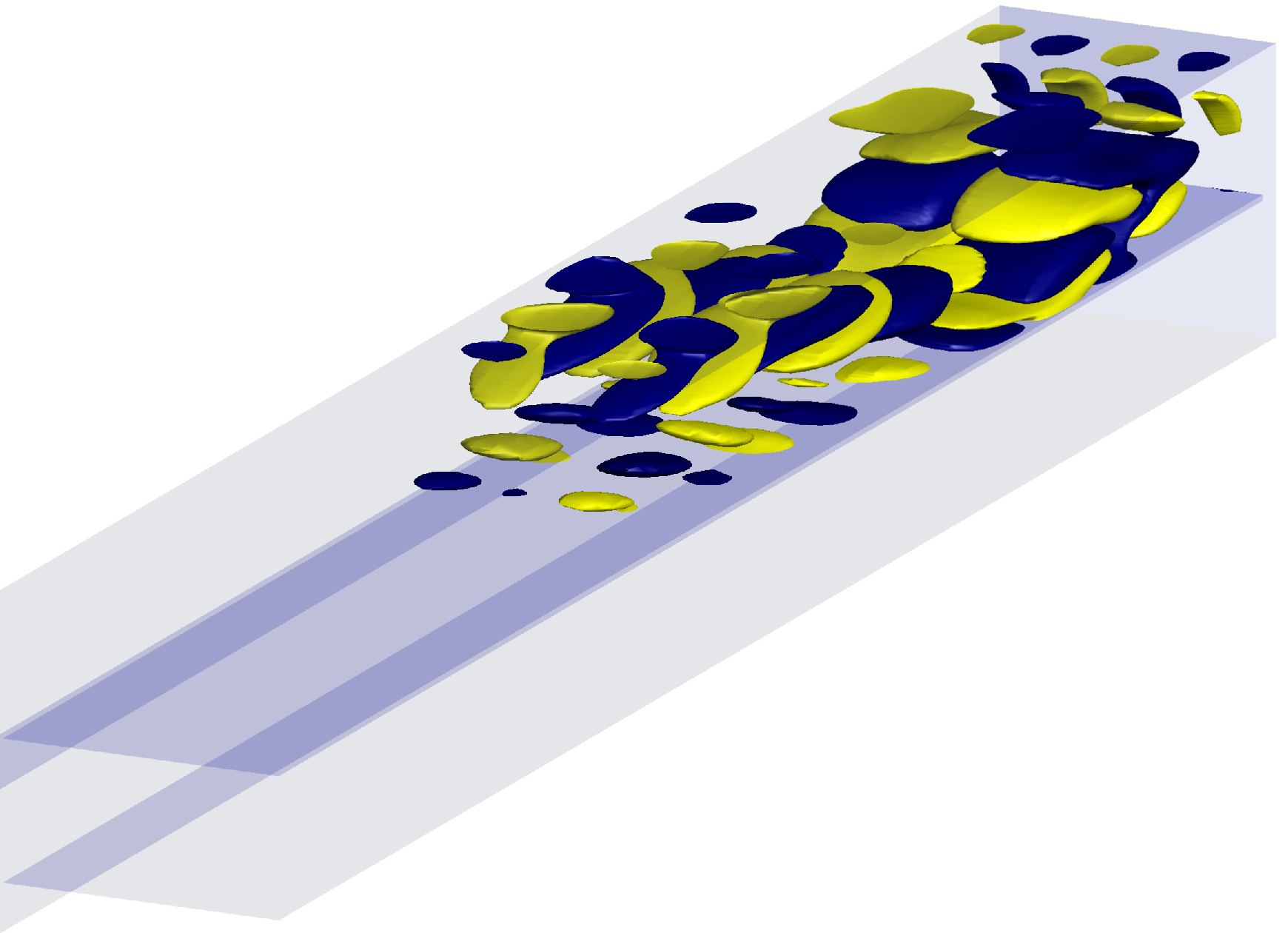}} \\
    (b) $\beta=1, \Rey=600, k=4.5$\\
    (i) & (ii) \\
    \multicolumn{1}{c}{\includegraphics[width=0.48\textwidth,keepaspectratio]{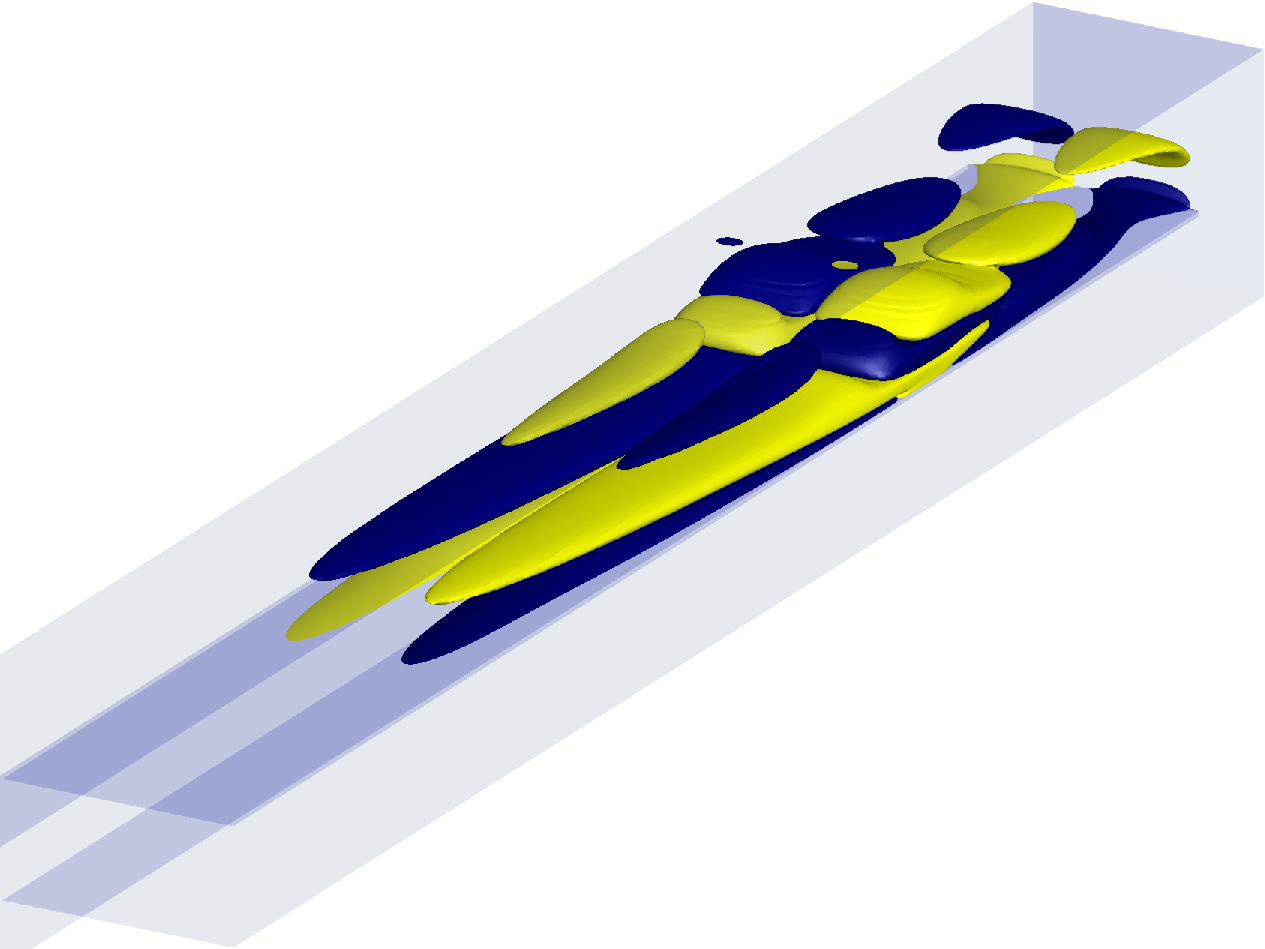}} &
    \multicolumn{1}{c}{\includegraphics[width=0.48\textwidth,keepaspectratio]{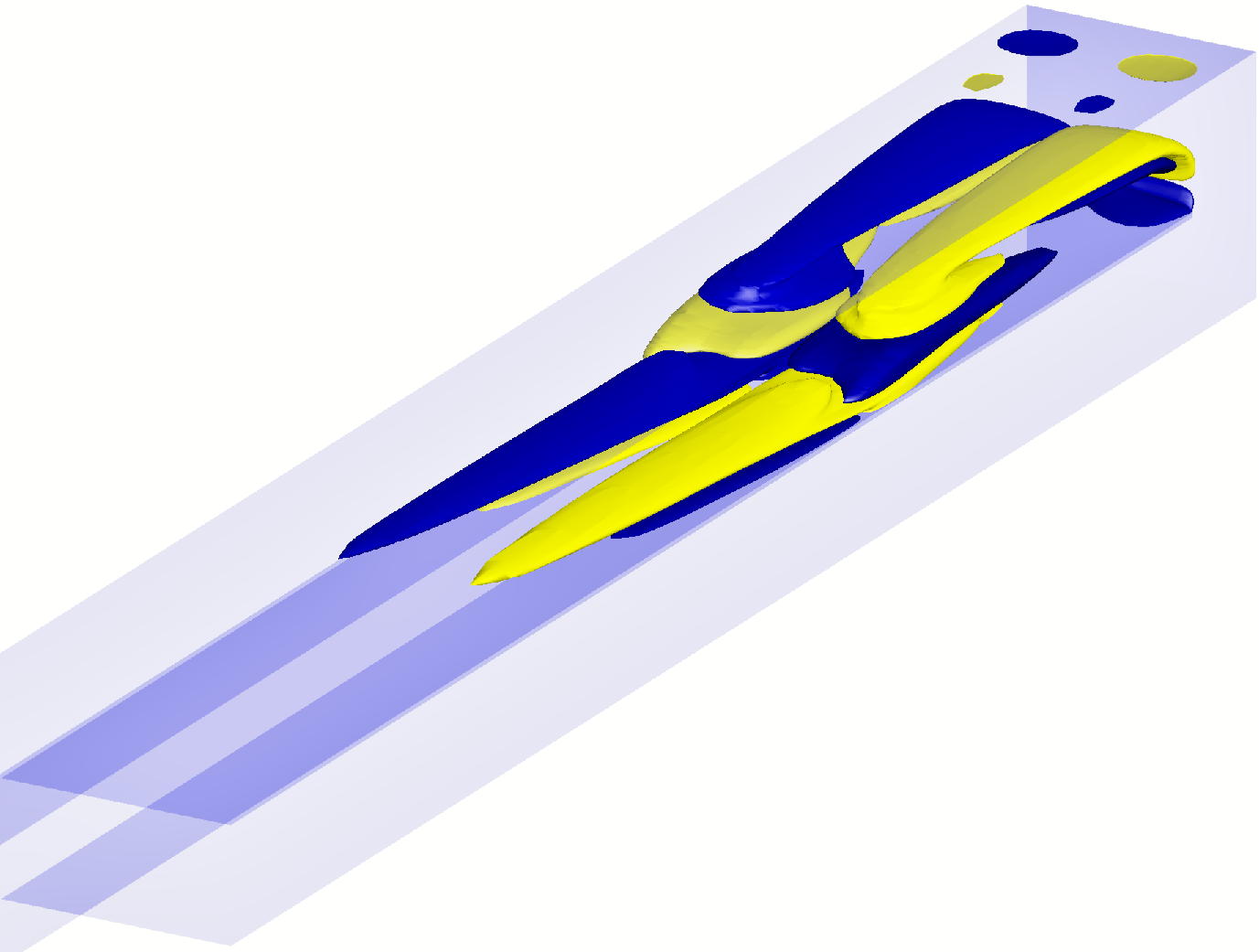}} \\
  \end{tabular}
  \caption{Visualisation of the \threed\ disturbances via iso-surface plots of the (streamwise) $x$-component of vorticity for (a) $\beta=0.2, \Rey=160, k=4$ and (b) $\beta=1, \Rey=600, k=4.5$. (i) shows the leading eigenmode predicted by the linear stability analysis, and (ii) shows the saturated state of a \threed\ DNS simulation. The saturated solution in (a ii) is oscillatory, and in (b ii) is steady-state.}
\label{fig:LSA_and_3D}
\end{figure}
\Threed\ simulations were subsequently performed at selected $\beta$ and $\Rey$ combinations. The spanwise wavenumber in each case is deliberately set to match the corresponding linear instability eigenmode. It is acknowledged that this choice excludes long-wavelength features that may or may not arise. However, it facilitates an isolation of the instability mode under scrutiny. Figure~\ref{fig:2D_to_3D} shows the time history of the spanwise velocity in these \threed\ simulations for (a) $\beta=0.2$, $\Rey=160$, $k=4$ and (b) $\beta=1$, $\Rey=600$, $k=4.5$. The oscillatory and the synchronous behavior in figure~\ref{fig:2D_to_3D}(a) and~(b), respectively, agree well with the behavior of the leading eigenmodes predicted using linear stability analysis (ref.~figure~\ref{fig:lsa_curve}).
From the time history of $w$-velocity, the growth rate of the perturbation can be calculated. Table \ref{tab:LSA_vs_3D} shows strong agreement between the growth rates of perturbation field obtained from the linear stability analysis and \threed\ DNS simulation for five different cases. $\beta=2$ and $k=4.5$ is chosen to demonstrate the accuracy of the predictions on flow with fast growing perturbations. Figure \ref{fig:LSA_and_3D} shows \threed\ isosurface plots of streamwise vorticity for the same parameters as in figure~\ref{fig:2D_to_3D}, respectively, comparing the predicted \threed\ eigenmode with the actual \threed\ state produced once the flow saturates following instability growth. The streamwise vorticity from the linear stability analysis (figure~\ref{fig:LSA_and_3D}(a-b)(i)) has a strong resemblance to those of the \threed\ DNS simulations (figure~\ref{fig:LSA_and_3D}(a-b)(ii)). The strong agreement seen between the predicted eigenmode structure and the resulting saturated \threed\ structure verifies that the linear stability analysis provides meaningful predictions of the \threed\ nature of the flow. The Reynolds numbers in figure \ref{fig:LSA_and_3D}(a) and (b) are $28\%$ and $51\%$ higher than the critical Reynolds numbers for $\beta=0.2$ and $1$, respectively. Both cases produce non-zero \Fourier\ mode energy at saturation in the magnitude of $10^{-2}$ relative to the base flow energy, which is very small. The smaller the disturbance energy compared to the base flow energy, the closer the saturated state will be to the predicted infinitesimal eigenmode because the contribution of nonlinear terms is weaker.

From the \threed\ DNS simulations, we found that the unsteady saturated state for $\beta=0.2$ persists to at least $\Rey=160$ \threed\ simulations were not conducted beyond $\Rey=160$, though it might be anticipated based on the close agreement between the saturated \threed\ flow structure and the corresponding predicted linear instability mode, that a similar behaviour would extend to higher Reynolds numbers: linear stability analysis was performed up to $\Rey=200$, continuing to capture this mode. Meanwhile, for $\beta=0.5$, as $\Rey$ increases, the saturated state changed from a steady at $\Rey=300$ and $400$ to an unsteady saturated state at $\Rey=500$ with $k=4.5$. The same observation is predicted in figure~\ref{fig:lsa_curve}, demonstrating the applicability of linear stability analysis to this flow, and the rich tapestry of flow regimes across the $\Rey$--$\beta$ parameter space.

\subsection{Stuart--Landau model analysis}
In this subsection, analysis of the nonlinear features of the instability mode evolution is performed using a truncated Stuart--Landau equation. The Stuart--Landau model is valid in the vicinity of the transition Reynolds number, and has found wide application for classification of the non-linear characteristics of bifurcations in fluid flows. Examples include analysis of the Hopf bifurcation from steady-state flow past a circular cylinder producing the classical K{\'a}rm{\'a}n vortex street \cite[][]{ProvMathBoyer1987, DusekLeGalFraunie1994, SchummBergerMonkewitz1994, AlbaredeProvansal1995, thompson2004stuart}, the regular (steady-to-steady) bifurcation breaking axisymmetry in the flow behind a sphere \cite[][]{thompson2001physical}, and three-dimensional transition behind a cylinder \cite[][]{henderson1996secondary, sheard2003coupled}, staggered cylinders \cite[][]{CarmoSherwinBearmanWillden2008jfm-staggered-cyl}, and rings \cite[][]{SheardThompsonHourigan2004, SheardThompsonHourigan2004ejmb}. The model describes the growth and saturation of perturbation as \citep{landau1976mechanics}
\BEQ\label{eq:complex-SL-eqn}
\frac{\mathrm{d}A}{\mathrm{d}t} = (\sigma+\mathrm{i}\omega)A-l(1+\mathrm{i}c)|A|^2 A+\dots,
\EEQ
where $A$ is the complex amplitude of the evolving instability as a function of time, and the right side of the equation represents the first two terms of a series expansion. The growth rate and angular frequency of the mode in the linear regime ($|A|\rightarrow 0$) are respectively denoted by $\sigma$ and $\omega$, while weakly non-linear properties are determined by the second term on the right hand side. The sign of $l$ dictates whether the mode evolution is via a supercritical ($l>0$) or subcritical ($l<0$) bifurcation, and any frequency shift is described by Landau constant $c$ \cite[][]{DusekLeGalFraunie1994, LeGalNadimThompson2001, thompson2001physical}. A common treatment \cite[][]{LeGalNadimThompson2001, thompson2001physical} is to decompose $A(t)$ into magnitude and phase components as
\BEQ\label{eq:SL-A-decompose}
A(t) = \rho(t)\exp{i\phi(t)},
\EEQ
where $\rho(t)=\left|A(t)\right|$ and $\phi(t)=\arg(A(t))$. Substitution of equation~(\ref{eq:SL-A-decompose}) into~(\ref{eq:complex-SL-eqn}), separation into real and imaginary parts, and simplification yields separate equations for amplitude and phase, \ie
\BER
\label{eq:SL-amplitude_eqn}\frac{\mathrm{d}\rho}{\mathrm{d}t} = \sigma\rho-l\rho^3,\\
\frac{\mathrm{d}\phi}{\mathrm{d}t} = \omega-lc\rho^2.
\EER
It is convenient to then manipulate (\ref{eq:SL-amplitude_eqn}) as
\BEQ
\frac{\mathrm{d}\!\left(\log\rho\right)}{\mathrm{d}t} = \frac{1}{\rho}\frac{\mathrm{d}\rho}{\mathrm{d}t} = \sigma - l\rho^2.
\EEQ
Hence a positive slope ($-l$) in a plot of $\mathrm{d}\!\left(\log{|A|}\right)/\mathrm{d}t$ against $|A|^2$ will indicate a subcritical bifurcation, while a negative slope corresponds to a supercritical bifurcation. Whether a supercritical bifurcation is of a pitchfork or Hopf type is dependent on whether the growing mode is synchronous or oscillatory.

\begin{figure}
\centering
    \begin{tabular}{ll}
    (a) & (b) \\
    \multicolumn{1}{c}{\includegraphics[width=0.48\textwidth,keepaspectratio]{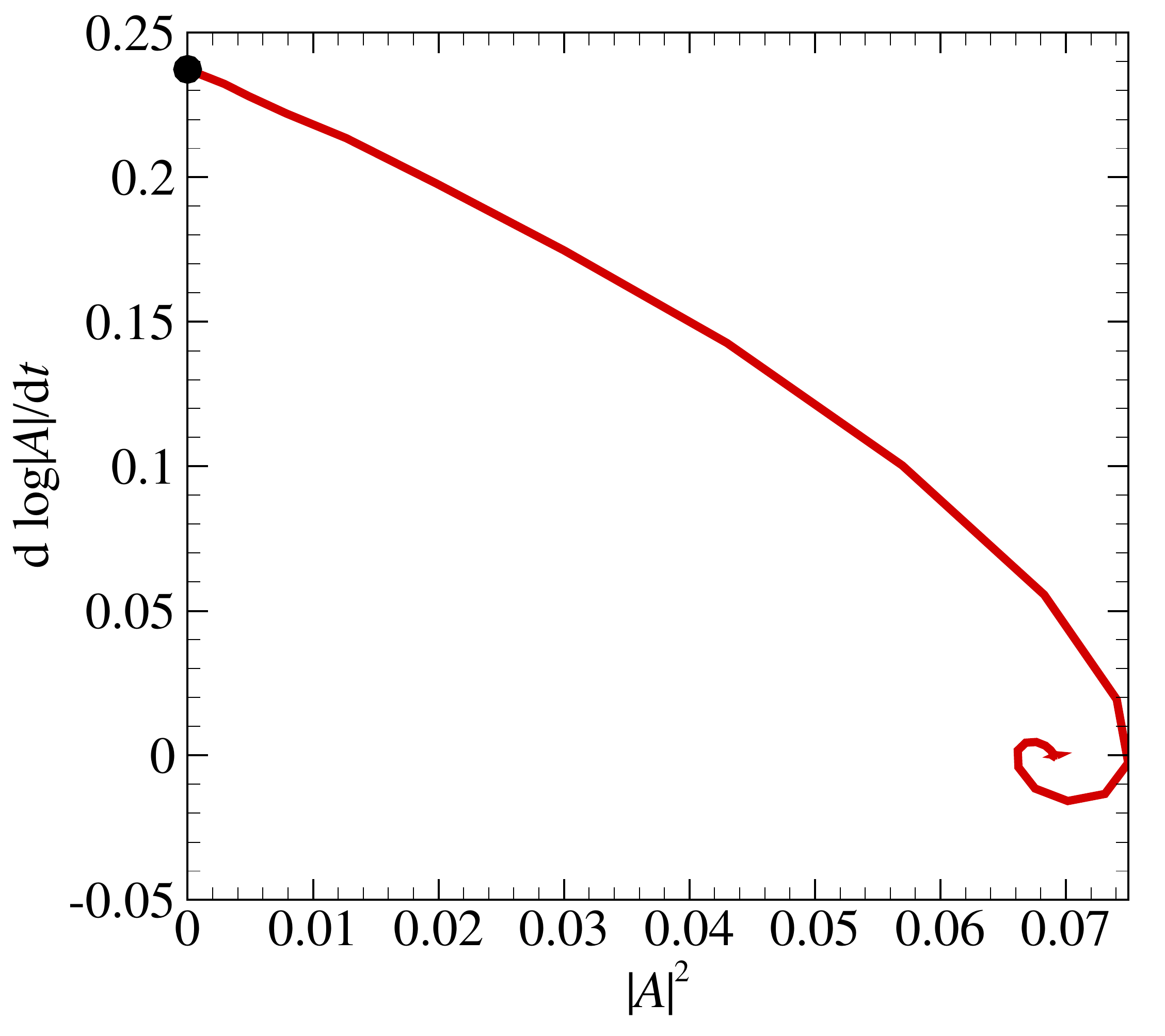}} &
    \multicolumn{1}{c}{\includegraphics[width=0.48\textwidth,keepaspectratio]{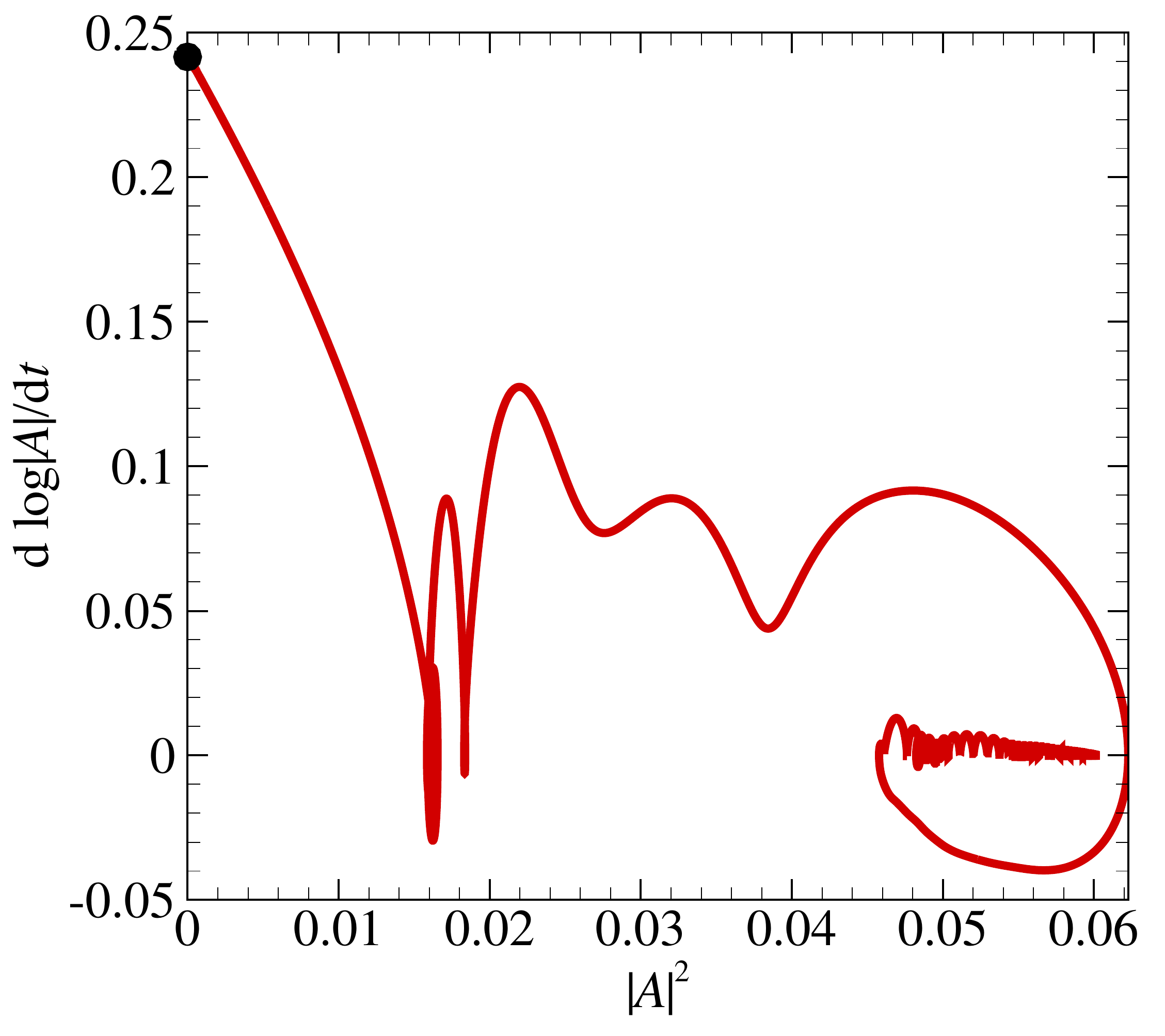}} \\
    \end{tabular}
  \caption{The time derivative of mode amplitude logarithm plotted against the square of the amplitude for (a) $\beta=0.2$, $\Rey=160$, $k=4$ and (b) $\beta=0.5$, $\Rey=400$, $k=3.5$ demonstrating supercritical behaviour. The solid circle symbol represents the linear stability analysis predicted growth rate.}
\label{fig:dlogA_A}
\end{figure}
The envelope of energy in the non-zero spanwise Fourier modes of the \threed\ simulations was taken as a global amplitude measure ($|A|^2$) of the growing three-dimensional instabilities. Figure~\ref{fig:dlogA_A} plots the time derivative of amplitude logarithm against the square of the amplitude for two different cases: (a) $\beta=0.2$, $\Rey=160$, $k=4$, which grows from an oscillatory mode, and (b) $\beta=0.5$, $\Rey=400$, $k=3.5$, which grows from a synchronous mode. The behaviour shown for $\beta=0.5$ is consistent with that found at larger $\beta$. The nearly linear variation with a negative gradient towards small $|A|^2$ shown in both plots indicates that the transition in both cases occurs through a supercritical bifurcation. This is consistent with other confined flow featuring recirculation bubbles such as the flow past a backward facing step \citep{kaiktsis1991onset}, through a sudden expansion in a circular pipe \citep{mullin2009bifurcation}, and past a sphere \citep{tomboulides2000numerical}. The oscillatory and synchronous behaviour of the modes in (a) and (b) reveal them to occur through supercritical Hopf and pitchfork bifurcations, respectively.

In addition to the aforementioned sub- and supercritical bifurcation scenarios, another possibility is that of a transcritical bifurcation. A dynamical system producing transcritical behaviour takes the form
\BEQ
\frac{\mathrm{d}\rho}{\mathrm{d}t} = \sigma \rho - l \rho^2,
\EEQ
which differs from the amplitude part of the Stuart--Landau model (\ref{eq:SL-amplitude_eqn}) by the replacement of $\rho^3$ with $\rho^2$ in the non-linear term on the right hand side. Under an analogous manipulation,
\BEQ
\frac{\mathrm{d}\!\left(\log{\rho}\right)}{\mathrm{d}t} = \sigma - l \rho = \sigma - l \sqrt{\rho^2}.
\EEQ
The $\sqrt{\rho^2}$ term indicates that an infinite gradient would present at $|A|^2=0$ in a plot of $\mathrm{d}\!\left(\log{|A|}\right)/\mathrm{d}t$ against $|A|^2$. The absence of such a behaviour in figure~\ref{fig:dlogA_A} rules out a transcritical bifurcation. Similarly, a transcritical dynamical system expressed in terms of the complex amplitude $A$,
\BEQ
\frac{\mathrm{d}A}{\mathrm{d}t} = \left(\sigma+i\omega\right)A - l\left(1+ic\right)A^2,
\EEQ
features a real component that simplifies to
\BEQ
\frac{\mathrm{d}\!\left(\log{\rho}\right)}{\mathrm{d}t} = \sigma - l \sqrt{\rho^2}\left[\cos{\phi}-c\sin{\phi}\right].
\EEQ
In the limit $\left|A\right|=\rho\rightarrow 0$, the oscillation described by the trigonometric term produces gradients in $\mathrm{d}\!\left(\log{|A|}\right)/\mathrm{d}t$ as a function of $|A|^2$ that approach infinity; the absence of this behaviour in figure~\ref{fig:dlogA_A} again supports the present classification of these bifurcations as being supercritical.

\section{Conclusions}
We conducted a linear stability analysis to characterise the onset of unsteadiness in the flow around a 180-degree sharp bend. We  considered a range of opening ratios $\beta$ spanning all regimes from a jet-like flow through a small aperture to flow topologies involving a recirculation within the turning part \citep{zhang2013influence}. In all cases, we found the base flow with steady bubble in the outlet to be unstable to infinitesimal perturbations at a finite critical Reynolds number $\Rey_\mathrm{c}$, that increases monotonically with the effective opening ratio $\beta_\textrm{eff}$. $\beta_\textrm{eff}$ measures the actual width of the main stream in the turning part. Consequently $\Rey_\mathrm{c}$ increases monotonically with the geometric opening ratio $\beta $ so long as the mean stream occupies the whole turning part (up to $\beta\sim1$). By contrast, for high opening ratios, $\Rey_c(\beta)$ decreases asymptotically to $\Rey_c(\beta_\textrm{eff}\simeq0.8)$, because the recirculation in the turning part reduce the effective width available to the main stream.

The linear stability analysis revealed three types of leading eigenmodes. In the subcritical range, the leading eigenmode is reminiscent of long-wave Taylor--G\"ortler vortices localised in the main stream between the two recirculation bubbles attached to either outlet walls. This mode was, however, never found to become unstable. As $\Rey$ was increased, two unstable branches emerged that were respectively associated a real eigenvalue and a complex-conjugate pair of eigenvalues. The former dominates for $\beta\geq0.3$. The corresponding perturbation has a spanwise wavenumber $k\simeq2$ and is confined within the first recirculation region, with maximum intensity where the back flow is most intense. For $\beta=0.2$, by contrast unsteadiness sets in via the second branch under the form of a spanwise oscillating mode, akin to that found in backward facing step flows with small opening ratios \citep{lanzerstorfer2012global}.

In all cases, critical modes were \threed. Accordingly, locally unstable \twod\ modes (\emph{\ie\ having zero spanwise wavenumber}) are only found at higher Reynolds numbers than \threed\ modes, but do not grow through a global instability. They drive a Kelvin--Helmholtz instability in the main stream between the two recirculating bubbles that is consistent with the DNS of \cite{zhang2013influence}.

Analysis of the non-linear evolution of dominant instability modes using \threed\ DNS and the Stuart--Landau equation demonstrated that transition from \twod\ to \threed\ flow consistently occurred through a supercritical bifurcation: a supercritical Hopf bifurcation at small $\beta$ and a supercritical pitchfork bifurcation at large $\beta$.\\

A.\,M.\,S.\ is supported by the Ministry of Education Malaysia and International Islamic University Malaysia. This research was supported by Discovery Grants DP120100153 and DP150102920 from the Australian Research Council, and was undertaken with the assistance of resources from the National Computational Infrastructure (NCI), which is supported by the Australian Government. A.\,P.\ acknowledges support from the Royal Society under the Wolfson Research Merit Award Scheme (grant WM140032).

\end{document}